\documentclass[a4paper,fleqn,usenatbib]{mnras}

\usepackage{newtxtext,newtxmath}
\usepackage[T1]{fontenc}
\usepackage{ae,aecompl}

\usepackage{graphicx}	% Including figure files
\usepackage{amsmath}	% Advanced maths commands
\usepackage{amssymb}	% Extra maths symbols
\usepackage{supertabular}
\usepackage[caption=false]{subfig}
\def\Lya{\mbox{Ly\,$\alpha$ }}

\newcommand{\gold}{{\it gold}}
\newcommand{\silver}{{\it silver}}
\newcommand{\bronze}{{\it bronze}}
\newcommand{\copper}{{\it copper}}

\newcommand{\RC}[1]{ #1}

\title[Strong lensing analysis on Abell 2744 - MUSE]{Strong lensing analysis of Abell 2744 with MUSE and Hubble Frontier Fields images.}

\author[G. Mahler et al.]{G. Mahler,$^{1}$\thanks{E-mail: guillaume.mahler@univ-lyon1.fr }
J. Richard,$^{1}$, 
 B. Cl\'ement$^{1}$, 
D. Lagattuta$^{1}$, 
K. Schmidt$^{2}$, 
V. Patr\'icio$^{1}$, 
\newauthor
G. Soucail$^{3}$, 
R. Bacon$^{1}$, 
R. Pello$^{3}$,
R. Bouwens$^{4}$, 
M. Maseda$^{4}$,
J. Martinez$^{1}$,
\newauthor
M. Carollo$^{5}$,
H. Inami$^{1}$,
F. Leclercq$^{1}$, 
L. Wisotzki$^{2}$
\\
$^{1}$Univ Lyon, Univ Lyon1, Ens de Lyon, CNRS, Centre de Recherche Astrophysique
de Lyon UMR5574, F-69230, Saint-Genis-Laval, France\\
$^{2}$AIP, Leibniz-Institut f\"ur Astrophysik Potsdam (AIP) An der Sternwarte 16, D-14482 Potsdam, Germany\\
$^{3}$IRAP (Institut de Recherche en Astrophysique et Plan\'etologie), Universit\'e de Toulouse, CNRS, UPS, Toulouse, France\\ 
$^{4}$Leiden Observatory, Leiden University, P.O. Box 9513, 2300
RA, Leiden, The Netherlands\\
$^{5}$ETH Zurich, Institute of Astronomy, Wolfgang-Pauli-Str. 27, CH-8093 Zurich, Switzerland
}	

\date{Accepted XXX. Received YYY; in original form ZZZ}

\pubyear{2017}

\begin{document}
\label{firstpage}
\pagerange{\pageref{firstpage}--\pageref{lastpage}}
\maketitle

% Abstract of the paper

\begin{abstract}
We present an analysis of MUSE observations obtained on the massive Frontier Fields cluster Abell 2744. This new dataset covers the entire multiply-imaged region around the cluster core. \RC{The combined catalog consists of 514 spectroscopic redshifts (with 414 new identifications). We use this redshift information to perform a strong-lensing analysis revising multiple images previously found in the deep Frontier Field images, and add three new MUSE-detected multiply-imaged systems with no obvious HST counterpart. The combined strong lensing constraints include a total of 60 systems producing 188 images altogether, out of which 29 systems and 83 images are spectroscopically confirmed, making Abell 2744 one of the most well-constrained clusters to date.
Thanks to the large amount of spectroscopic redshifts we model the influence of substructures at larger radii, using a parametrisation including two cluster-scale components in the cluster core and several group-scale in the outskirts. The resulting model accurately reproduces all the spectroscopic multiple systems, reaching an rms of 0.67\arcsec\ in the image plane.  
The large number of MUSE spectroscopic redshifts gives us a robust model, which we estimate reduces the systematic uncertainty on the 2D mass distribution by up to $\sim2.5$ times the statistical uncertainty in the cluster core. In addition, from a combination of the parametrisation and the set of constraints, we estimate the relative systematic uncertainty to be up to 9\% at 200kpc.}
\end{abstract}
% Select between one and six entries from the list of approved keywords.
% Don't make up new ones.

%%%%%%%%%%%%%%%%%%%%%%%%%%%%%%%%%%%%%%%%%%%%%%%%%%

%%%%%%%%%%%%%%%%% BODY OF PAPER %%%%%%%%%%%%%%%%%%

\section{Introduction}
Cluster of galaxies represent a natural merging process over large scales, and as such gather many valuable observables for our Universe. From a statistical point of view they can constraint various physical processes, such as structure formation or cosmological parameters \citep{Schwinn2016,Jullo2010}. By measuring cluster mass distributions we also gain insight into cluster-specific properties, such as dark matter content (\citealt{Bradac2008}, \citealt{Umetsu2009}). 
Furthermore, offsets between the location of baryonic and dark matter profiles can be used to test the nature of dark matter (e.g., its self-interacting cross section \citealt{Markevitch2004, Harvey2015}).

Strong gravitational lensing precisely measures the enclosed mass of a cluster at a given radius, making it a powerful tool for studying dark and luminous matter. The effect occurs when the curvature of spacetime is large enough near the cluster center to make different light paths from the same source converge on the field of view of the observer.

With the first spectroscopic confirmation of a giant arc in Abell 370 \citep{Soucail1988}, 
the use of the strong lensing effect has evolved into a valuable technique for measuring the total mass of a cluster (both luminous and non-luminous components, e.g. \citealt{Limousin2016}). By refining the mass model of clusters it is possible to calibrate them as cosmic telescopes and quantify the magnification of background sources to study the high-redshift Universe \citep{Coe2013,Atek2014,Alavi2016,Schmidt2016}.

The correct identification of multiply-imaged background sources is crucial to lens modeling because these objects can precisely probe the mass distribution in the cluster core. This requires the high spatial resolution of the Hubble Space Telescope (HST) to ascertain their morphologies and properly match the different lensed images to the same source. By combining observations in multiple HST bands, \citet{Broadhurst2005} were able to identify 30 multiply-imaged systems in the massive cluster Abell 1689 based on similarities in their colours and morphologies. This idea was further pursued in the Cluster Lensing And Supernovae survey with Hubble (CLASH, \citealt{Postman2012}). Using photometry from shallow observations ($\sim$1 orbit) of 25 clusters in 16 bands, \citet{Jouvel2014} finely sampled the Spectral Energy Distribution (SED) of galaxies, obtaining accurate photometric redshifts. In the same set of data, \citet{Zitrin2015} identified from 1 to 10 multiple-image systems per cluster.

More recently, the Hubble Frontiers Field initiative (HFF, \citealt{Lotz2016}) combined very deep HST observations (\textasciitilde180 orbits per target) of six clusters in seven bands. The HFF observed six massive clusters (typical $\sim10^{15}$ $M_\odot$) at $z=0.3-0.6$ selected for their lensing ability. In particular, their capability of strongly magnifying very distant ($z>6$) galaxies. The deep images revealed a remarkable collection of hundreds of multiple images in each of the six clusters observed(\citealt{Lotz2016} \citealt{Jauzac2014}). 

To tackle this wealth of data, several teams have recently engaged in an effort to accurately model the  mass of the cluster cores (e.g., \citealt{Lam2014}, \citealt{Jauzac2014}, \citealt{Diego2016}). Such a large number of multiple images leads to very precise mass estimates: for example, \citet{Jauzac2014,Jauzac2015} obtained $<1\%$ statistical error on the integrated mass at 200 kpc radius in the clusters MACS0416 and Abell 2744, and \citet{Grillo2015} measured $<2\%$ error on the integrated mass at 200 kpc radius of MACS0416. However, the disagreement between models of the same cluster is typically ($\geq$10\%), significantly larger than the statistical uncertainty (see e.g. the mass profiles presented in \citealt{Lagattuta2016}).  Therefore, the next step in further improving the accuracy of the mass estimates is to better understand the sources of systematic uncertainties. While two main drawbacks in strong lensing analysis are the potential use of incorrectly-identified multiple image systems and the lack of redshifts for the sources (used to calibrate the geometrical distance), spectroscopic confirmation of these systems is the best leverage to tackle both issues.

Spectroscopic observations have greatly improved the quality of cluster mass models, as demonstrated by \citet{Limousin2007}, where a large spectroscopic campaign on the cluster Abell 1689 provided redshift measurements for 24 multiple systems and enabled the rejection of incorrect multiple-image candidates in the process. However, multi-object slit spectroscopy is very costly when targeting multiple images in cluster cores due to the small number of objects (typically below 50) that can be targeted in a single observation. 
As demonstrated by \citealt{Grillo2015} in CLASH clusters. Other initiatives such as the Grism Lens-Amplified Survey from Space (GLASS, \citet{GLASS2}, \citet{GLASS1}) offers a valuable alternative to the slit-spectroscopy by observing spectra over the entire image using a grism. The main benefit of slit-less spectroscopy is the blind search for emission in the field of view, but it is limited by low spectral resolution (typically R$\sim$200) and strong overlap of the spectra on the detector.

 A more recent alternative makes use of the Multi Unit Spectroscopic Explorer (MUSE; \citealt{Bacon2010}) instrument on the Very Large Telescope. MUSE is a large integral field spectrograph, providing spectra in the optical range (between 4800 and 9300 \AA) over its entire 1\arcmin$\times$1\arcmin\ field of view using the technology of image slicers. This provides both a large multiplexing capability and a high sensitivity, on top of a good spectral resolution (R$\sim$3000). Not only does MUSE provide an efficient follow-up of faint HST sources in very crowded regions, it also performs very well in the detection of very faint emission lines, especially Lyman $\alpha$ emission at high redshift (\citealt{Bacon2015}, \citealt{Drake2016}, \citealt{Bina2016}). Overall, these capabilities make MUSE an ideal instrument for the spectroscopic follow-up of cluster cores: its field-of-view is well-matched with the size of the multiply-imaged region and it can easily isolate line emission embedded inside the bright continuum emission of cluster members (\citealt{Caminha2016}, \citealt{Karman2016}, \citealt{Jauzac2016b}, \citealt{Grillo2015}).

As part of the MUSE Guaranteed Time Observing (GTO) program on lensing clusters, the powerful combination of MUSE spectroscopy and the lensing efficiency of clusters is used to achieve a number of science goals: to observe the resolved properties of highly-magnified distant galaxies \citep{Patricio2016}, to build reliable mass models  \citep{Richard2015} and challenge the Frontiers Fields modeling with dozens of images \citep{Lagattuta2016}, or to constrain the Lyman $\alpha$ luminosity function at faint luminosities \citep{Bina2016}.

In this paper, we present a MUSE-GTO spectroscopic survey and strong lensing analysis of the HFF cluster Abell 2744 (\citealt{CouchNewell1984,Abell1989}, $\alpha=00^{\rm h} 14^{\rm m} 19.51^{\rm s}$, $\delta=−30^{\rm o} 23\arcmin 19.18\arcsec$, $z=0.308$). This massive ($M(<\rm{1.3\ Mpc})=2.3\pm0.1 \ 10^{15}\ M\protect_{\odot}$, \citealt{Jauzac2016}), X-ray luminous ($L_X=3.1\ 10^{45}$ erg\  s$^{-1}$, \citealt{Allen1998}) merging cluster shows concentrated X-ray emission near its core and extending to the north-west \citep{Owers2011,Eckert2015}. 

 Abell 2744 has been well-studied  for its complex galaxy dynamics \citep{Owers2011}, and its strong lensing properties, both through free-form \citep{Lam2014} and parametric mass modeling \citep{Richard2014,Johnson2014,Jauzac2015}, as well as the combination of strong and weak lensing (\citealt{Merten2011, Jauzac2016}, hereafter J16).  In their weak-lensing analysis, using both the Canada-France-Hawaii Telescope (CFHT) and the Wide Field Imager (WFI) on the MPG/ESO 2.2-m, J16 recently identified several group-scale substructures located $\sim$ 700 kpc from the cluster core, each of them having masses ranging between 5 and 8 $\times 10^{13}$ M$_{\sun}$. Yet, despite the careful attention given to this cluster, it has suffered from a lack of spectroscopic redshifts. The most recent strong-lensing study \citep{Wang2015} used only 7 multiply-imaged sources with spectroscopic redshifts, combined with 18 photometric redshift systems,
 to model the mass of the cluster core. 

The deepest data obtained in the MUSE GTO cluster program covered Abell 2744 with a mosaic totaling an exposure time of 18.5 hours. This deep coverage makes it possible for us to obtain an incredible amount of data over the entire field-of-view (FoV) and even confirm or reject multiply-imaged systems. 
In addition, we supplement this dataset with LRIS observations from Keck.  Using all of this spectroscopic data, we are able to dig deeper into the nature of the cluster and advance our understanding of systematic uncertainties.

The paper is organised as follows. In Section 2 we give an overview of the data. In Section 3 we describe the data processing to compute a redshift catalog. In Section 4 we detail the strong lensing analysis. In Section 5 we summarise the main results of the mass modeling. In section 6 we discuss systematic uncertainties in the analysis, the influence of the outskirts and compare our results with other groups. 
Throughout this paper we adopt a standard $\Lambda$-CDM cosmology with $\Omega_m =0.3$, $\Omega_{\Lambda}=0.7$ and $h=0.7$. All magnitudes are given in the AB system \citep{Oke1974}.

\section{Data description}
\subsection{Hubble Frontier Fields images}
The HFF observations of Abell 2744 (ID: 13495, P.I: J. Lotz) were taken between 2013 Oct 25 and 2014 Jul 1 in 
seven different filters, three with the Advanced Camera for Surveys (ACS; F435W, F606W,
F814W) and four taken with the Wide Field Camera 3 (WFC3; F105W, F125W, F140W, and F160W). In total 280 orbits were devoted to Abell 2744 reaching in each filter a 5-$\sigma$ limiting magnitude AB$\sim$29.
The self-calibrated data provided by STScI\footnote{\url{https://archive.stsci.edu/missions/hlsp/frontier\\
/abell2744/images/hst/}},(version v1.0 for WFC3 and v1.0-epoch2 for ACS) with a pixel size of 60 mas are used in this study.

\subsection{MUSE observations}
\begin{figure}
	\includegraphics[width=\columnwidth]{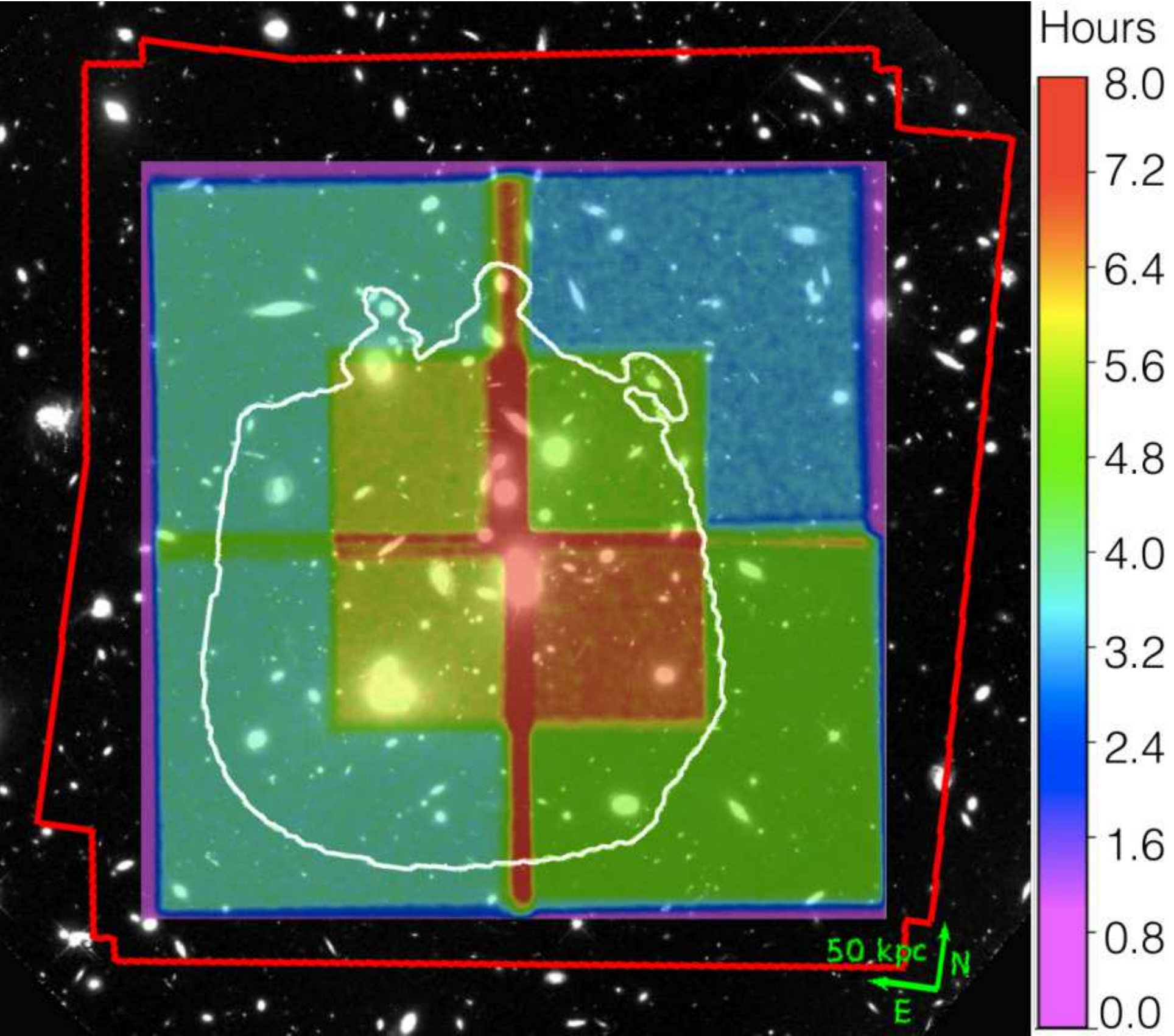}
    \caption{Full MUSE mosaic overlaid on the HFF F814W image. The shaded colour regions highlight our observing strategy, showing the total exposure time devoted to each section of the cluster. The region where multiple images are expected is marked by the white countour, and the red region shows the outline of the HFF WFC3 image mosaic.}
    \label{fig:obsFOV}
\end{figure}

Abell 2744 was observed with the Multi Unit Spectrographic Explorer (MUSE) between September 2014 and October 2015 as part of the GTO Program 094.A-0115 (PI: Richard).
A 2$\times$2 mosaic of MUSE pointings was designed to cover the entire multiple image area, centered at $\alpha = 00^{\rm h}14^{\rm m}20.952^{\rm s}$ 
and $\delta =  -30^{\rm o} 23\arcmin 53.88 \arcsec$. 
The four quadrants were observed for a total of 3.5, 4, 4 and 5 hours, in addition to 2 hours at the center of the cluster. 
Each pointing is split into 30 minutes individual exposures with a 90 degrees rotation applied in between, to minimise the 
striping pattern caused by the IFU image slicers.  Figure \ref{fig:obsFOV} details the MUSE exposure map overlaid on top of an HFF RGB image. The full MUSE mosaic is contained within all 7 HFF bands (ACS and WFC3).

\subsection{MUSE data reduction}

The data reduction was performed with the MUSE ESO pipeline \citep{Weilbacher2012,Weilbacher2014}   up to the mosaic combination. This comprises bias subtraction, flat fielding (including illumination and twilight exposures), sky subtraction, flux calibration and telluric correction. The last two steps were performed with calibration curves derived from the median response of 6 suitable standard stars observed in the MUSE GTO Lensing Clusters program. After basic corrections we align
individual exposures to a common WCS with {\sc SCAMP} \cite{Scamp}, shifting each frame relative to a reference image, in this case, the F814W HFF data. No correction for rotation was applied since only a maximum rotation offset of $0.03^{\circ}$ was observed. We then transform the realigned images into data cubes, resampling all pixels onto a common 3-dimensional grid with two spatial and one spectral axis.

Sky residuals were removed using the Zurich Atmosphere Purge  ({\sc ZAP}; \citealt{Soto2016}), which uses principal component analysis to characterise the residuals and remove them from the cubes.  Objects above a 3$\sigma$ threshold, measured on an empty region on the white light of a previously combined cube, were masked during the process of residual estimation. The individual cubes were then combined in the mosaic using median absolute deviation (MAD) statistics to compare exposures and reject pixels deviating by more than 3 (Gaussian-equivalent) standard deviations. To correct for variations in sky transmittance during the observations, we calculated the average fluxes of bright sources in each cube with {\sc sextractor}. The frame with the highest flux was then taken as a reference to scale individual exposures during combination. The final combined cube was once more cleaned with {\sc ZAP} and the background was corrected by subtracting the median of the 50 spectral-neighbouring wavelength planes (masking bright objects) to each spatial row and column of the cube.

The final product is a 2$'\times2'$ MUSE field of view mosaic with 1.25 \AA\ spectral sampling and 0.2$''$ spatial sampling. The PSF size was estimated by convolving the HST F814W image with a moffat kernel and correlating it with a filter matched MUSE image. We obtained a moffat FWHM of 0.58$''$ in this filter for a $\beta$ parameter of 2.5. Comparing the fluxes of the HST PSF matched image with the MUSE image we estimate that the MUSE photometry is accurate up to $\sim7\%$. These steps were performed using the MUSE Python Data Analysis Framework {\sc mpdaf}\footnote{\url{https://git-cral.univ-lyon1.fr/MUSE/mpdaf.git}} software. 
A final version of the cube is publicly available 
for download\footnote{\url{http://muse-vlt.eu/science/a2744/}}.
 
\subsection{Keck/LRIS spectroscopy}

We observed Abell 2744 using the Low Resolution Imager and Spectrograph (LRIS) on the Keck-I telescope, during the night of December 7th 2015. One single spectroscopic mask covered seven multiple images selected in the cluster core: 1.1, 10.3, 25.3, 35.1, 37.1, 39.1 and 57.2 over 4.8 ksec and 4.5 ksec in the blue and red arms of the instrument, respectively. 
The blue arm was equipped with the 400 lines/mm grism blazed at 3400 \AA, while the red arm was equipped with the 400 lines/mm grism blazed at 8500 \AA.  The light for both arms was separated using the 6800 \AA\ dichroic.

This configuration provided nearly complete coverage of the wavelength range $3500<\lambda<9700$ \AA, with a spectral resolution of 5.2 \AA\ and 4.8 \AA\ in the blue and red arms respectively. Each slit was individually reduced using standard IRAF procedures for bias subtraction, flat-fielding, wavelength and flux calibration.

We inspected each 2D reduced slit for faint emission lines and identify clear emission in the spectrum of images 35.1 and 37.1, centered at 4446 and 4438 \AA\ respectively. The absence of any other strong emission line in the wavelength range gives a secure identification of Lyman-$\alpha$ at similar  redshifts: $z=2.656$ for image 35.1 and $z=2.650$ for image 37.1. No strong spectral feature was found in any the other multiple images included in the mask.

\section{Data analysis}
Since MUSE is most sensitive to emission line objects, very faint (m$_{F814W}\geq$25) sources lacking emission lines can be hard to detect.  Therefore, in order to extract the maximum number of sources possible, we applied three complementary detection methods over the entire field:

 \begin{enumerate}

 \item Forced spectral extraction at the location of known faint sources detected in deep (m$_{\rm lim} \sim$ 30) HFF imaging.
 \item Emission line detection of sources based on a narrow-band filtering of the MUSE cube mosaic.
 \item A few manual extractions of sources not captured by i) and ii) and found through visual inspection of the datacube  (see, e.g., the special case of multiply-imaged system 2 explained in the appendix table \ref{tab:comparison_redshifts}).
 \end{enumerate} 

We then searched the combined list of objects extracted with methods (i)-(iii) for spectral features, measuring redshifts which we compared to ancillary redshift catalogs of Abell 2744.  This process is described in the following sub-sections.

\subsection{HST photometric catalog}
\label{sect:phot}

\begin{figure*}
\includegraphics[width=\textwidth]{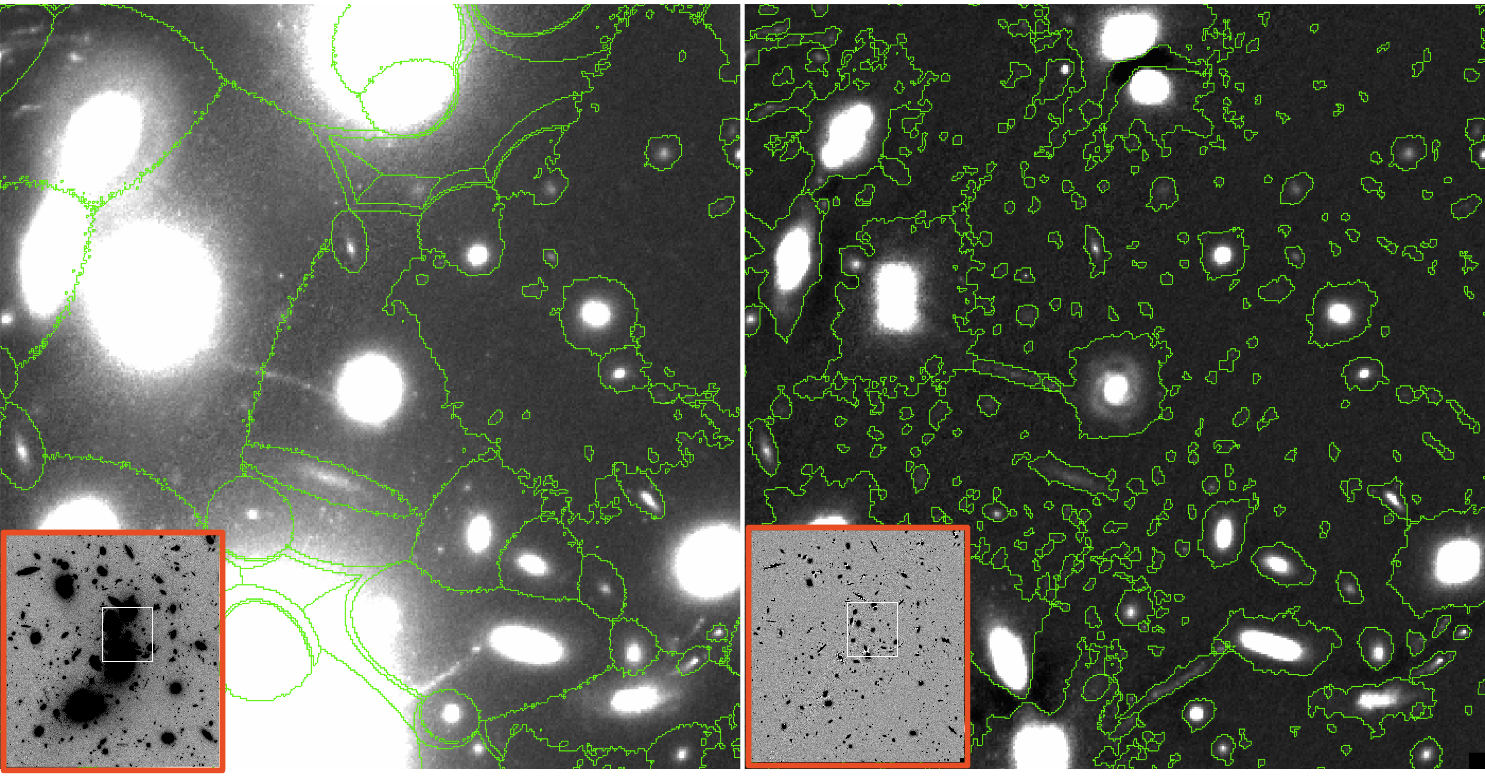}	
    \caption{Example of the procedure used to subtract the intra-cluster light (ICL). Each panel is 23$\arcsec$ (105 kpc at z$=$0.308) on a side. The white rectangles in the inserted panels show the location of the zoomed area. On the left, a region in the original HST F814W filter. On the right the same region and filter with the median removed, as described in Section \ref{sect:phot}. The scale and colour-levels used in the two panels are the same. The median filter is calculated in a 21x21 pixel running window. The ICL and wings of bright cluster members are largely removed, leading to an increased contrast around small and faint sources, improving their detectability. The green contours show segmentation maps from identical detection-setups.}
    \label{fig:med_sub}
\end{figure*}

Our MUSE spectral extraction (method (i) described above) relies on apertures defined using a photometric catalog.  We build this catalog taking full advantage of the depth and high spatial resolution of the HFF images to detect as many objects as possible. However, diffuse intracluster light (ICL) is an important and significant component of the core of the clusters and affects the detection of faint sources in the vicinity of cluster members, which is usually the case for multiple images (e. g. \citealt{Montes2014,Livermore2016,ASTRODEEP1}). For the current study, we remove the ICL and cluster member wings in each filter by subtracting the results of a running median, calculated within a window of $\sim 1.3\arcsec$ (21 pixels with  60 mas pixel scale HST images). Figure \ref{fig:med_sub} illustrates the improvement of our filtering procedure on the extraction of faint objects in a heavily crowded region near the cluster core. The ICL-subtracted images were weighted by their inverse-variance map and combined into one deep image. To perform a consistent photometric analysis {\sc SExtractor} \citep{SEx} was used in dual-image mode, with objects detected in the combined image and their fluxes measured from the individual median subtracted images.

By using the median-subtraction process, we inevitably underestimate the total flux of individual galaxies. To measure the level of underestimation, we compare photometric data between images with and without median subtraction. For consistency, we use identical detection-setups on both images. We find that the total flux is underestimated by about 50\%
for bright objects (m$_{F814W}\sim$20) and by $\sim$15\% for faint objects (m$_{F814W}\sim$27). However, the contrast and detectability of faint and peaky objects is also increased by $\sim$15\%. 
The {\sc SExtractor} parameters used to construct this catalog are provided for reference in the published catalog.

\subsection{Extracting spectra}

\begin{figure}
	\includegraphics[width=\columnwidth]{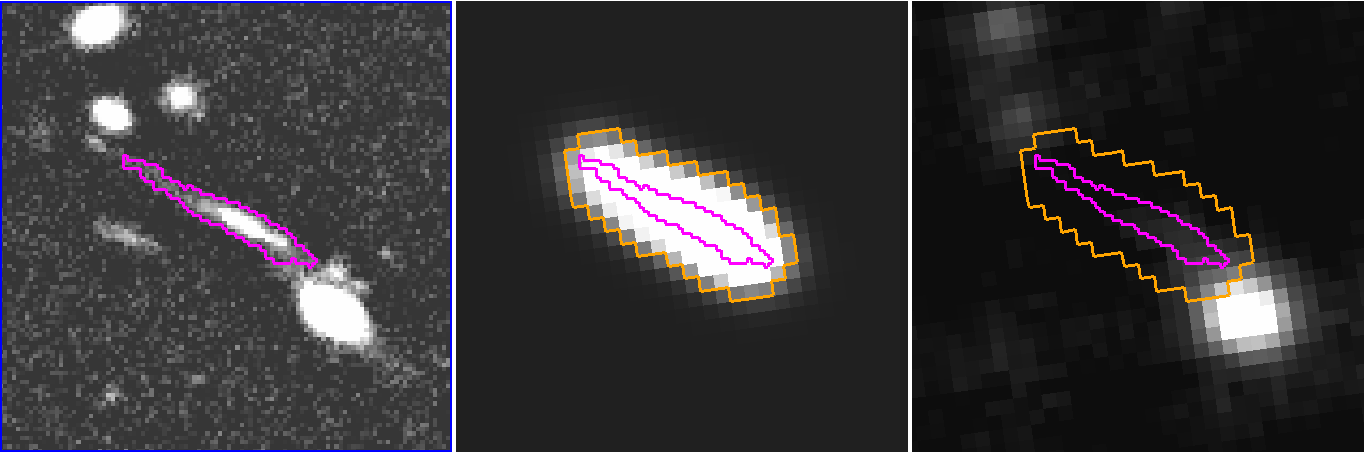}
    \caption{From left to right: 1.) Combined HST image used for source detection in the photometric catalog. 2.) Associated \textsc{SExtractor} segmentation map, convolved to the MUSE seeing level. 3.) MUSE data, collapsed over all wavelengths.  The magenta contour represents the HST-based detection, while the orange contour represents the 10\% cutoff level of normalised flux, after convolving the segmentation map to the MUSE seeing.\RC{Each panel is $\sim 6.25 \arcsec$ on a side}}
    \label{fig:conv}
\end{figure}

The resolution and sensitivity of the HFF images give morphological information of continuum emission, enabling us to deblend close pairs of objects. Based on the deblended source catalog, an associated extraction area was used to extract spectral information from the MUSE datacube according to the largest PSF measured ($\sim$0.7$''$), which appeared to be on the bluest part of the cube. The extraction area is based on a {\sc SExtractor} segmentation map of each individual object broadened by a Gaussian convolution with a FWHM matched to this PSF.
The resulting mask is rebinned to match the MUSE spatial sampling (0.2\arcsec/pixel) and the area of the mask is cut off at 10\% of the maximum flux. Figure \ref{fig:conv} highlights steps of the masking process. MUSE pixels within the mask are combined in each wavelength plane, weighting each pixel by the signal-to-noise ratio.  For further details of the method see \cite{Horne86}. 
We note that our chosen set of detection parameters led {\sc SExtractor} to deblend the most extended sources, such as giant arcs, into multiple objects in the catalog. In these few cases, spectra were extracted after visual inspection and manual merging of the segmentation regions.

\subsection{Automatic line detection}

Complementing the extraction method based on HST continuum levels, we search the MUSE datacube for emission lines using the dedicated software {\sc MUSELET}\footnote{{\sc MUSELET} is an analysis software released by the consortium as part of the MPDAF suite \url{http://mpdaf.readthedocs.io/en/latest/muselet.html}}. This analysis tool produces a large number of pseudo-narrow band images over the entire wavelength range of the MUSE cube, summing the flux over 5 wavelength bins (6.25 \AA ) and subtracting the corresponding median-filtered continuum estimated over two cube slices of 25 \AA\ width each. 

{\sc SExtractor} is then used on each of these narrow-band images to detect the flux excess due to emission lines. All {\sc SExtractor} catalogs are then matched and merged to produce a list of line emissions which may or may not be associated with strong continuum flux. When multiple emission lines are identified for a single source, the redshift is automatically provided, otherwise the remaining lines are visually inspected to identify [\ion{O}{ii}]$_{\lambda\, \lambda\, 3727,\, 3729}$, \Lya or another line.

\subsection{Catalog construction}
\label{sect:cata}
\begin{figure*}
	\includegraphics[width=\textwidth]{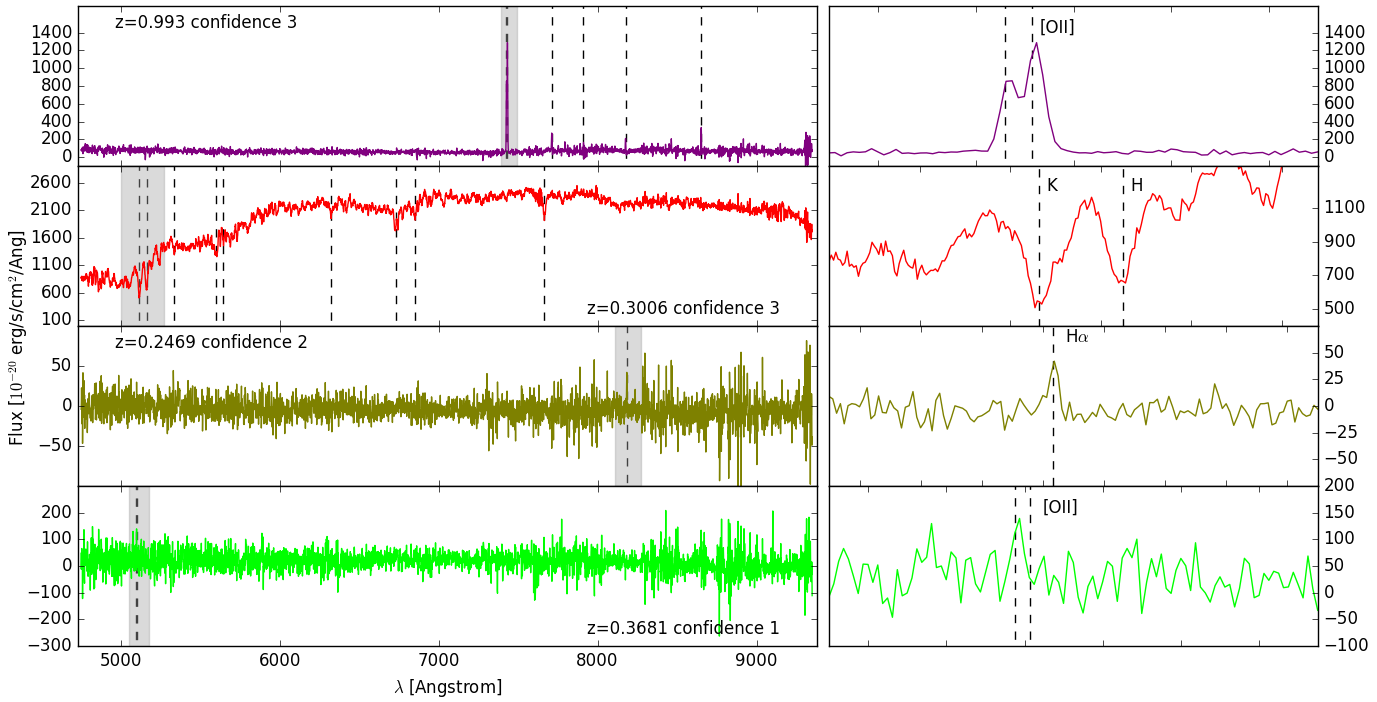}
    \caption{Examples of 1D spectral identification. The 4 rows highlight the grading process in terms of confidence level. Panels on the left show the complete spectrum, while panels on the right show the zoomed-in region marked by the gray shaded area. Spectra are graded into three levels of confidence, from 1 (tentative), to 3 (secure). See Section \ref{sect:cata} for details. From top to bottom, we show: a confidence 3 spectrum identified by multiple emission line features (marked by the vertical dashed lines),  
a confidence 3 spectrum based on absorption features,  
a confidence 2 spectrum based on a single line detection, and a confidence 1 spectrum with a tentative, faint emission line feature identified as [\ion{O}{ii}].  }
    \label{fig:rank}
\end{figure*}

Redshift assessment was performed independently by six authors (GM, JR, BC, DL, VP, and JM), using several methods. 
We systematically reviewed all HST-based extracted sources down to a signal-to-noise in the continuum where no secure redshift relying on continuum or absorption features were able to be assessed. This empirically corresponds to an HST magnitude of $m_{\rm F814W} = 24.4$. Each of these spectra was at least reviewed by one of the authors. The redshift catalog was completed with information from the emission line finder MUSELET where reviewers also checked every line suggested by the software. Multiply-imaged systems already recorded throughout the literature (\citealt{Jauzac2015},\citealt{Zitrin2015}, \citealt{Kawamata2015}, \citealt{Johnson2014}, \citealt{Lam2014} and \citealt{Richard2014}) were carefully vetted by the same six authors in order to increase confidence in the redshift assessment. 
We assigned each measured redshift a confidence level based on the strength of spectral features according to the following rules:

\begin{itemize}
\item Confidence 3 : secure redshift, with several strong spectral features.
\item Confidence 2 : probable redshift, relying on 1 spectral feature or several faint absorption features.
\item Confidence 1 : tentative redshift
\end{itemize}
Examples of spectra assigned confidence 1, 2, and 3 are shown in Fig.\,\ref{fig:rank}. 

We next construct a master redshift catalog, including only spectra with a confidence level of 2 or 3.  The only exceptions are made for multiply imaged systems ranked as very secure photometric candidates by HFF lens modelers (see Sect. \ref{sect:SLanalysis} for more details). The master redshift catalog was compared to entries in the NASA/IPAC Extragalactic 
Database (NED, \url{https://ned.ipac.caltech.edu}), the publicly available redshift catalog from the 
GLASS collaboration\footnote{\url{https://archive.stsci.edu/prepds/glass/}} and the redshifts presented by \citet{Wang2015}, and corrected as needed. The details of this comparison is presented in Table \ref{tab:comparison_redshifts} of Appendix \ref{sect:redshifts_comparison}

\begin{figure*}
    \centering
    \includegraphics[scale=0.59]{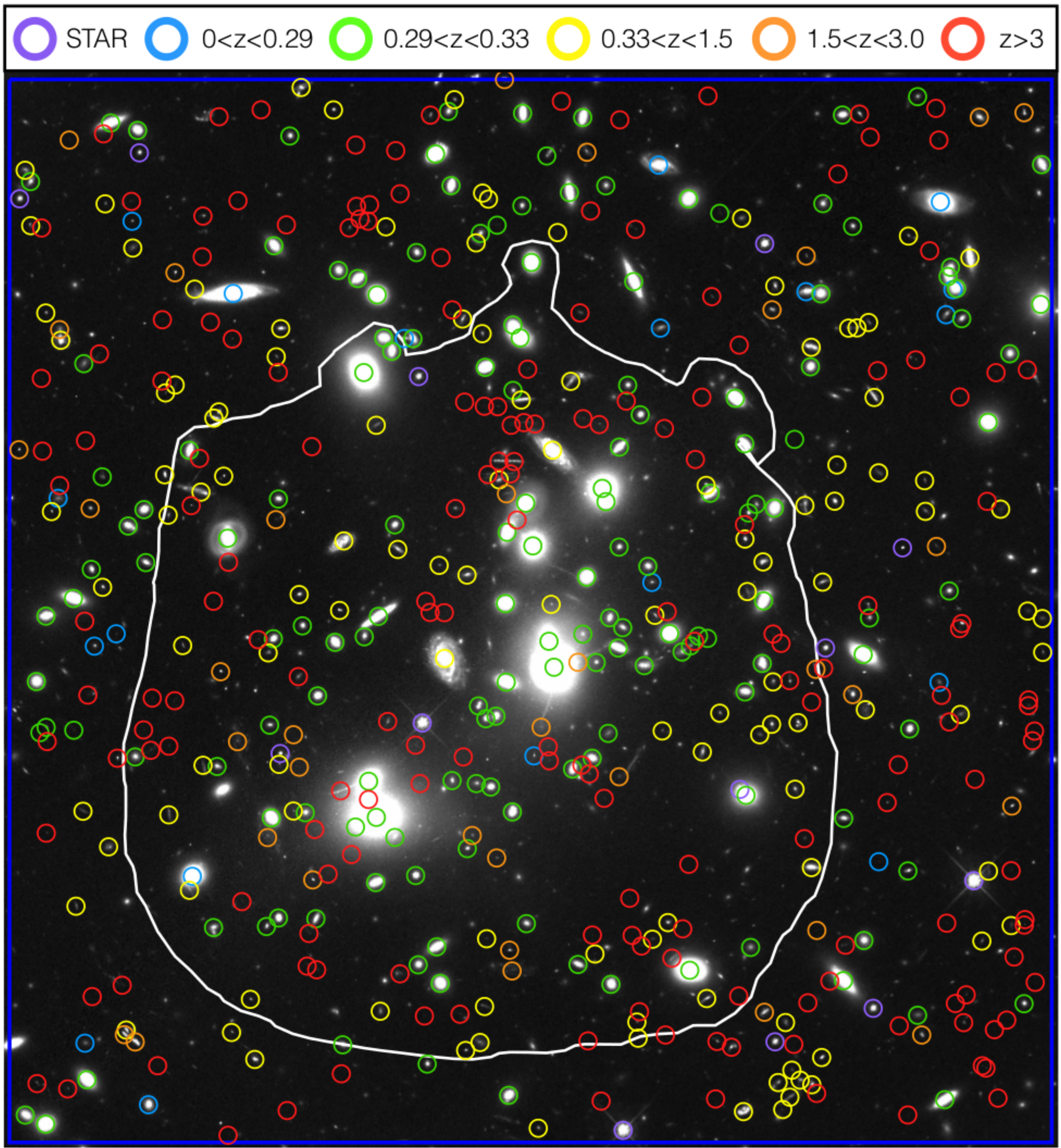}
    \vskip-0.2ex
    \includegraphics[width=\textwidth]{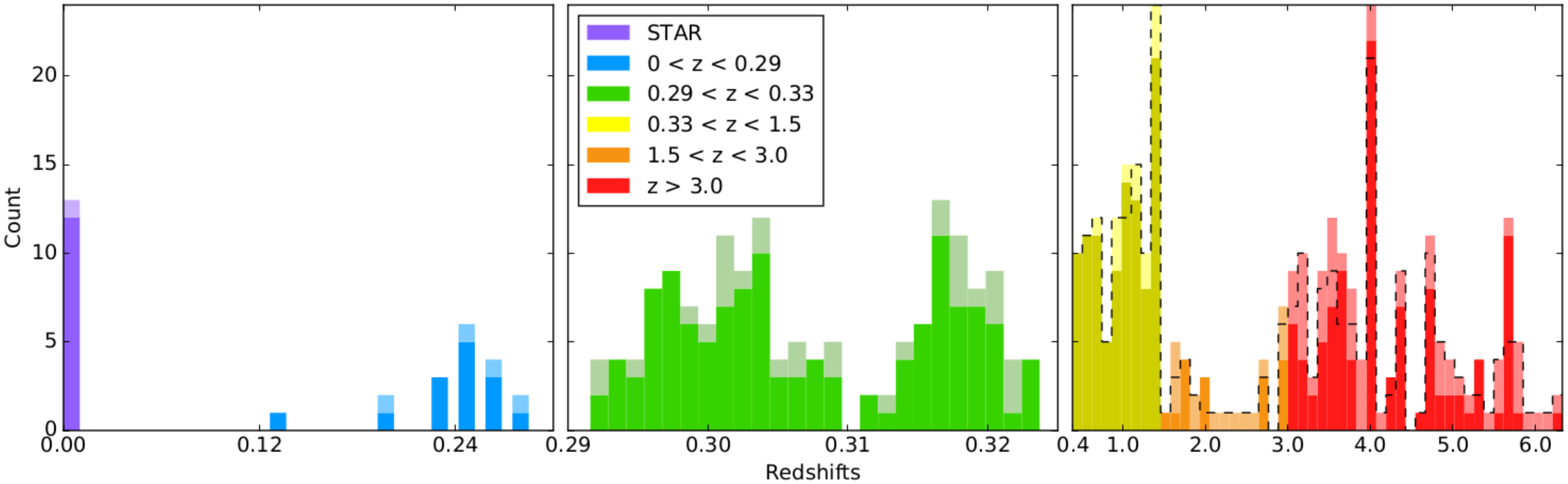}
    \vskip-0.2ex
    \caption{The top panel represents the spatial distribution of all secure redshifts, superimposed on an RGB HST image. The dark blue box represents the full extent of the \RC{$2\arcmin x 2\arcmin$} MUSE mosaic, while the white line encloses the multiple image area for objects with $z \leq 10$. The lower panels represent the redshift histogram of the same sources. The darker colour represents confidence 3 objects and the lighter colour represents confidence 2 objects. The lower left panel presents the foreground redshifts with respect to the cluster. The lower middle panel shows the cluster redshifts distribution. The lower right panel shows the redshift distribution of background sources. The black dashed line shows the number of independent background sources (corrected from the multiplicity due to lensing). Note that the bin sizes differ in the three bottom panels ($\Delta$z$\approx$0.0165, 0.001, and 0.119, respectively)
    }%bin size is truly = 0.0165 0.001248 0.11873
    \label{fig:z_full}
\end{figure*}

The final catalog contains 514 redshifts, including 10 with confidence 1 and 133 with confidence 2 and 371 with confidence 3.  The spectral and spatial distributions of this catalog can be seen in Fig.\,\ref{fig:z_full}. 
Table \ref{tab:head_cata} presents the very first entries of the catalog and the full version is available in the online version\footnote{available at \url{http://muse-vlt.eu/science/a2744/}}. 

\begin{table*}
	\centering
	\caption{First six lines of the redshift catalog released with this work. The columns ID, RA, DEC and $z$ represent the identification number, the right ascension, the declination and the redshift of each entry. The column CONFID represents the confidence level of the detection, from 3 for very secure down to 1 for less secure identifications according to our grading policy, see section \ref{sect:cata}. TYPE represents the classification of the object based on the system used for the MUSE-UDF analysis (Bacon et al.\, in prep.): TYPE$=$0 are stars, TYPE$=$2 are [\ion{O}{II}] emitters, TYPE$=$3 are absorption line galaxies, TYPE$=$4 are \ion{C}{III]} emitters and TYPE$=$6 are Lyman $\alpha$ emitters (the other MUSE-UDF TYPE do not match any entries of this catalog). The MUL column shows the multiple image ID if it is reported in our strong lensing analysis. Columns named FXXXW and FXXXW\_ERR present the photometry and its error in the seven HST filters used in this study. MU and MU\_ERR represent the magnification ratio and its error computed from our lensing mass model. Objects MXX are only detected in the MUSE cube as they do not match any entry from our photometric catalog.
    }
	\label{tab:head_cata}
	\begin{tabular}{lccccccccccccc} 
		\hline
        \hline
ID & RA & DEC & $z$ & CON- & TYPE & MUL & F435W & F435W & ... & F160W & F160W & MU & MU \\
   &    &     &   & FID &      &      &       & \_ERR & ... &       & \_ERR &     & \_ERR \\
   & [deg] & [deg] & & & & & [mag] & [mag] & & [mag] & [mag] \\
      \hline
M39 & 3.5889097 & -30.3821391 & 6.6439 & 2 & 6 & "" & "" & "" & ... & "" & "" & 2.221 & 0.061 \\
2115 & 3.5938048 & -30.4154482 & 6.5876 & 2 & 6 & "" & $>$29.44 & 99.0 & ... & 26.70 & 0.0383 & 3.575 & 0.09 \\
M38 & 3.5801476 & -30.4079034 & 6.5565 & 2 & 6 & "" & "" & "" & ... & "" & "" & 2.958 & 0.084 \\
M37 & 3.5830603 & -30.4118859 & 6.5195 & 2 & 6 & "" & "" & "" & ... & "" & "" & 2.868 & 0.07 \\
10609 & 3.598419 & -30.3872993 & 6.3755 & 2 & 6 & "" & $>$30.39 & 99.0 & ... & 30.00 & 0.3039 & 1.768 & 0.051 \\
5353 & 3.6010732 & -30.4039891 & 6.3271 & 3 & 6 & "" & $>$29.57 & 99.0 & ... & 28.04 & 0.0938 & 3.821 & 0.133 \\
... \\
\end{tabular}
\end{table*}

We compared the MUSE redshift catalog presented here to the NED database, checking in particular the redshifts presented by the GLASS team \citep{Wang2015}. In Appendix \ref{sect:redshifts_comparison} we list corrections made to redshifts published in the literature based on the MUSE data.

\section{Strong Lensing analysis}
\label{sect:SLanalysis}
In this section, we provide a brief summary of the gravitational lensing analysis technique used in this work. 
We refer the reader to \cite{Kneib1996}, \cite{Smith2005}, \cite{Verdugo2011} and \cite{Richard2011} for more details.

\subsection{Methodology}
\label{Methodology}
Although many different analysis methods exist throughout the literature, they can generally be classified into two broad categories.
The first category, known as parametric methods, use analytic profiles for mass potentials and rely on a range of parameters to describe the entire cluster mass distribution. The second category, referred to as non-parametric methods, make no strong assumption on the shape of the mass profile. Instead, the mass is derived from an evolving pixel-grid minimisation. In this study, we take a parametric approach, using {\sc Lenstool} \citep{Jullo2007} to model the cluster mass distribution as a series of dual pseudo-isothermal ellipsoids (dPIE, \citealt{Eliasdottir2007}), which are optimised through a Monte Carlo Markov Chain minimisation. 

To model the cluster mass distribution, Dark Matter (hereafter DM) dPIE clumps are combined to map the DM at the cluster scale. Galaxy scale DM potentials are used to describe galaxy scale substructure. Considering the number of galaxies in the cluster, including several hundreds in the core alone, it is not feasible to optimise the parameters of every potential, as the large parameter space will lead to an unconstrained minimisation. Moreover, individual galaxies contribute only a small fraction to the total mass budget of the cluster, so their effects on lensing are minimal at most. To reduce the overall parameter space we scale the parameters of each galaxy to a reference value, using a constant mass-luminosity scaling relation given by the following equations:

\begin{equation}
\begin{array}{l}
\sigma_0=\sigma_{0}^{*}\left(\frac{L}{L^{*}}\right)^{1/4},\\
r_{\rm core}= r_{\rm core}^{*}\left(\frac{L}{L^{*}}\right)^{1/2},\\
r_{\rm cut}=r_{\rm cut}^{*}\left(\frac{L}{L^{*}}\right)^{1/2}
\end{array} 
\label{eq:ML}
\end{equation}
where $\sigma_{0}^{*}$, $r_{\rm core}^{*}$, and $r_{\rm cut}^{*}$ are the parameters of an $L^{*}$ galaxy. The $r_{\rm core}^{*}$ is fixed at $0.15$kpc as $r_{\rm core}^{*}$ is expected to be small at galaxy scales and also degenerate with $\sigma_{0}^{*}$.

Some galaxies in the FoV are not expected to follow this relation, based on their unique properties or formation histories. As a result, we remove these objects from the scaling relation to avoid biasing the results. One prominent example is the Brightest Cluster Galaxy (BCG) which will have a significantly different mass-to-light ratio and size since it is the center point of the merging process. 
As advised by \citet{Newman1,Newman2} 
the two BCGs of Abell 2744 are modeled separately. 
In addition, bright (therefore massive) galaxies behind the cluster can also contribute to the lensing effect near the core, so we include them in the galaxy sample, but model them separately from the scaling relation.  In order to normalise the effects of these galaxies on the model, we rescale their total masses based on their line-of-sight distance from the cluster.  These ``projected-mass'' galaxy potentials are then optimised.

Given the complexity of the cluster, the strong lensing models are optimised iteratively, starting with the most obvious strong lensing constraints (as discussed in Section \ref{sect:reliability}). After the initial run concludes,  parameters are then adjusted and the set of constraints can be reconsidered. Once these changes are made, another minimisation is started and the model is revised according to the new results. This offers the possibility of testing different hypotheses, such as adding DM clumps or 
including an external shear field. Throughout this process, 
multiple image constraints can be paired differently and new counter-image positions can be 
identified by their proximity to the model predictions. 
Ending this iterative process is not obvious and an arbitrary level of satisfaction is needed to stop. In this work, the $\chi^{2}$ value and RMS statistics measured with respect to the observed positions of multiply-imaged galaxies are used to rank different models and priors.

\subsection{Selection of cluster members}

To construct a catalog of cluster members, we start with the colour-colour selection from \cite{Richard2014}: all galaxies that fall within 3$\sigma$ of a linear model of the cluster red sequence in both the (m$_{F606W}$-m$_{F814W}$) vs m$_{F814W}$ and the (m$_{F435W}$-m$_{F606W}$) vs m$_{F814W}$ colour-magnitude diagrams. However, we 
limit ourselves to only those galaxies contained within the WFC3 FoV.  This is because the WFC3 field approximately matches the MUSE FoV, allowing us to focus on modeling the cluster core (see \citealt{Jauzac2015} and reference therein).  As mentioned in the previous section, cluster members included in the mass model are scaled through a mass-to-light relation. In order to better fit the scaling relation to the selected galaxies, we take magnitudes from the ASTRODEEP photometric catalog (see \citealt{ASTRODEEP1} and \citealt{ASTRODEEP2} for a complete view of the catalog making process). When available, we use the ASTRODEEP magnitudes for our objects, since they assume a Sersic model fit of galaxy photometry. Compared to our photometric catalog, a major difference can be seen in bright objects.  This is due to the broad limit between galaxy wings and ICL, which we remove with our median filtering. In cases where an F814W magnitude is not available from ASTRODEEP, we substitute it with the photometry of the catalog detailed in Sect. \ref{sect:cata}. Because faint cluster galaxies far from lensed arcs only have a small lensing effect, only galaxies brighter than $0.01$ $L_*$ are included in the final galaxy selection (m$_{F814W}<$24.44; M$\approx1.5\times10^{9}\,{\rm M}\odot$, \citealt{Natarajan2017}). The global effect of missing cluster members will be degenerate with the total mass in the large-scale DM clumps.

Additionally, galaxies that match the initial colour selection but have confirmed redshifts outside of the cluster range [$0.29 < z < 0.33$] (see Fig.\,\ref{fig:z_full}) are removed from the cluster member catalog (8), while non-colour-matched galaxies with a confirmed cluster redshift are included (21).  After all of this, we are left with 246 cluster galaxies out of which a large fraction (156) have spectroscopic redshifts. As described in Sect.\,\ref{sect:dynamics} this large sample of cluster members provide vital information about the cluster dynamics.

\subsection{Strong lensing constraints}
This section describes our methodology of categorizing multiply-imaged systems and details the reviewing of all known multiple systems used and reported in the strong lensing analyses of Abell 2744. Table \ref{tab:sys_comparison} summarises the number of systems, images and spectroscopic redshifts from each study.

Prior to the FF observations, early lens models by \citet{Merten2011}, \citet{Richard2014}, and \citet{Johnson2014} constructed a catalog of 55 multiple systems, including three secure spectroscopic redshifts for systems 3, 4 and 6 \citep{Johnson2014}. Later work by \citet{Jauzac2014}, \citet{Lam2014}, \cite{Ishigaki2015}, and \citet{Kawamata2016} proposed $\sim$185 additional images from the analysis of the HFF data. This includes spectroscopic redshifts of 7 lensed sources found by the GLASS team \citep{Wang2015} measured for images 1.3, 2.1, 3.1 and 3.2, 4.3 and 4.5, 6.1, 6.2 and 6.3, 18.3, 22.1. 
The spectroscopic measurement for system 55 are associated with the same sources as system 1 (see \citealt{Wang2015} for details). The existing numbers of multiple imaged systems (N$_{sys}$) and the total number of source images in these (N$_{im}$) as well as the fraction of spectroscopically confirmed redshifts are summarised in Table \ref{tab:sys_comparison}.

\begin{table}
	\centering
	\caption{Number of images and systems reported in the strong lensing analyses of Abell 2744 to date. N$_{\rm{sys,z}}$ gives the number of systems having at least one image confirmed with a spectroscopic redshift and used in the model, N$_{\rm im,z}$ the number of images confirmed with a redshift in these systems, compared to the total number of systems (N$_{\rm sys}$) and images (N$_{\rm im}$) presented.}
	\label{tab:sys_comparison}
	\begin{tabular}{lcccc} % four columns, alignment for each
		\hline
		\hline

      Study & N$_{\rm sys,z}$ & N$_{\rm im,z}$ & N$_{\rm sys}$ & N$_{\rm im}$ \\
      \hline
      \hline
      pre-HFF   &    &   &   &   \\
      \hline
      \citealt{Merten2011}       &0	&0	&11	&34 \\
      \citealt{Richard2014}	   &2	&2	&18 &55 \\
      \citealt{Johnson2014}	   &3	&3	&15	&47\\
      \hline
      post-HFF       &    &   &   &   \\
      \hline
      \citealt{Lam2014}	    &4	&4	&21	&65  \\
      \citealt{Zitrin2014}     &4  &4  &21 &65\\
      \citealt{Ishigaki2015}   &3	&3	&24	&67\\
      \citealt{Jauzac2015}     & 3	&8	&61	&181   \\
      \citealt{Wang2015}       &3	& 8	&57	&179\\
      \citealt{Kawamata2016}     &5	&5	&37	&111  \\
      &&&&\\
      \textbf{This work}      &\textbf{29}  &\textbf{83}	&\textbf{60}	&\textbf{188} \\
      \hline
      \end{tabular}
\end{table}

\subsubsection{Incorporating MUSE spectroscopic constraints}

We use all Confidence levels 2 and 3 MUSE redshifts to check the multiplicity and the reliability of each multiple system. 
While \citet{Wang2015} report a detection of H$\alpha$ line at $z=1.8630$ for image 1.3 with good confidence (Quality 3), the analysis of the stacked MUSE spectrum of system 1 leads to a  secure redshift $z=1.688$ based on multiple features (see in the Appendix \ref{sect:redshifts_comparison} for details). As in their study we also consider system 55 and system 1 belong to the same source such as system 56 and system 2.

\begin{figure*}
	\includegraphics[width=\textwidth]{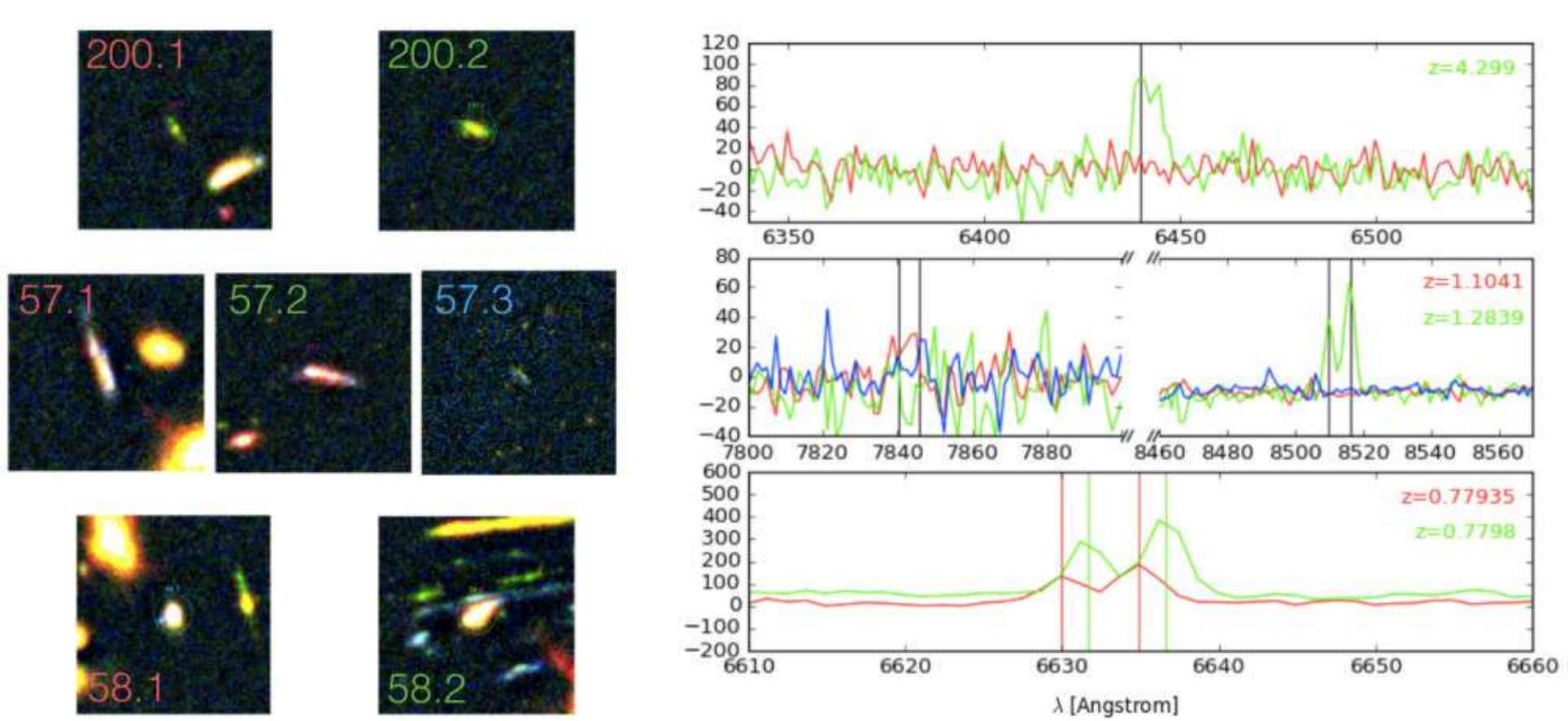}

   \caption{The three multiply-imaged candidate systems downgraded to single images in this study. The top row presents system 200, where we are only able to measure a redshift for image 200.2. Using the location of the object and its measured redshift, our model predicts that it is not multiply-imaged. The middle row presents system 57, where we are able to measure redshifts of images 57.1 and 57.2.  From the spectra in the right-hand panel, we can see that these two images have very different redshift values, meaning that they do not come from the same source.  Finally, the bottom row presents system 58.  While the redshifts of the two images are closer than those in system 57, they are still different enough that we reject them as a multiply-imaged pair. \RC{Each image panel is $\sim 5 \arcsec$ on a side}}
    \label{fig:single}
\end{figure*}

We reject the multiplicity assumption for five candidates: 57.1, 57.2, 58.1, 58.2 and 200.2 which are measured at a redshifts of 1.1041, 1.2839, 0.779, 0.78 and 4.30 respectively. No redshifts were measured for images 200.1 and 57.3. Figure \ref{fig:single} gives an overview of the rejected images. 
%KAWAMATA+16 62

In our inspection of the MUSE datacube  we discovered Ly$\alpha$ emitters corresponding to three new multiply-imaged systems. No photometric counter-part in the HST images could securely be associated with their Ly$\alpha$ emission (see systems 62, 63 and 64 in the list of multiple images).

\subsubsection{Reliability of multiply-imaged systems}
\label{sect:reliability}

The secure identification of multiple-image systems is key in building a robust model of the mass of the cluster. Because of the nature of lensing, constraints can only probe the total mass within an Einstein radius corresponding to the unique position and redshift of the source. Increasing the number of constraints at different positions and various redshifts thus makes it possible to map the mass distribution over the entire cluster. To maximise our coverage we consider two categories of constraints: hard and soft.

Hard constraints occur when both the position of images and the redshift are known accurately. Thus the mass potential parameters have to reproduce the correct position of the multiply-imaged systems at the given redshift. Soft constraints occur when the position is known but not the redshift. In that case, the redshift is considered to be a free parameter and the model has to optimise the redshift that best predicts the multiple-image positions. Soft constraints introduce a large degeneracy between redshift and enclosed mass, that will only be broken if a large number of such constraints are used. 

In order to test the reliability of our multiple-image identifications, we compute a SED $\chi^{2}$ statistic to quantify the similarity of the photometry in each pair of images within a given system:

\begin{equation} \hspace{1cm}
  \footnotesize
 \chi^{2}_{\nu}=\frac{1}{N-1} \min_{\alpha}\left(\sum\limits_{i=1}^{N} \frac{ ( f_{i}^{A}- \alpha f_{i}^{B})^{2} } {{\sigma_{i}^{A}}^{2}+ \alpha^{2} {\sigma_{i}^{B}}^{2} }\right)
\label{chi2}
 \end{equation}
 Where $N$ is the total number of filters, ($f^{X}_{i}$, $\sigma^{X}_{i}$) the flux estimate and error in filter i for images $A$ and $B$ considered to compute the $\chi^{2}$.
The conservation of colours between two lensed images make their photometry similar up to an overall flux ratio $\alpha$ which is minimised in this equation. As shown by Mahler et al. (in prep.) this statistic quantifies the probability of two images to come from the same sources. It shows some similarities with the approach used by \citet{Wang2015} and \citet{Hoag2016}, expect  for their use of colours and a normalisation per pair of filters in their calculation. Combining all HFF filters, we found acceptable values for $\chi^{2}$ (0 to 3) for almost all images, with slightly higher values typically being observed for sources whose photometry is compromised by bright nearby galaxies or suffer from ''over-deblending''

The good $\chi^{2}$ value of system 7 ($\chi^{2} \sim$1.2) promote the system to secure system and the poor agreement between the flux ratio and the predicted amplification ratio by three order of magnitude demote the counter image 10.3 to less reliable constraint.

We divide constraints into four different types of multiply-imaged constraints, according to their confidence.
\begin{itemize}
\item The most reliable constraints, dubbed {\it gold}, consists of hard constraints (i.e. having spectroscopic redshifts).
{\it Gold} systems 
do not include counter-images without a spectroscopic redshift, except for system 2 which has a very distinct morphology. 
83 images belonging to 29 systems are marked as {\it gold}.
\item The second set of constraints, dubbed {\it silver}, are the most photometrically convincing images and systems in addition of  {\it gold} constraints, 
following mostly the (unofficial) selection of Frontier Fields challenge modelers. By adding 22 images and 9 systems, this brings the total number of constraints to 105 images over 38 systems.
\item The third set, dubbed {\it bronze}, includes less reliable constraints. The {\it bronze} set contains 143 images of 51 systems.
\item The fourth set, dubbed {\it copper}, include images 3.3, 8.3, 14.3, 36.3, 37.3, 38.3 because they were previously in disagreement among previous studies (see \citealt{Lam2014} and \citealt{Jauzac2015} as an example of disagreement). {\it Copper} set of constraints include as well all the remaining counter images and systems reported bringing the total number of images to 188 belonging to 60 systems.

\end{itemize}

The multiple images used in this study are shown in Fig.\,\ref{fig:mul_img}. The full list of multiply images is provided in Table \ref{multipletable} in Appendix \ref{apendix:mul}. Spectral identification of each gold image is presented in Appendix \ref{sect:img_mul}.

\begin{figure*}
    \includegraphics[width=\textwidth]{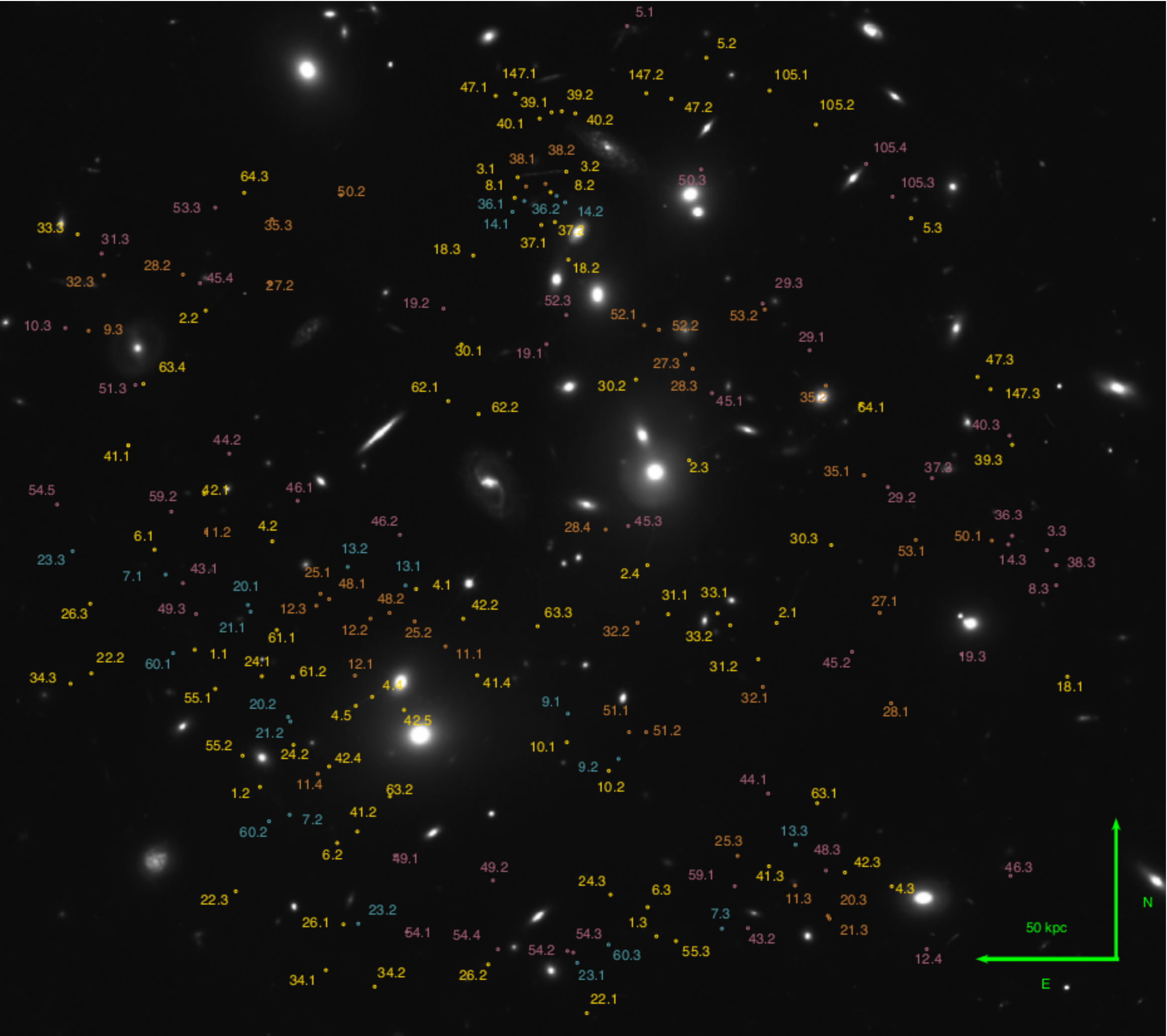}
    \caption{The gold, silver, bronze and copper circles match different sets of constraints called \gold, \silver, \bronze, and \copper. Each of the constraints matches its corresponding colour. To avoid any mismatching, \silver\ constraints appear bluer and \copper\ constraints appear pinker. See in the Appendix Tab\, \ref{multipletable} for details.}
    \label{fig:mul_img}
\end{figure*}

\section{Lens modeling results}
\label{sect:lens_result}
In this section we construct lens models and describe their properties, along with the details of individual strong lensing features. 

\subsection{Mass distribution in the cluster core }
\label{sect:mass-distri}
To investigate improvements on currently known mass models, we test several assumptions using a series of different model configurations.
\RC{We quantitatively compare the quality of models with two criteria. The first one is the rms, which describes how well the model reproduces the positions of the constraints. The second one is the Bayesian Information Criterion (BIC) which is a statistical measurement based on the model Likelihood $\mathcal{L}$, penalised by the number of free parameters $k$ and the number of constraints $n$: 
\begin{equation}
{\rm BIC} = -2 \times \log(\mathcal{L}) + k \times \log(n),
\end{equation}}
\RC{The rms will give indication good indication of the global distance between your predicted images position in comparison with the measured one. Thus we seek to reduce the rms as much as possible. The BIC will quantify the balance between the improvement of the model likelihood and the addition of parameters and constraints. Thus we seek to see the best improvement of the likelihood while keeping the lowest BIC value possible.}

For our initial model, we start with a parametrisation similar to \citet{Jauzac2015}, namely: two dark matter clumps representing cluster-scale potentials and two small-scale background galaxies (MUSE$_{9778}$ and MUSE$_{7257}$), in addition to identified cluster members (246). We also optimise the two primary BCGs separately from the mass-to-light scaling relation (see section \ref{Methodology}). While the \citet{Jauzac2015} model achieves an rms of 0.69\arcsec, our model -- which includes 24 new systems with secure redshifts from MUSE and Keck data -- has an rms of 1.87\arcsec\ \RC{and a BIC $=$ 4893}.
The higher rms is expected: by increasing the number of spectroscopic constraints, the model can no longer adjust the redshifts of these systems to better fit the model. However systems 5 and 47 (as defined in \citealt{Jauzac2015}) contribute the most to the rms (system 5: rms$=3.24\arcsec$, system 47 rms$=1.71\arcsec$). \RC{Since there is a chance these objects might be incorrectly identified} 
and because they affect mainly the Northern part of the cluster core we 
temporarily remove these two systems \RC{for our next test.}

In an attempt to improve the model further, we add a third cluster-scale clump $\sim$20\arcsec\ north  of the northern BCG, free to vary in position. We choose this location due to the significant number of cluster galaxies in the area. After running two models, one with and one without the third clump, the resulting global rms is 0.77\arcsec\ in both cases 
\RC{, whereas the 2 clumps hypothesis has a BIC$=$332 which is slightly lower than the 3clumps BIC$=$362}. Note that at this stage system 5 and 47 are still not included as constraints.

\RC{However,} we next test the same assumptions but we revise the positions of systems 5 and 47 
\RC{by adjusting them} to the centroid of the Lyman alpha emission. 
\RC{Additionally,} thanks to the MUSE blind identification of the extended Lyman alpha emission of these two sources, we are able to add 
two new constraints: system 105 and system 147 which function as separate substructures of system 5 and 47, respectively. The mean rms (BIC) for the two different configurations increases from $0.77\arcsec$ to $0.86\arcsec$ \RC{(from 332 to 419)} for the 2-clump assumption and from $0.77\arcsec$ to $0.96\arcsec$ \RC{(from 362 to 511)} for the 3-clump assumption. 
This significant improvement on the models, \RC{compared to the initial one, }
is consistent with the observation of a diffuse gaseous component around the two galaxies sources of systems 5 and 47. The study of the physical properties of all background sources behind Abell 2744 will be presented in a forthcoming paper (De la Vieuville et al. in prep).

\begin{figure*}
    \includegraphics[width=\textwidth]{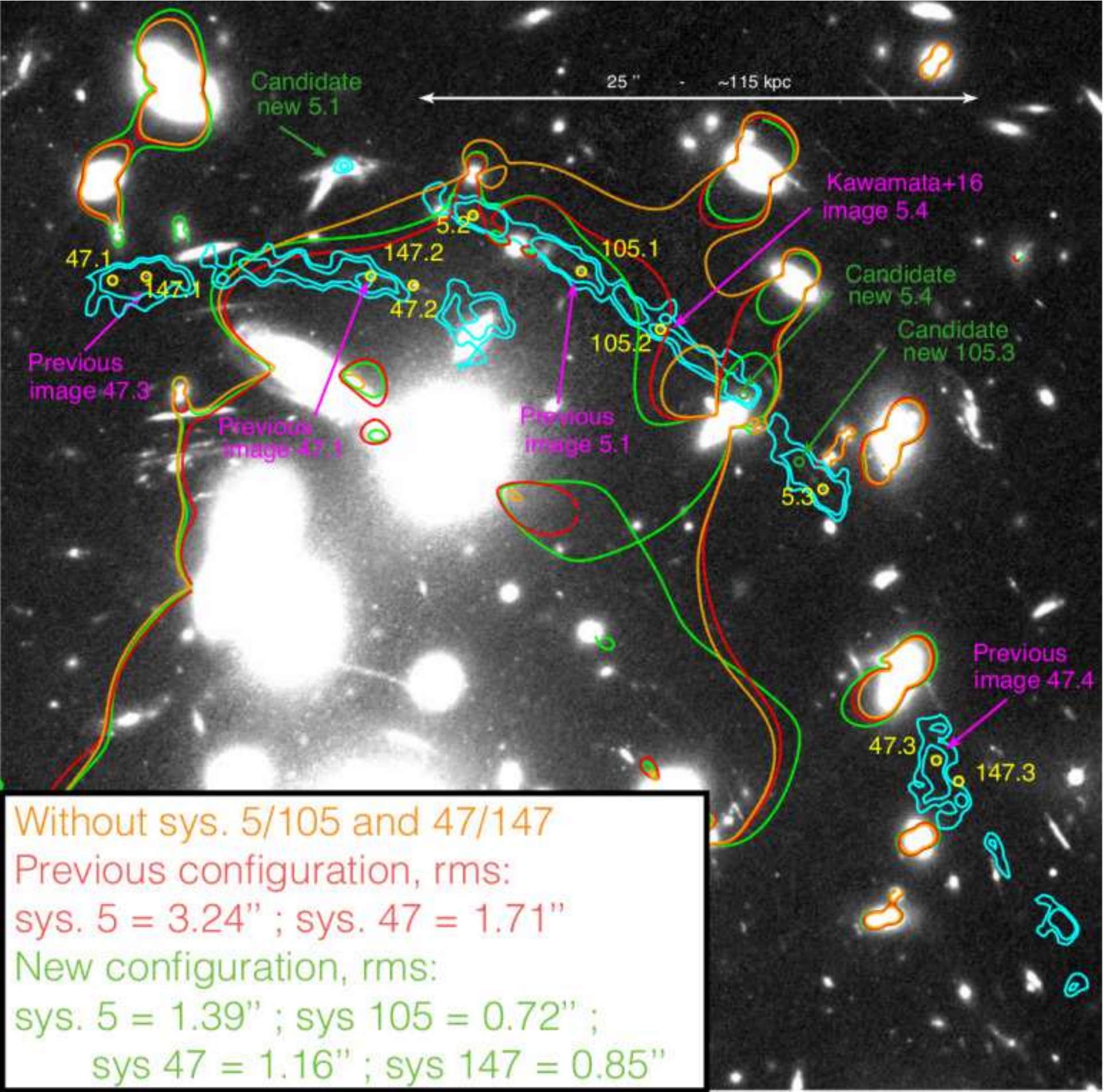}
    \caption{Differences between models based on the assumptions developed in section \ref{sect:lens_result}. The blue contour shows the Lyman $\alpha$ emission at the redshift of systems 5, 105, 47, and 147 ($z=4.0225$). The orange, red and green lines show the tangential critical curve at the same redshift for three different models.  While all models use the same parametrisation for the mass components, each makes a different assumption about systems 5 and 47.  As a reference, the orange line traces the critical curve when the two systems are removed from the model entirely.  Conversely, the red line shows the critical curve assuming the previous constraint configuration: namely, that systems 5 and 47 are each an individual multiply-imaged object.  Finally, the green line represents the critical curve measured when system 5 and 47 are both divided in two distinct components (system 5-> 5 and 105 and system 47 -> 47 and 147). This new configuration better matches the observed Lyman $\alpha$ emission.}
    
    \label{fig:2DM-3DM}
\end{figure*}

Since the addition of a third clump at best leaves the rms unchanged, we favor the simpler 2-clump model moving forward. At the same time, we keep the new constraint configuration of systems 5/105 and 47/147, since this reduces the rms from the original model.  Differences in models are shown in Fig.\,\ref{fig:2DM-3DM}.

\subsection{Influence of the cluster environment}
\label{sect:MJclumps}

The weak-lensing analysis of J16 reported the identification of six cluster substructures at large radii ($\sim$700 kpc) with a significance level above 5$\sigma$. We expect these complex, large-scale structures to have an effect on the location of multiple images in the cluster core.

To first order the influence from these mass substructures 
can be approximated as a shear field.
To test this possibility we include the influence of an external (constant) shear field in our model, described by the following two parameters: the strength of the shear $\gamma$ and the position angle $\theta$.
\RC{We use a broad uniform prior on the external shear: $0<\gamma<0.2$ for the strength and $-90.<\theta<90.$ deg. for the angle.}
The resulting model rms is 0.78\arcsec\ \RC{and BIC$=$384} (compared to 0.86\arcsec\ \RC{and BIC$=$419} before) with best-fit parameters $\theta=-36\pm1$ deg and a strength $\gamma=0.17\pm0.01$. The effect of the external shear is global on the cluster core and not specifically targeted to a single location. \RC{From the analysis of the posterior probability distribution of each parameter, we notice a small correlation for each of the external shear parameters. The shear angle slightly correlates with the core radius and velocity dispersion parameters both from the southern DM clumps whereas the shear strength slightly anti-correlates with the two same parameters. These small correlations will not affect our results.} 

While adding an external shear improves our mass model, in some ways it is not physical, because it does not rely on specific masses. Therefore, we construct an alternative model which includes the J16 substructures as individual mass components. 
We exclude the substructure on the West side (labeled as Wbis in J16) because it is 
behind the cluster.  We model the other clumps using dPIE potentials. Because the J16 weak lensing analysis does not provide analytic parameters for mass profiles, we place priors on the dPIE parameters. In order to make the model clumps recreate the J16 total mass values as closely as possible, we look for the best scaling relation parameters ($\sigma^*$, r$_{\rm cut}^*$) matching the J16 masses for each substructure based on the following criteria: 
\begin{enumerate} 
\item the enclosed mass in a radius of 150 kpc from the clump centre, 
\item the enclosed mass in a radius of 250 kpc from the clump, and 
\item the overall smoothness of the J16 cluster mass contours  
\end{enumerate}
Due to the different amount of light associated to the substructures as reported by J16 we separated the six potentials in two different scaling relations.
To maintain the same number of parameters as the model with an external shear, we only optimise the values of $\sigma^*$ of the two scaling relations. The resulting masses of the clumps are reported in Table \ref{tab:mass_comparison}, following the nomenclature from J16. The resulting rms is 0.67\arcsec\, which is comparable to the rms of the model including external shear (0.78\arcsec). We will discuss the comparison of these two models in more details in section \ref{sect:extshear}.

\begin{table}
	\centering
	\caption{
    Comparison of the masses of the individual mass-clumps used in this study and in \citealt{Jauzac2016}.  Figure \ref{fig:MJ_clumps} shows the location of each of the clumps.}
	\label{tab:mass_comparison}
	\begin{tabular}{lcc} % four columns, alignment for each
		\hline
		\hline
Clump &This study  & J16 \\
&M(10$\protect^{13}$) M$\protect_{\odot}$&M(10$\protect^{13}$) M$\protect_{\odot}$\\
\hline
 N & 9.86&6.10$\pm$0.5\\
 NW & 13.22 &7.90$\pm$0.60\\
 S1 & 4.61 &5.00$\pm$0.40\\
 S2 & 5.00 &5.40$\pm$0.50\\
 S3 & 12.4 &6.50$\pm$0.60\\
 S4 & 5.68 &5.50$\pm$1.20\\
	\hline
\end{tabular}
\end{table}

\subsection{Dependence on the constraints}
\label{sect:constraints}
To this point, we have tested several model parametrisations while limiting our constraints to the {\it gold} set. We now reverse the process and look into the effect of using other sets of constraints ({\it silver, bronze}, and {\it copper}, see \ref{sect:reliability}), while keeping a fixed set of parameters. For these tests we use the model parametrisation which includes substructures in the outskirts.

For each set of constraints we optimise the model and the best-fit parameters are presented in Table \ref{tab:best-fit}.
There is an apparent improvement on the rms from the \gold-constrained (0.67\arcsec)  to the \silver-constrained (0.59\arcsec) model. However the higher BIC (400) of the \silver-constrained compared to the \gold-constrained model (332) suggests that the penalty of adding new constraints outweighs the improvement in the fit, despite the lower rms. In other words, the BIC indicates that the additional photometric candidates do not bring new information to the constraints that already exist in the \gold\ sample. 

Looking into the \bronze-constrained model we can see the rms has increased relative to the \silver-constrained model, returning to the same level as the original \gold-constrained model. However the penalty of adding the additional constraints is clearly seen since the BIC is significantly larger than either the \gold- or \silver-constrained model values.

The considerably larger rms value for the model with \copper constraints is mainly due to systematics, such as including the wrong (non-spectroscopic) counterimage to systems which have spectroscopic redshifts.  This may include images 10.3 and 37.3, which provide some of the largest rms errors on the model (rms$_{37.3}$ = 9.62\arcsec; rms$_{10.3}$ = 4.91\arcsec), see the rms columns in Table \ref{multipletable}

\begin{table*}
\centering
\caption{Lens models and best-fit parameters for each dPIE clump. From left to right: central coordinates (measured relative to the position ($\alpha=00^{\rm h} 14^{\rm m} 20.7022^{\rm s}$, $\delta=-30^{\rm o} 24\arcmin 00.6264\arcsec $), ellipticity (defined to be $(a^2-b^2) / (a^2+b^2)$, where $a$ and $b$ are the semi-major and semi-minor axes of the ellipse), position angle, central velocity dispersion, cut and core radii. Quantities within brackets are fixed parameters in the model.
DM1 and DM2 refer to the large scale dark matter halo while BCG1 and BCG2 refer to the first and second brightest galaxy in the cluster core. NorthGal and SouthGal are two background galaxies that are projected into the lens plane to be modeled%.
\RC{ as they could influence locally the position of multiple images.} J16 clumps A and J16 clumps B divide the six cluster substructures detected in J16 into two groups \RC{: clumps A (N, NW and S3 in J16) have bright luminous counterparts, while clumps B (S1, S2 and S4 in J16) have faint luminous counterparts}.
}
\label{tab:best-fit}
\begin{tabular}{lrcccccccc}
\hline
Model name & Component & $\Delta\alpha$ & $\Delta\delta$ & $\varepsilon$ & $\theta$ & $\sigma_0$ & r$_{\rm cut}$ & r$_{\rm core}$\\
(Fit statistics) & -- & (\arcsec) & (\arcsec) &   & ($\deg$) & (km\ s$^{-1}$) & (kpc) & (kpc)\\
\hline
Gold constraints & DM1 & -2.1$^{+0.3}_{-0.3}$ & 1.4$^{+0.0}_{-0.4}$ & 0.83$^{+0.01}_{-0.02}$ & 90.5$^{+1.0}_{-1.1}$ & 607.1$^{+7.6}_{-0.2}$ & [1000.0] & 18.8$^{+1.2}_{-1.0}$\\ 
rms = 0.67\arcsec\  & DM2 & -17.7$^{+0.2}_{-0.3}$ & -15.7$^{+0.4}_{-0.3}$ & 0.51$^{+0.02}_{-0.02}$ & 45.2$^{+1.3}_{-0.8}$ & 742.8$^{+20.1}_{-14.2}$ & [1000.0] & 10.7$^{+1.1}_{-0.5}$\\ 
$\chi^2/\nu$ = 1.7 & BCG1 & [0.0] & [0.0] & [0.2] & [-76.0] & 355.2$^{+11.3}_{-10.2}$ & [28.5] & [0.3]\\ 
$\log$($\mathcal{L}$) = -113& BCG2 & [-17.9] & [-20.0] & [0.38] & [14.8] & 321.7$^{+15.3}_{-7.3}$ & [29.5] & [0.3]\\ 
BIC = 332& NorthGal & [-3.6] & [24.7] & [0.72] & [-33.0] & 175.6$^{+8.7}_{-13.8}$ & [13.2] & [0.1]\\ 
 & SouthGal & [-12.7] & [-0.8] & [0.3] & [-46.6] & 10.6$^{+43.2}_{-3.6}$ & 1.5$^{+20.6}_{-0.7}$ & [0.1]\\ 
   & $L^{*}$ Galaxy & -- & -- & -- & -- & 155.5$^{+4.2}_{-5.9}$ & 13.7$^{+1.0}_{-0.6}$ & [0.15]\\ 
   & J16 clumps A & -- & -- & -- & -- & 209.6$^{+5.8}_{-6.1}$ & [300.0] & [0.0]\\ 
   & J16 clumps B & -- & -- & -- & -- & 82.7$^{+8.6}_{-9.3}$ & [600.0] & [0.0]\\ 
\hline 
Silver constraints & DM1 & -1.4$^{+0.3}_{-0.4}$ & 3.9$^{+0.0}_{-0.4}$ & 0.83$^{+0.01}_{-0.01}$ & 92.1$^{+1.0}_{-1.0}$ & 553.6$^{+17.1}_{-13.4}$ & [1000.0] & 16.5$^{+1.9}_{-1.4}$\\ 
rms = 0.59\arcsec\  & DM2 & -17.4$^{+0.3}_{-0.3}$ & -16.0$^{+0.3}_{-0.4}$ & 0.46$^{+0.02}_{-0.02}$ & 44.2$^{+1.1}_{-1.1}$ & 732.2$^{+15.3}_{-16.6}$ & [1000.0] & 9.9$^{+0.7}_{-0.5}$\\ 
  $\chi^2/\nu$ = 1.4 & BCG1 & [0.0] & [0.0] & [0.2] & [-76.0] & 335.8$^{+11.1}_{-10.1}$ & [28.5] & [0.3]\\ 
 $\log$($\mathcal{L}$) = -120 & BCG2 & [-17.9] & [-20.0] & [0.38] & [14.8] & 305.3$^{+9.3}_{-8.9}$ & [29.5] & [0.3]\\ 
 BIC = 400 & NorthGal & [-3.6] & [24.7] & [0.72] & [-33.0] & 180.8$^{+9.3}_{-14.0}$ & [13.2] & [0.1]\\ 
& SouthGal & [-12.7] & [-0.8] & [0.3] & [-46.6] & 81.9$^{+52.7}_{-7.8}$ & 1.4$^{+20.7}_{-0.7}$ & [0.1]\\ 
 & $L^{*}$ Galaxy & -- & -- & -- & -- & 163.5$^{+4.9}_{-4.8}$ & 13.3$^{+0.8}_{-0.5}$ & [0.15]\\ 
 & J16 clumps A & -- & -- & -- & -- & 218.7$^{+4.4}_{-6.4}$ & [300.0] & [0.0]\\ 
 & J16 clumps B & -- & -- & -- & -- & 105.7$^{+7.3}_{-9.9}$ & [600.0] & [0.0]\\ 
\hline 
Bronze constraints & DM1 & -1.1$^{+0.2}_{-0.3}$ & 3.8$^{+0.0}_{-0.5}$ & 0.87$^{+0.01}_{-0.02}$ & 94.8$^{+0.9}_{-0.7}$ & 591.5$^{+17.4}_{-15.7}$ & [1000.0] & 26.4$^{+1.6}_{-1.5}$\\ 
rms = 0.67\arcsec\  & DM2 & -16.5$^{+0.2}_{-0.1}$ & -15.4$^{+0.2}_{-0.2}$ & 0.49$^{+0.01}_{-0.02}$ & 43.6$^{+0.7}_{-0.7}$ & 765.8$^{+9.5}_{-7.2}$ & [1000.0] & 11.0$^{+0.5}_{-0.2}$\\ 
 $\chi^2/\nu$ = 1.8& BCG1 & [0.0] & [0.0] & [0.2] & [-76.0] & 353.7$^{+6.9}_{-9.5}$ & [28.5] & [0.3]\\ 
$\log$($\mathcal{L}$) = -192  & BCG2 & [-17.9] & [-20.0] & [0.38] & [14.8] & 328.1$^{+5.5}_{-5.9}$ & [29.5] & [0.3]\\ 
 BIC = 622 & NorthGal & [-3.6] & [24.7] & [0.72] & [-33.0] & 203.2$^{+6.0}_{-9.1}$ & [13.2] & [0.1]\\ 
 & SouthGal & [-12.7] & [-0.8] & [0.3] & [-46.6] & 83.2$^{+16.0}_{-50.3}$ & 4.2$^{+8.6}_{-9.9}$ & [0.1]\\ 
 & $L^{*}$ Galaxy & -- & -- & -- & -- & 181.9$^{+0.8}_{-0.9}$ & 12.8$^{+0.3}_{-0.3}$ & [0.15]\\ 
 & J16 clumps A & -- & -- & -- & -- & 211.5$^{+3.0}_{-3.3}$ & [300.0] & [0.0]\\ 
 & J16 clumps B & -- & -- & -- & -- & 95.3$^{+6.1}_{-7.5}$ & -- & [0.0]\\ 
\hline
Copper constraints & DM1 & -2.2$^{+0.1}_{-0.1}$ & 1.1$^{+0.3}_{-0.4}$ & 0.9$^{+0.01}_{-0.0}$ & 91.5$^{+0.5}_{-0.4}$ & 607.1$^{+4.5}_{-4.0}$ & [1000.0] & 21.2$^{+0.4}_{-0.5}$\\ 
rms = 1.48\arcsec\  & DM2 & -17.3$^{+0.0}_{-0.0}$ & -14.8$^{+0.1}_{-0.1}$ & 0.6$^{+0.01}_{-0.01}$ & 46.4$^{+0.5}_{-0.4}$ & 785.0$^{+4.4}_{-5.1}$ & [1000.0] & 12.2$^{+0.2}_{-0.2}$\\ 
 $\chi^2/\nu$ = 43.2 & BCG1 & [0.0] & [0.0] & [0.2] & [-76.0] & 390.5$^{+6.3}_{-7.8}$ & [28.5] & [0.3]\\ 
$\log$($\mathcal{L}$) = -4511  & BCG2 & [-17.9] & [-20.0] & [0.38] & [14.8] & 387.3$^{+4.4}_{-4.7}$ & [29.5] & [0.3]\\ 
 BIC = 9325 & NorthGal & [-3.6] & [24.7] & [0.72] & [-33.0] & 192.8$^{+4.3}_{-8.3}$ & [13.2] & [0.1]\\ 
 & SouthGal & [-12.7] & [-0.8] & [0.3] & [-46.6] & 186.6$^{+7.6}_{-8.5}$ & 0.4$^{+0.9}_{-0.8}$ & [0.1]\\ 
   & $L^{*}$ Galaxy & -- & -- & -- & -- & 166.0$^{+1.0}_{-2.4}$ & 12.8$^{+0.2}_{-0.2}$ & [0.15]\\ 
   & J16 clumps A & -- & -- & -- & -- & 196.3$^{+1.8}_{-1.7}$ & [300.0] & [0.0]\\ 
   & J16 clumps B & -- & -- & -- & -- & 75.5$^{+9.8}_{-5.1}$ & -- & [0.0]\\ 
\hline

\end{tabular}

\end{table*}

\section{Discussion}
We discuss here the overall structure of the cluster Abell 2744 in the context of the new MUSE data. We investigate the dynamics of cluster members and the influence of the environment of the cluster on our models. 

\subsection{Dynamics of the cluster core}
\label{sect:dynamics}
\citet{Owers2011} performed the largest spectroscopic survey of cluster members to date in the Abell 2744 field, using the AAOmega spectrograph on the Anglo-Australian Telescope (AAT). They measured redshifts for 343 members within a 3 Mpc projected radius from the cluster core. Their analysis of the cluster dynamics clearly preferred a model including 3 dynamical components, with two distinct clumps (A and B) centered around the cluster core and a separate LOS velocity distribution encompassing the rest of the cluster. The strong lensing region we model in this paper is referred to as the southern compact core in their study.

\begin{figure*}
	\includegraphics[width=\textwidth]{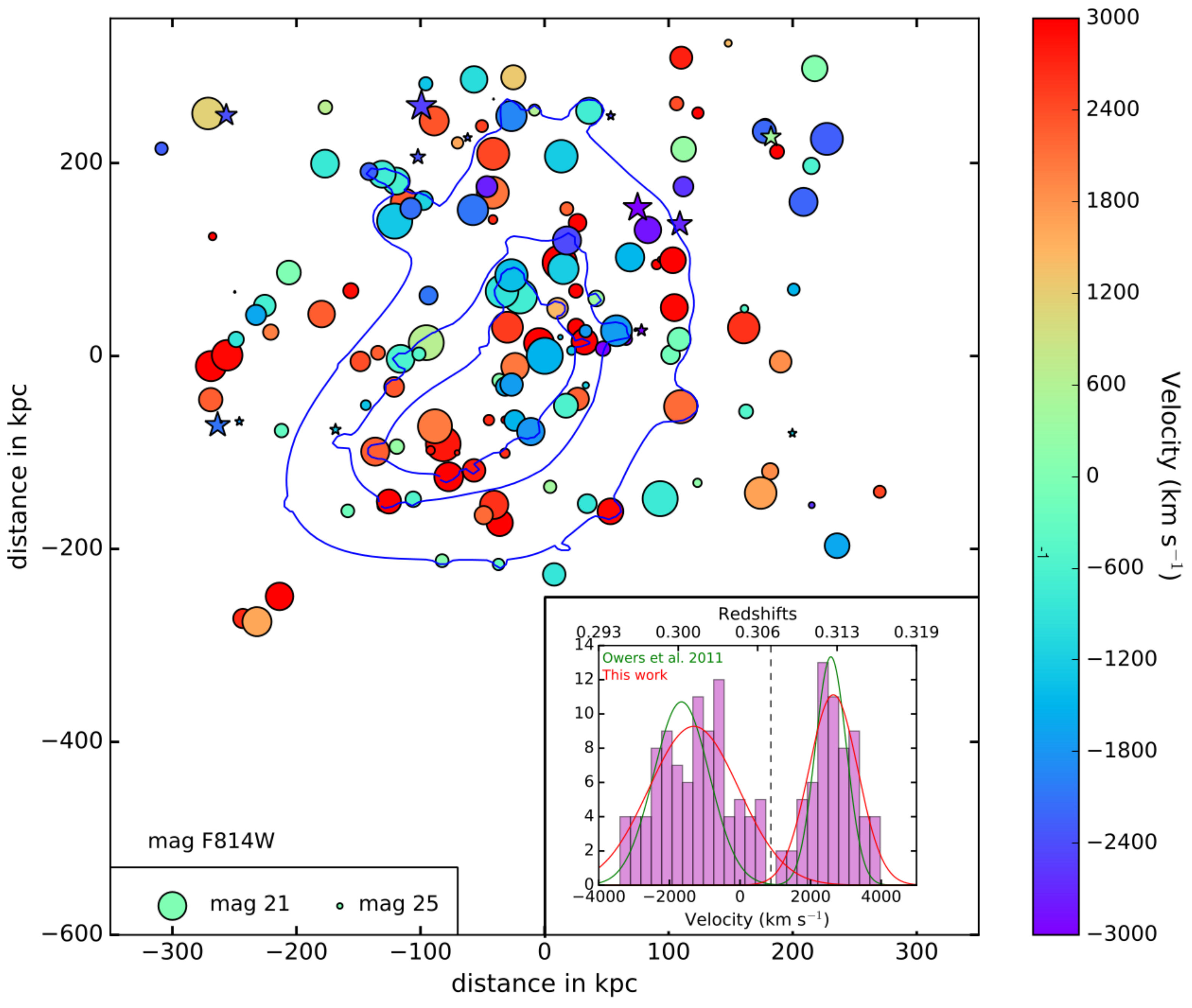}
\caption{Spatial map of the velocity shift of cluster members, relative to the systemic cluster velocity.  Circle symbols represent galaxies with no emission lines and star symbols represent galaxies with strong emission lines.  The histogram of velocities is shown in the inset, overplotted with the velocity peaks found in \citet{Owers2011} and our study.  For reference, we also show mass contours of the cluster (blue lines) at $1 \times 10^9$ M$_\odot\,{\rm kpc}^{-2}$, $1.5 \times 10^9$ M$_\odot\,{\rm kpc}^{-2}$, and $2 \times 10^9$ M$_\odot\,{\rm kpc}^{-2}$.}
    \label{fig:velocity}
\end{figure*}

Despite covering a smaller region around this core ($r<550$ kpc), we measure redshifts for 156 cluster members using the new MUSE observations (Fig.\,\ref{fig:z_full}, middle panel). To get a more robust estimate of their relative velocities we refine these redshifts using the Auto-Z \citep{Baldry2014} software, cross-correlating each spectrum with template spectra consisting of both passive and star-forming galaxies. In Fig.\,\ref{fig:velocity} we present a 2D map of the LOS velocities relative to the cluster redshift $z=0.3064$, defined as the mean redshift in \citet{Owers2011}. The colour scheme used in the figure reflects the relative velocity of each cluster member, while the symbol size is scaled according to the brightness in the ACS/F814W band  from our photometric catalog. 

Figure \ref{fig:velocity} reveals a clear dichotomy in the distribution of velocities, with two groups of cluster members at low and high velocities centered around the NW and SE regions, respectively. Star forming cluster members (represented by star symbols in Fig.\,\ref{fig:velocity} and selected from [\ion{O}{ii}] emission) tend to be located in the outskirts of the field of view ($>100$ kpc radius), where the surface mass density of the cluster drops below $1.5 \times 10^9\ M_\odot$. 

 The bimodality in the distribution of velocities appears clearly in the redshift histogram (Fig.\,\ref{fig:velocity}, inset). 
We adjust each of the two components (separated at $v=870$ km\,s$^{-1}$) with a Gaussian distribution and find the parameters ($v_{\rm center}=-1308\pm161$ km\,s$^{-1}$, $\sigma=1277\pm189$ km\,s$^{-1}$) and ($v_{\rm center}=2644\pm72$ km\,s$^{-1}$, $\sigma=695\pm76$ km\,s$^{-1}$). These values
are remarkably close to the parameters found by \citet{Owers2011} for Clump B ($v_{\rm center}=-1658$ km\,s$^{-1}$, $\sigma=789$ km\,s$^{-1}$) and Clump A ($v_{\rm center}=2574$ km\,s$^{-1}$, $\sigma=441$ km\,s$^{-1}$) respectively, as seen in Fig.\,\ref{fig:velocity}. The main difference is a small excess in the distribution of galaxies in the velocity range [-200,800] km\,s$^{-1}$. This excess could be due to the presence of cluster members that do not belong to the A or B clump.

Due to the clear gap in velocities between the two clumps A and B ($\sim 4000$ km\,s$^{-1}$) the simplest hypothesis would suggest a pre-merger phase between those two components (e.g. \citealt{Maurogordato2011}). 
These clumps are separated by $\sim$ 75 kpc in projection, along a similar SE - NW direction as the two DM mass clumps of the mass model.
This projected distance is small but significant: the two clumps are therefore separated both spatially and dynamically. This small offset strengthens the assumption of a merging process along the line-of-sight.  However, the relative complexity of the X-ray emission in the cluster \citep{Owers2011,Jauzac2016} suggest a more complicated scenario. A joint analysis of the temperature of the gas and the dynamics of galaxies would shed light into the cluster merging history, but is out of the scope of the paper.

\subsection{External shear effect}
\label{sect:extshear}

\begin{figure*}
	\includegraphics[width=\textwidth]{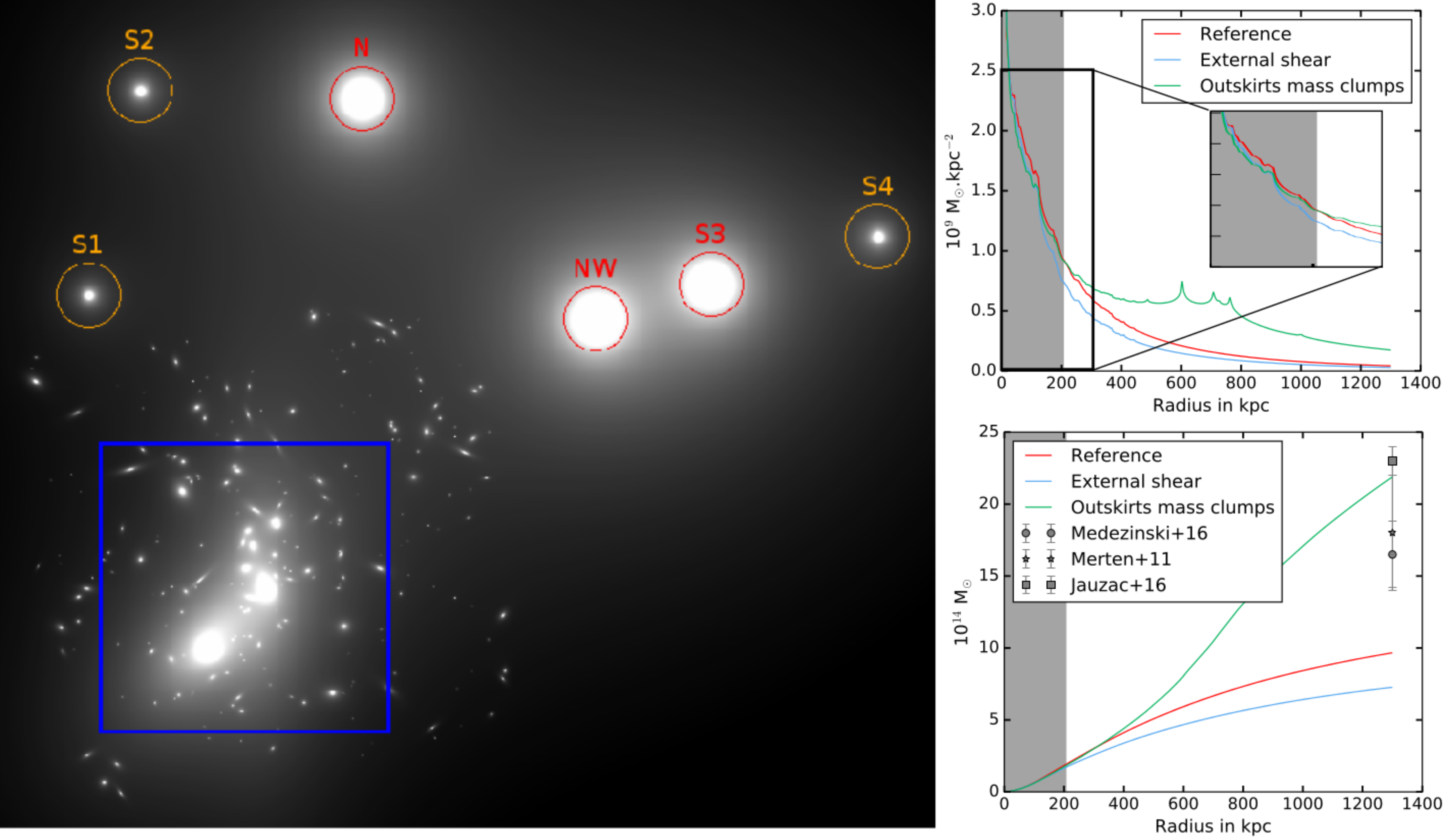}
    \caption{Left: mass map of the model included mass clump in the outskirts following J16 nomenclature. The blue box is $\sim$290 kpc large and corresponds to the region where the shear comparison was made (Fig.\,\ref{fig:shear1}). Circles show the outskirts mass potentials which correspond to the two scaling relations, red circles are associated with high luminosity counterparts while orange circles group the other potentials (See sect \ref{sect:MJclumps}). Top right:
differential mass profile of three models. ``Reference'' is the model using only spectroscopic constraints and clumps in the WFC3 FoV, ``External shear'' adds a constant external shear to the previous model and the ``Outskirts potentials'' model replaces the constant shear with the mass clumps in the outskirts, which are represented by the circles in the left-hand panel. Lower right: integrated mass profiles of each of the previous models. The three points represent the mass at 1.3 Mpc found by three weak lensing studies. The mass profile is computed on a circular radius centered on the first BCG in the southern cluster core.  }
    \label{fig:MJ_clumps}
\end{figure*}

Here we discuss the relevance of including external shear in the model, first mentioned in section \ref{sect:MJclumps}. To probe for shear effects, we build three models using only the \textit{gold} spectroscopic constraints. The first model, called the ``reference'' model, is built according to the methodology reported in section \ref{sect:SLanalysis}, only including mass potentials which are in the WFC3 FoV. After optimising this model, we find a best-fit rms of 0.86\arcsec. Next, we add a constant external shear field to the reference model which reduces the global rms (0.78\arcsec) during optimisation.  This model is known as the ``external shear'' model.  Finally, we replace the external shear field with the optimised J16 weak-lensing mass clumps as described in \ref{sect:SLanalysis}, creating what we call the ``outskirt mass clumps'' model.  We find the optimised rms of this model to be 0.67\arcsec.
Comparing the overall model properties, we find that the external shear model is the least massive of the three. This arises from the fact that a pure external shear field has no intrinsic mass in our modelling scheme. 
The difference is small though, as shown by the two mass profiles in the right-hand panels of figure \ref{fig:MJ_clumps} and the total mass contained within a radius R = 1.3Mpc, which differs by only $\sim$7\%.

\begin{figure*}
    \includegraphics[width=\columnwidth]{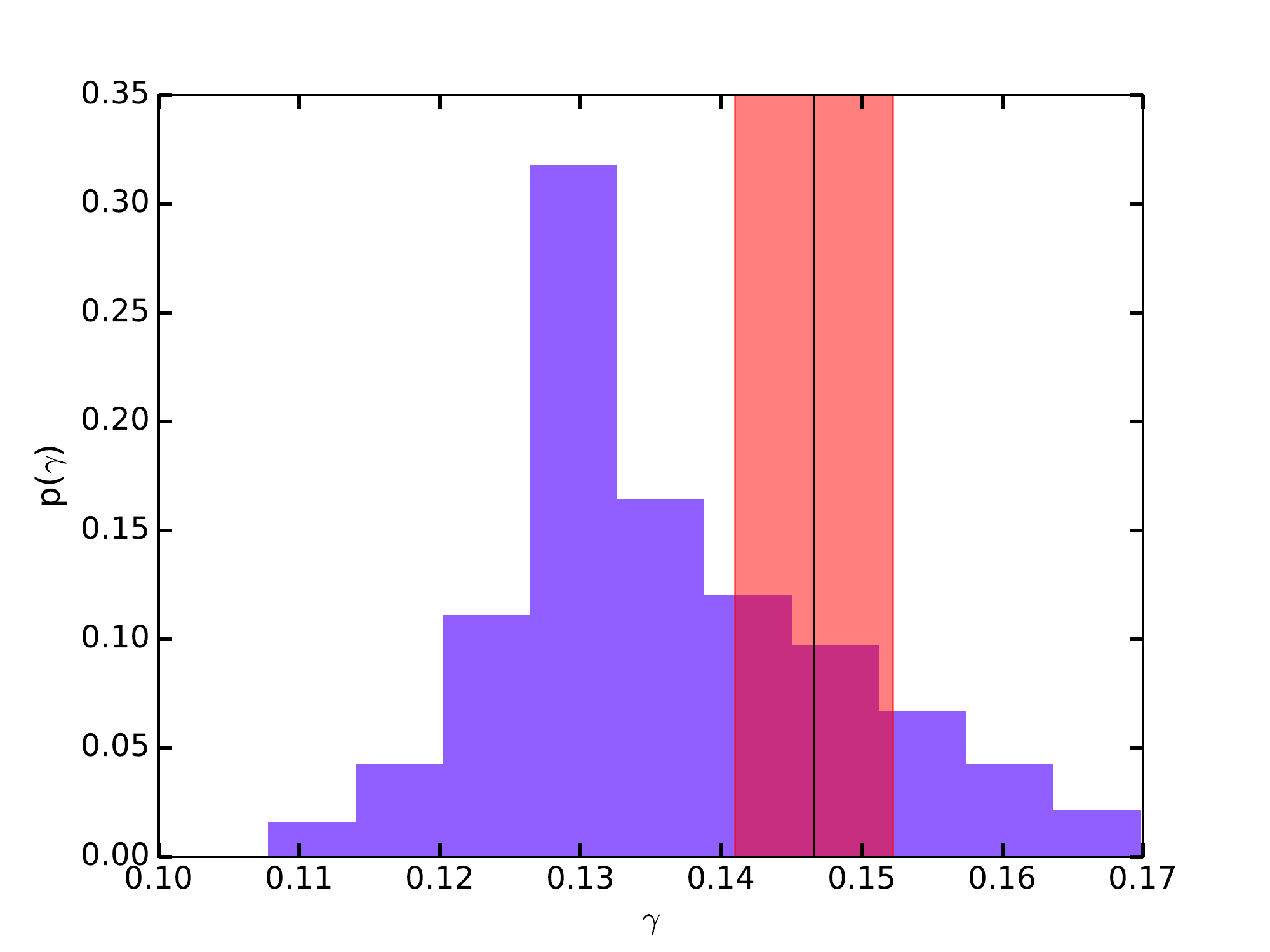}
    \includegraphics[width=\columnwidth]{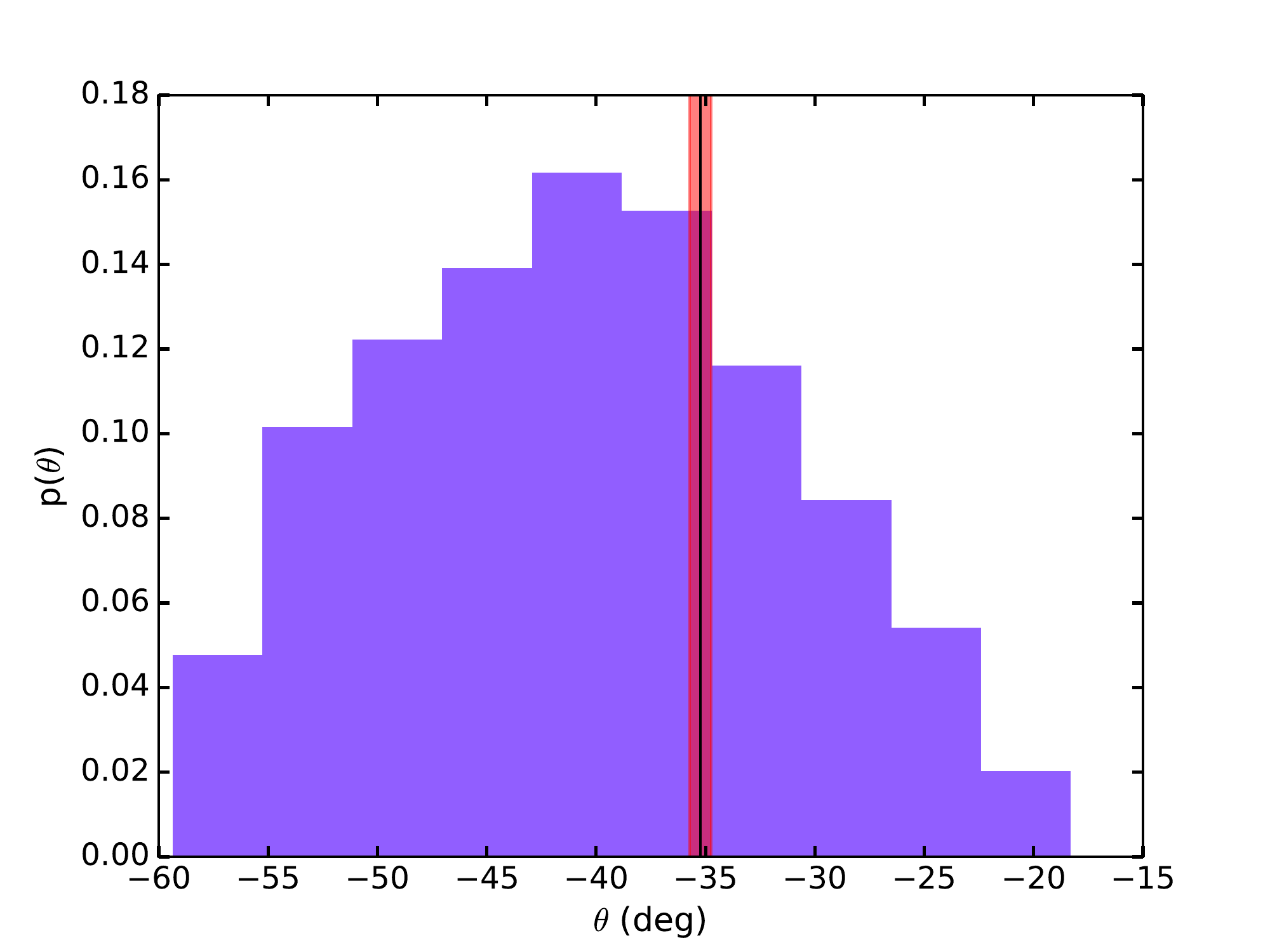}
    \caption{ The left panel shows the histogram of the $\gamma$ value for the models included mass clumps in the outskirts, values are computed into a square region enclosing all the constraints and mark as a blue box in Fig\,\ref{fig:MJ_clumps}. The black line and the shaded red region represent the mean and its 1 sigma uncertainties for the value of the $\gamma$ in the external shear model. The left panel is similar and compare with the shear angle value $\theta$. }
    \label{fig:shear1}
\end{figure*}

Since the main influence on the strong lensing region by the outskirts mass clumps is a shear effect, we can compare the shear fields produced by this model and the pure external shear model. 
To do this, we generate a grid of shear values over the entire Abell 2744 field, and measure the induced ellipticity in a region encompassing all of the strong-lensing constraints.  
The histograms in Fig.\,\ref{fig:shear1} show the comparison between the external shear and the shear induced by the outskirts mass clumps. We see that the shear fields are entirely consistent in terms of strength $\gamma$ and orientation $\theta$ for both models. For both parameters, the black line represent the mean value and the red area the 1-$\sigma$ error computed from all the models sampled during the external shear model optimisation process. In both histograms, the blue distribution is computed from the outskirts mass model taking the shear value within the central region described before and shown in Fig.\,\ref{fig:MJ_clumps}. The good agreement between this distribution (initially detected based on weak-lensing effect at large radius) and the external shear parameters highlights the need for including environmental effects to better model the cluster core.

Moreover, we note that systems 15 and 16 \citep{Richard2014} are close enough to the N and NW clumps, that their positions are likely significantly affected by these masses.  However, for simplicity -- since both systems lack spectroscopic redshift -- we did not use them as constraints in the outskirts mass model (see section \ref{sect:MJclumps}).  It should be mentioned, though, that the masses shown in Table \ref{tab:mass_comparison} are large enough to produce multiple images.

Overall, thanks to numerous constraints with spectroscopy we are able to reach the level of sensitivity where mass clumps in the environment influence our model. The improvement in rms combined with the comparison between the external shear model and the outskirts mass model show that distant clumps ($\sim$700 kpc) contribute to the mass reconstruction in the vicinity of the cluster core by their induced shear.
However, the mass profiles from both models tend to only separate one from each other at the end of the multiply imaged region ($\sim$200 kpc).

\subsection{The overall mass profile}
\label{sect:mprofiles}
We now compare the two best mass profiles found in this study (external shear and outskirt masses) to similar profiles derived from other post-HFF strong lensing models of Abell 2744. To do this, we construct the azimuthally-averaged radial mass profile centered on the first BCG.  The mass maps are generated by HFF modeling teams and are publicly available in the Mikulski Archive for Space Telescopes (MAST) \footnote{\url{https://archive.stsci.edu/prepds/frontier/lensmodels/}}.

\begin{figure*}
    \includegraphics[trim={0 2cm 0 1cm},width=\textwidth]{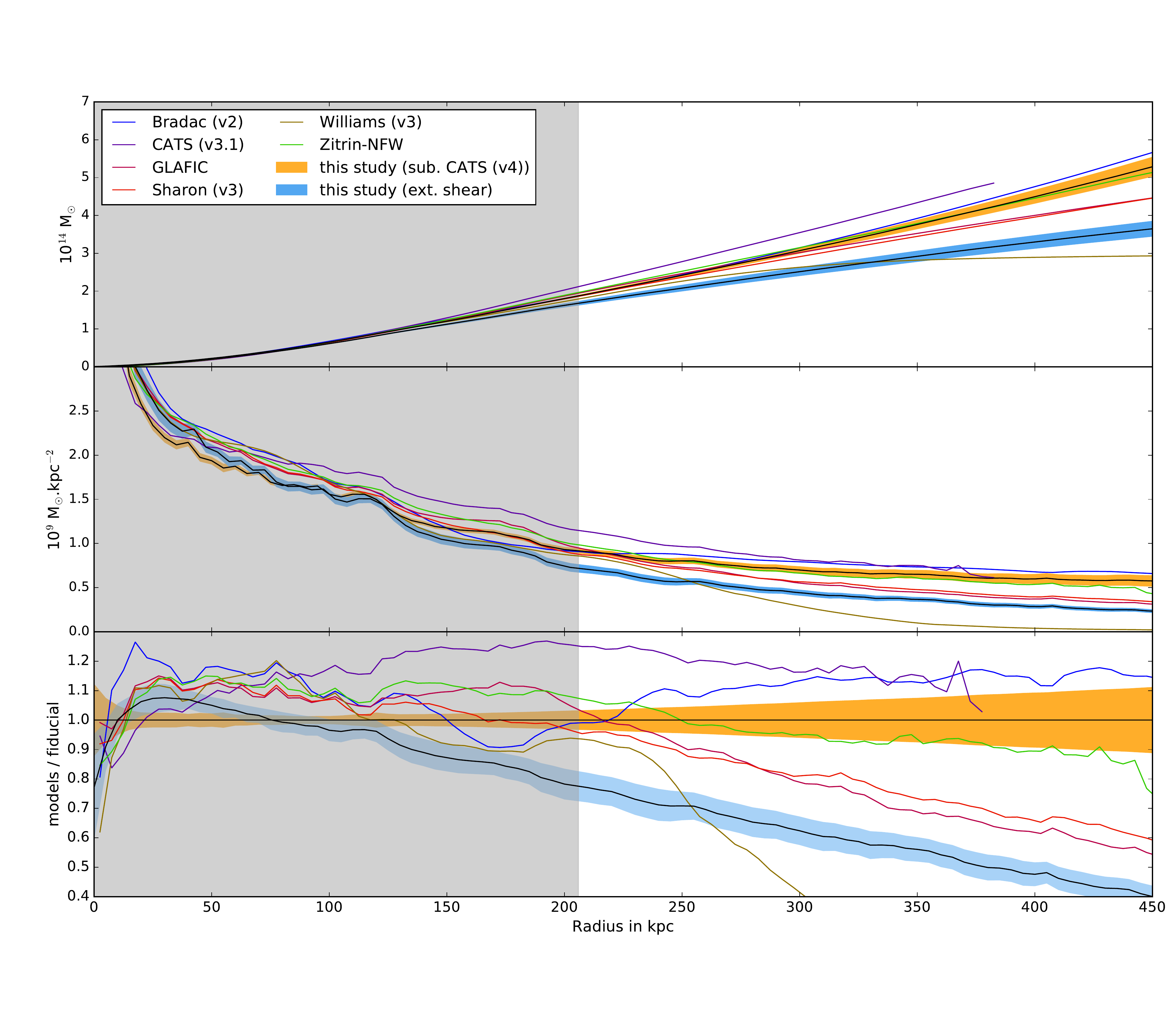}
        \caption{\RC{The integrated (upper panel), differential (middle panel) and ratio (lower panel)} mass profiles of the cluster from different studies. 
 \RC{The ratio is computed over our fiducial model, the one with clumps in the outskirts and only \gold\ sample}. The black line surrounded by the blue shaded region represents the mass profile for the external shear and 3$\sigma$ statistical uncertainties. The black line surrounded by the shaded orange region represents mass profile for the fiducial model 
 and the 3$\sigma$ statistical uncertainties. The gray area represents the region where multiple images used as constraints are located.}
    \label{fig:mass_other}
\end{figure*}

Figure \ref{fig:mass_other} shows the mass profiles derived from all the studies, including our new analysis. The differential mass profile (middle panel) shows that within $\sim$100 kpc both of our mass profiles are lower than any other study (except the central 20 kpc of the CATS (v3.1) model) and our 3$\sigma$ statistical uncertainties do not cover the difference in mass.
The gray area represents the area where multiple images are expected, and corresponds to the region where our constraints are located.
Between 100 kpc and the end of the gray area different mass profiles tend to diverge one from another. Most of the models agree more with the outskirts mass model, which we will use as our fiducial model in the rest of the paper.
Mass profiles are extrapolated beyond the edge of the gray area, since no hard strong-lensing constraints are found at these distances. Here, a clear separation between models appears: Bradac (v2), CATS (v3.1), Zitrin-NFW (v3), and our fiducial model profiles tend to keep a high density at large radii, while GLAFIC (v3), Sharon (v3), and and our external-shear model generate a lower mass profile. At very large distances, the profile from Williams (v3) drops considerably lower than all others. These effects are seen in both panels of Fig\,\ref{fig:mass_other}. 

The discrepancies in the inner core could be due to the new hard constraints we add to our mass models, while the discrepancies outside the multi-imaged region may be related to different aspects of the modeling technique and their sensitivity to environmental effects.  By probing the overall discrepancies in the mass profiles between different analyses we can begin to understand the magnitude of systematic uncertainties and their overall effects on the modeling.

Compared to other parametric models (v3 from CATS, GLAFIC, and Sharon), our fiducial model reaches a similar, or slightly higher rms (0.6\arcsec). The main difference is a higher ratio between the number of constraints ($k$) and the number of free parameters ($n$), which can be used as a metric on the level of constraints available. Thanks to the large number of spectroscopic redshifts available for the \gold\ systems, we obtain $k/n=134/30=4.5$ compared to $70/30=2.33$ for CATS (v3), and $146/100=1.46$ for GLAFIC (v3) \citep{Kawamata2016}. The ratio between constraints and free parameters in our model is comparable to the Sharon (v3) model ($k/n=108/27=4$, K.Sharon private comm.) but they use multiple emission knots in each galaxy as constraints, making them not strictly independent. We note that our current model is able to accommodate a large number of constraints with only a small number of free parameters, which means that we are reaching a limit (in rms) with this method.

\subsection{Estimating the level of systematics in mass models}

Our fiducial model, containing mass clumps in the outskirts, constrained with the \gold\ set of multiple images and released as CATS (v4) as part of the Frontier Fields model challenge, shows a \textit{statistical} error of $\sigma_{\rm stat}\sim1\%$ on the mass density \RC{profile} in the cluster core (see Fig.\ref{fig:mass_other} and Sect.\ref{sect:mprofiles}). This error is comparable to the estimates from the previous CATS (v3) model. \RC{However, these uncertainties only arise from the statistical fluctuation of models during the minimisation procedure and do not reflect inherent \textit{systematic} uncertainties. Such uncertainties}
can arise from the choice of constraints, the model parametrisation, and scatter on the position of multiples images due to unaccounted structures within the cluster or over the line of sight.
 
\RC{Thanks to our large sample of spectrosopic redshifts, we are able to significantly improve our model compared to previous work.  This is mainly thanks to our ability to unambiguously identify multiply-imaged systems used as model constraints, reducing the overall systematic uncerteinty on the model.  Specifically, we no longer misidentify the configuration of systems, which has been problematic in the past.  A key example of this can be seen in the reconfiguration of previously-identified systems 5 and 47, where we subdivide these objects into two distinct segments.  As highlighted in sect\,\ref{sect:mass-distri} and in Fig\,\ref{fig:2DM-3DM}, models with the wrong configuration do not converge toward a good model, instead straying far from the true spectroscopic redshift and maintaining a high final rms (1.87\arcsec).
In addition to the mass profile, misidentification can also affect magnification predictions and lead to a biased measurement on the properties of lensed galaxies.  However, such topics are outside of the scope of this paper.}
\citet{Johnson2016} studied systematic uncertainties on the lensing mass reconstruction for the simulated lensing cluster ARES \citep{Meneghetti2016}. ARES, and its companion HERA, are simulated lensing clusters designed by the HFF project 
as a way to fairly compare mass models from different teams.  
From their work on ARES, \citet{Johnson2016} investigate the effect of systematic uncertainties arising from the choice of lensing constraints, dividing all constraints in two categories: spectroscopic and non-spectroscopic. By testing a series of combinations of constraints with and without spectroscopy, they conclude that 25 spectroscopic systems (among the 66 available in the ARES cluster) are required to get the true rms of the cluster and reach the systematic level of uncertainty. In their work they also discover that constraints distributed inhomogeneously (either spatially or in redshift) lead to strongly disfavored models.  We believe that with our 29 systems
evenly distributed around the cluster, our level of uncertainty drops very close to the systematic limit.

\begin{figure}
	\includegraphics[width=\columnwidth]{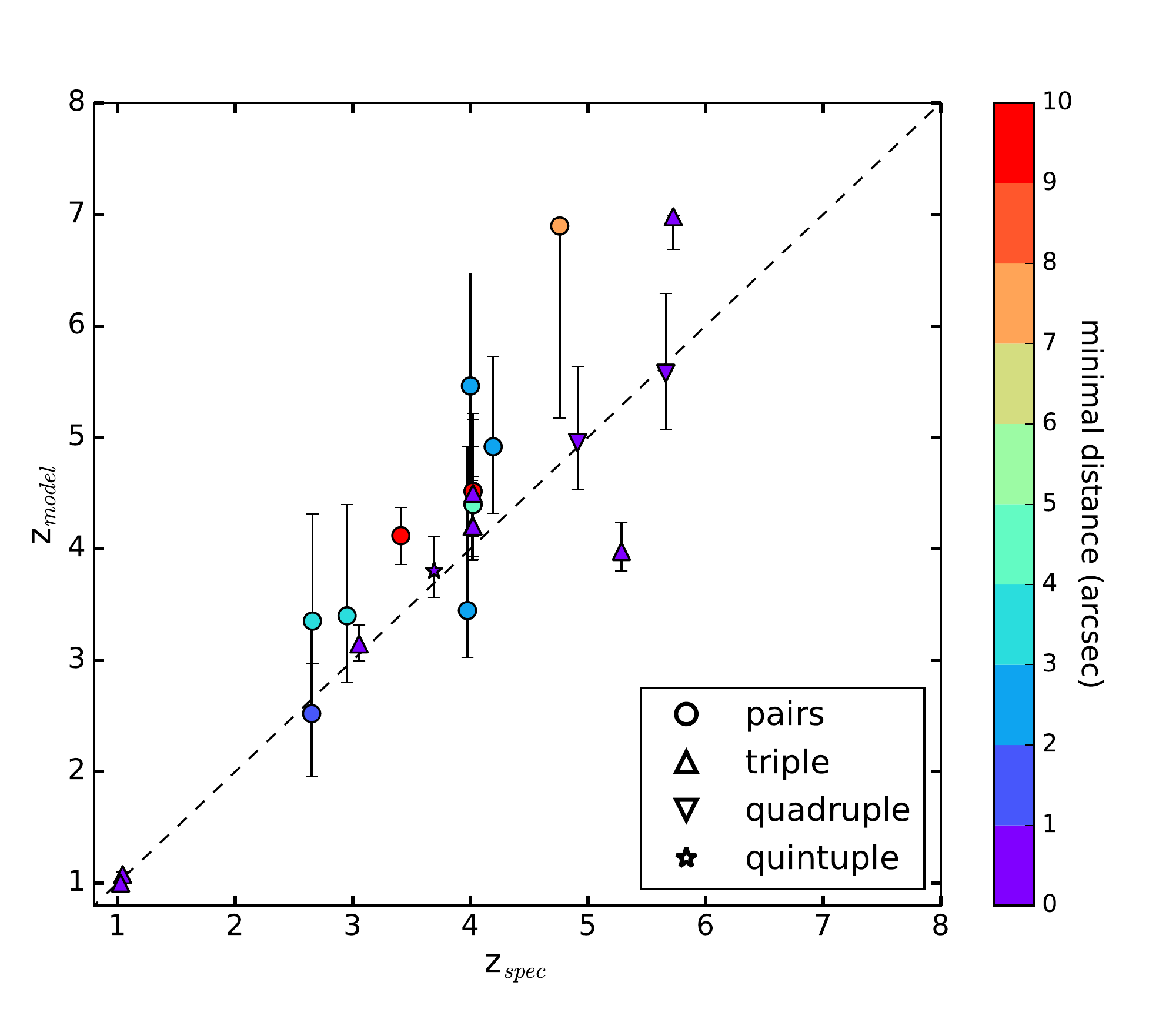}
        \caption{\RC{Comparison between model-predicted redshifts and the new  spectroscopic measurements. The models used for prediction is based on the fiducial model (mass clumps in the outskirts plus \gold\ sample) but newly measured redshifts compared to \citet{Wang2015} are optimised during the minimisation procedure. The colour codes for the minimal distance between two images from the same systems whereas symbols show how many images compose the systems.} }
    \label{fig:zmodelVSzspec}
\end{figure}

As the systematics likely arise from different contributions we try to estimate their level using complementary techniques. 
\RC{First, we try to highlight the reduced level of systematic error brought about by the addition of new spectroscopic redshifts. To do this, we modify our fiducial model, keeping only systems identified by \citet{Wang2015} as objects with fixed redshifts (though we update all redshifts to their current MUSE values.)  We leave the redshifts of the other systems as free parameters, which are then optimised during minimisation.  To characterise the systematic differences, we compare the final model-optimised redshifts to the spectroscopic values used in the fiducial model.  This comparison is shown in Fig\,\ref{fig:zmodelVSzspec}. From the figure, we see that the optimised redshifts are systematically higher than the values measured from spectroscopy.  This impacts the overall mass distribution of the cluster and can affect our interpretation of the final results.}

\begin{figure}
	\includegraphics[width=\columnwidth]{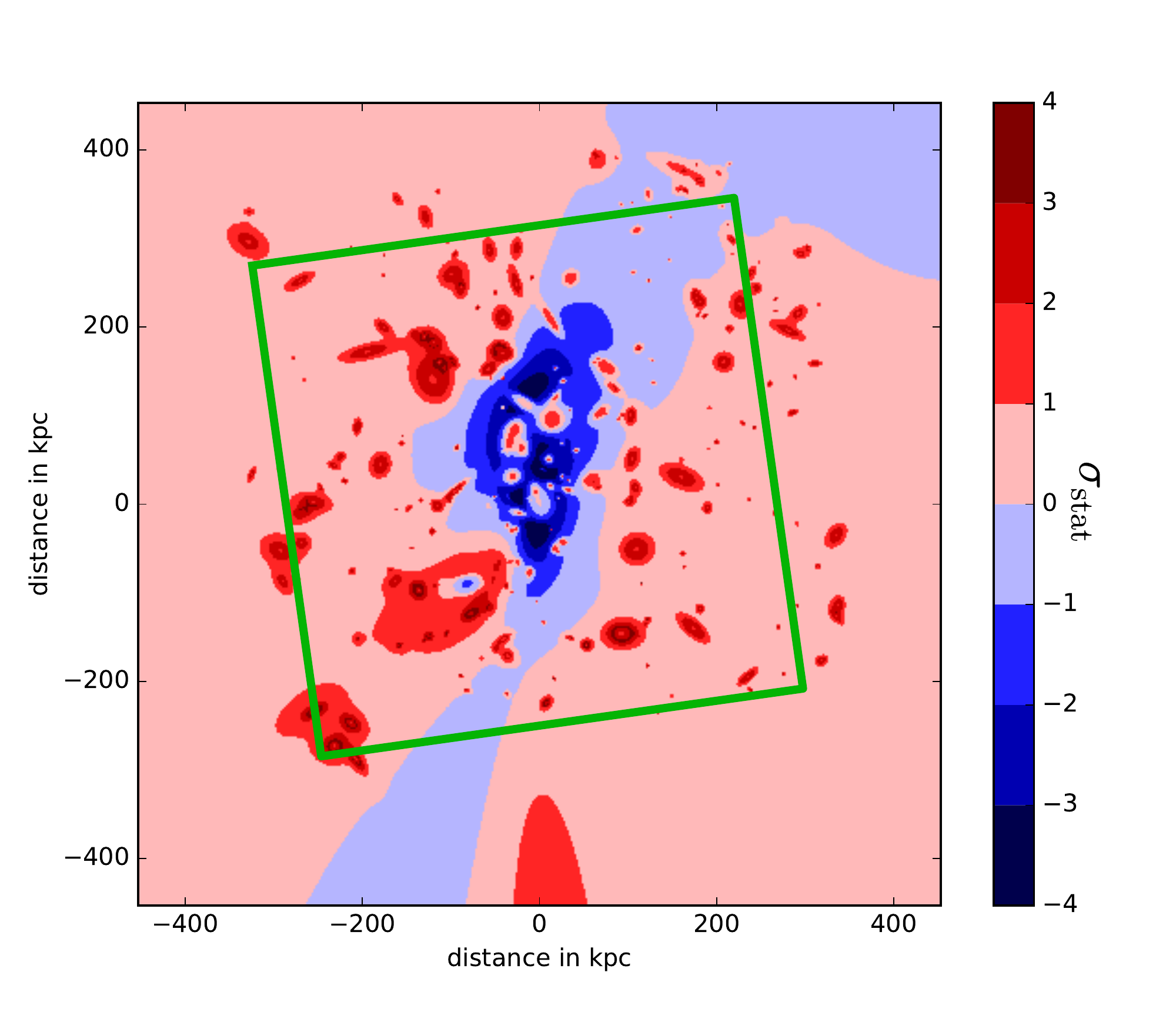}
        \caption{\RC{Relative variation in the 2D surface mass distribution, in units of the statistical uncertainty $\sigma{\rm stat}$, between two models with (fiducial) and without the MUSE spectroscopic constraints. These maps can be related to systematics uncertainties, with negative values referring to underestimated mass and positive value overestimated mass. The green box represents the 2\arcmin x2\arcmin MUSE-mosaic field of view. This highlights the benefit of a deep and wide spectroscopic coverage such as this study. }}
    \label{fig:systematic_ratio}
\end{figure}

\RC{Additionally, we investigate differences in the 2D mass distribution between the two models, to see if they are larger than the level of statistical uncertainty. Globally, enclosed within large radii, the total mass is almost unchanged.  However, we do notice differences at smaller scales, so we probe these local discrepancies to see if the significance of this signal is important. To do so we compute the difference in $\kappa$ (convergence) between the two models divided by the standard deviation of the modified (free-redshift) model. This is shown in Fig\,\ref{fig:systematic_ratio}.  From the figure we see that systematic variations in the very center and at the outer limit of the core are in opposite directions but both have magnitudes higher than 2 times $\sigma{\rm stat}$, a significant source of systematic variation. 
By doing this we note that we are only probing the systematics due to the addition of new redshift information.}

\RC{Secondly, }
we compare the mass density profiles of the models when changing the sets of constraints, keeping a fixed parametrisation (Sect. \ref{sect:constraints}). 
This comparison can be done within the cluster core ($r<200$ kpc) between the \gold, \silver\ and \bronze-constrained models which have a similar rms overall. We find that by including these less reliable constraints, the \silver-constrained model gives systematically higher mass densities by $\sim$2\% (2 $\sigma_{\rm stat}$) compared to the \gold\ model, 
while the \bronze-constrained model is $\sim$3\% higher (3 $\sigma_{\rm stat}$).
\RC{This test highlights a relative systematic in the mass distribution. Specifically, we see that including less reliable systems can alter the measured mass distribution.  This is mainly due to two reasons. 
First, by adding extra (misidentified) multiple image constraints, the model will add additional mass where it is not needed.  Second, by adding extra (correct) multiple image constraints, the model becomes more sensitive to additional regions of space which can contain, for example, local mass substructures.}

\RC{Thirdly, }
we can compare the results of different model parameterisation, this time keeping the constraints fixed to the \gold\ set, as in Sect.\,\ref{sect:extshear}. Specifically, we compare discrepancies between the external shear model and the outskirts mass clump model. At a distance of 200 kpc from the cluster center, we measure a typical variation of $\sim$6\% between the \RC{integrated} mass profiles of the two models (upper panel of Fig \ref{fig:mass_other}), giving another estimate of systemic uncertainty on the mass profiles. 

Similarly, we can compare our models to other HFF models which do not follow the same parametric approach. \RC{It is not fair to compare modeling techniques here because} 
latest published models (v3) do not include the same number of spectroscopic constraints.
\RC{However it is interesting to compare mass profiles from previous models with our more robust models, probing the scatter due to their lack of redshifts. From Fig.\,\ref{fig:mass_other} we can clearly see in the top panel that our fiducial model in orange is one the most massive models. More interestingly, due to the redshift contribution there is a significant reduction of the mass in the inner core, between 20kpc and 100kpc, by 10\%. }

From Fig.\,\ref{fig:mass_other}, we can see that the CATS (v3) model mass profile is significantly higher than other models in the cluster core, while this is no longer the case in our fiducial model. We note that a similar discrepancy was also reported by \citet{Priewe2017} as well as Bouwens et al. (in prep.) on the CATS (v3) model regarding systematically higher magnification values than other v3 models. Indeed, systematic uncertainties on the mass profiles also reflect on the magnification of background sources, especially for high values of $\mu$. Compared with our fiducial (\gold\ constrained) model, the magnification of a $z=6$ source typically vary by 5-6\% in the low magnification region ($\mu<10$) and between 10-20\% in the high magnification region ($\mu>10$), when using the \silver\  and \bronze\-constrained models instead. 

\begin{figure}
	\includegraphics[width=\columnwidth]{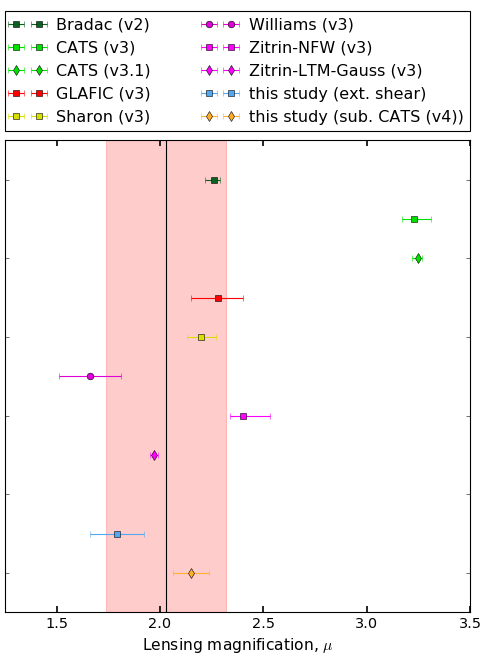}
    \caption{
    Comparison of the observed lensing magnification for the supernovae HFF14Tom to predictions from lens models. The vertical black line shows the constraints from the supernovae reported in \citealt{Rodney2015}, the shaded region marking the total uncertainty. Markers with horizontal error bars show the median magnification and 68\% confidence region from each models.   }
    \label{fig:SNHFF14Tom}
\end{figure}

Finally, we can look at the magnified supernova SN HFF14 Tom ($\alpha=00^{\rm h}14^{\rm m}17.87^{\rm s}$, $\delta=30^{\rm o}23^{\arcmin}59.7^{\arcsec}$) discovered at $z=1.3457\pm0.0001$ behind Abell 2744 \citep{Rodney2015}.  As a Type Ia supernova, the intrinsic luminosity of SN Tom is known from its light curve.  However, its observed luminosity is 0.77 $\pm$ 0.15 magnitudes brighter than expected (as compared to known unlensed Type Ia SNe at similar redshift), implying a lensing magnification of $\mu_{obs}$ = 2.03 $\pm$ 0.29.  Therefore, rather than using the supernova magnification as a constraint, we instead set it as a benchmark value to be derived from each model.  Our fiducial model gives a value $\mu=2.149\pm0.029$. 
On the other hand, the model including external shear 
gives a magnification of 1.789$\pm$0.045, but is probably lacking some mass (and magnification) at this radius. 
While these two values are significantly different from one another, they both fall within the overall uncertainty envelope as defined by \citealt{Rodney2015}, which is another probe of systematics.  A comparison of our magnifications values to those derived from other studies can be seen in Fig,\, \ref{fig:SNHFF14Tom}.  Here again, large-scale differences between models are likely due to systematics.

\section{Conclusions}@ARTICLE{Jauzac2016,
   author = {{Jauzac}, M. and {Eckert}, D. and {Schwinn}, J. and {Harvey}, D. and 
	{Baugh}, C.~M. and {Robertson}, A. and {Bose}, S. and {Massey}, R. and 
	{Owers}, M. and {Ebeling}, H. and {Shan}, H.~Y. and {Jullo}, E. and 
	{Kneib}, J.-P. and {Richard}, J. and {Atek}, H. and {Cl{\'e}ment}, B. and 
	{Egami}, E. and {Israel}, H. and {Knowles}, K. and {Limousin}, M. and 
	{Natarajan}, P. and {Rexroth}, M. and {Taylor}, P. and {Tchernin}, C.
	},
    title = "{The Extraordinary Amount of Substructure in the Hubble Frontier Fields Cluster Abell 2744}",
  journal = {\mnras},
archivePrefix = "arXiv",
   eprint = {1606.04527},
 keywords = {Gravitational Lensing, Galaxy Clusters, Individual (Abell 2744)},
     year = 2016,
    month = sep,
      doi = {10.1093/mnras/stw2251},
   adsurl = {http://adsabs.harvard.edu/abs/2016MNRAS.tmp.1357J},
  adsnote = {Provided by the SAO/NASA Astrophysics Data System}
}
In this paper, we use ultra-deep imaging data from the HFF program in combination with spectroscopic data from the VLT/MUSE to build a new strong lensing mass model for the HFF cluster Abell 2744. Our main conclusions are as follows:
\begin{itemize}
\item Thanks to the 18.5 hours of MUSE coverage we perform a spectroscopic analysis and construct a redshift catalog of a total of 514 objects \RC{(414 new identifications),} including \RC{326 background sources and} 156 cluster members.
The cluster members largely fall into two distinct groups, matching the velocities of the two large-scale structures found by \cite{Owers2011}.  We note that many of the galaxies detected in \cite{Owers2011} are measured at significantly larger radii from the cluster core than the MUSE coverage.

\item  We review every multiple image recorded in previous studies, spectroscopically confirming 78 of them and adding 8 new images. We grade the other multiple images based on their photometry and their compatibility with our lens modeling.

\item Thanks to the numerous constraints our modelling successfully includes the impact of neighboring substructures found by the weak-lensing analysis of \citet{Jauzac2016}, mainly through their shear effect on the images.

 \item Overall our fiducial mass model, only constrained with spectroscopic redshifts, gives a statistical error $\sim1\%$ on the mass profile in the cluster core. By testing the dependence on the choice of constraints \RC{(3\% relative systematic uncertainties)} and the parametrisation of the model \RC{(6\% relative systematic uncertainties) we find that our fiducial model error might suffer from a relative systematic uncertainty of up to $\sim9\%$ on the mass profile in the multiply-imaged region.}
 
\RC{\item We estimate the level of systematic error from the addition of new redshifts identified by MUSE to be up to $\sim2.5$ $\sigma_{\rm stat}$, lowering the mass in the cluster core while increasing the mass in the outer part.}

\item We use the background SN Ia Tomas \citep{Rodney2015} as a test on our magnification estimates and find a good agreement between the observed lensing magnification (2.03$\pm$0.29), and the magnification predicted by our fiducial model (2.149$\pm$0.029).
\end{itemize}

In the end the deep spectroscopic coverage of this cluster allows us to improve the overall accuracy of the lensing reconstruction, mainly by placing an unprecedented number of constraints on the mass profile of the cluster core.  This illustrates the usefulness of obtaining deep, complete spectroscopic coverage of lensing clusters.

The mass models, as well as the associated mass and magnification maps, have been publically released as CATS (v4) through the Frontier Fields mass modelling challenge, and the resulting spectroscopy was shared among the other lensing teams. Accurate magnification estimates will be particularly useful for high redshift studies constraining the faint-end slope of the luminosity function \citep{Atek2015,Bouwens2016}.

Compared to previous lensing works, the large increase in the number of multiple systems with confirmed redshifts sets a new challenge which we believe will help the overall lensing community to better understand the complexity of this cluster. The parametric approach used in our models can reproduce all strong-lensing constraints with a good rms (typically 0.6\arcsec), but ultimately new techniques will be needed to fully account for all the strong lensing information and further improve the quality of the models. One example of such a method is a hybrid model, combining parametrically-constructed cluster members with a free-form large-scale mass distribution.

\section*{Acknowledgments}

GM, JR, DL, BC, VP, JM  acknowledge support from the ERC starting grant 336736-CALENDS. 
RB, HI, FL  acknowledge support from the ERC advanced grant 339659-MUSICOS. 
RJB gratefully acknowledges support from TOP grant TOP1.16.057 from the
Nederlandse Organisatie voor Wetenschappelijk Onderzoek.
LW acknowledges support by the Competitive Fund of the Leibniz Association through grant SAW-2015-AIP-2. We acknowledge fruitful discussions with Mathilde Jauzac, Keren Sharon, Jean-Paul Kneib, Eric Jullo and Marceau Limousin. This work utilises gravitational lensing models produced by Brada{\v c}, Natarajan \& Kneib (CATS), Merten \& Zitrin, Sharon, and Williams, and the GLAFIC and Diego groups. This lens modeling was partially funded by the HST Frontier Fields program conducted by STScI. STScI is operated by the Association of Universities for Research in Astronomy, Inc. under NASA contract NAS 5-26555. The lens models were obtained from the Mikulski Archive for Space Telescopes (MAST). 
Based on observations made with ESO Telescopes at the La Silla Paranal Observatory under programme ID 094.A-0115. Based on observations obtained with the NASA/ESA Hubble Space Telescope, retrieved from the Mikulski Archive for Space Telescopes (MAST) at the Space Telescope Science Institute (STScI). STScI is operated by the Association of Universities for Research in Astronomy, Inc. under NASA contract NAS 5-26555. Also based on data obtained at the W.M. Keck Observatory, which is operated as a scientific partnership among the California Institute of Technology, the University of California and the National Aeronautics and Space Administration. The Observatory was made possible by the generous financial support of the W.M. Keck Foundation. The authors wish to recognise and acknowledge the very significant cultural role and reverence that the summit of Mauna Kea has always had within the indigenous Hawaiian community.  We are most fortunate to have the opportunity to conduct observations from this mountain. 
This works makes use of MPDAF, the MUSE Python Data Analysis Framework, an open-source (BSD licensed) Python package developed and maintained by CRAL and partially funded by the ERC advanced grant 339659-MUSICOS.

%%%%%%%%%%%%%%%%%%%%%%%%%%%%%%%%%%%%%%%%%%%%%%%%%%

%%%%%%%%%%%%%%%%%%%% REFERENCES %%%%%%%%%%%%%%%%%%

% The best way to enter references is to use BibTeX:

\bibliographystyle{mnras}

\bibliography{biblio_A2744}

\begin{thebibliography}{}
\makeatletter
\relax
\def\mn@urlcharsother{\let\do\@makeother \do\$\do\&\do\#\do\^\do\_\do\%\do\~}
\def\mn@doi{\begingroup\mn@urlcharsother \@ifnextchar [ {\mn@doi@}
  {\mn@doi@[]}}
\def\mn@doi@[#1]#2{\def\@tempa{#1}\ifx\@tempa\@empty \href
  {http://dx.doi.org/#2} {doi:#2}\else \href {http://dx.doi.org/#2} {#1}\fi
  \endgroup}
\def\mn@eprint#1#2{\mn@eprint@#1:#2::\@nil}
\def\mn@eprint@arXiv#1{\href {http://arxiv.org/abs/#1} {{\tt arXiv:#1}}}
\def\mn@eprint@dblp#1{\href {http://dblp.uni-trier.de/rec/bibtex/#1.xml}
  {dblp:#1}}
\def\mn@eprint@#1:#2:#3:#4\@nil{\def\@tempa {#1}\def\@tempb {#2}\def\@tempc
  {#3}\ifx \@tempc \@empty \let \@tempc \@tempb \let \@tempb \@tempa \fi \ifx
  \@tempb \@empty \def\@tempb {arXiv}\fi \@ifundefined
  {mn@eprint@\@tempb}{\@tempb:\@tempc}{\expandafter \expandafter \csname
  mn@eprint@\@tempb\endcsname \expandafter{\@tempc}}}

\bibitem[\protect\citeauthoryear{{Abell}, {Corwin}  \& {Olowin}}{{Abell}
  et~al.}{1989}]{Abell1989}
{Abell} G.~O.,  {Corwin} Jr. H.~G.,   {Olowin} R.~P.,  1989, \mn@doi [\apjs]
  {10.1086/191333}, \href {http://adsabs.harvard.edu/abs/1989ApJS...70....1A}
  {70, 1}

\bibitem[\protect\citeauthoryear{{Alavi} et~al.,}{{Alavi}
  et~al.}{2016}]{Alavi2016}
{Alavi} A.,  et~al., 2016, \mn@doi [\apj] {10.3847/0004-637X/832/1/56}, \href
  {http://adsabs.harvard.edu/abs/2016ApJ...832...56A} {832, 56}

\bibitem[\protect\citeauthoryear{{Allen}}{{Allen}}{1998}]{Allen1998}
{Allen} S.~W.,  1998, \mn@doi [\mnras] {10.1046/j.1365-8711.1998.01358.x},
  \href {http://adsabs.harvard.edu/abs/1998MNRAS.296..392A} {296, 392}

\bibitem[\protect\citeauthoryear{{Atek} et~al.,}{{Atek}
  et~al.}{2014}]{Atek2014}
{Atek} H.,  et~al., 2014, \mn@doi [\apj] {10.1088/0004-637X/786/1/60}, \href
  {http://adsabs.harvard.edu/abs/2014ApJ...786...60A} {786, 60}

\bibitem[\protect\citeauthoryear{{Atek} et~al.,}{{Atek}
  et~al.}{2015}]{Atek2015}
{Atek} H.,  et~al., 2015, \mn@doi [\apj] {10.1088/0004-637X/800/1/18}, \href
  {http://adsabs.harvard.edu/abs/2015ApJ...800...18A} {800, 18}

\bibitem[\protect\citeauthoryear{{Bacon} et~al.,}{{Bacon}
  et~al.}{2010}]{Bacon2010}
{Bacon} R.,  et~al., 2010, in Ground-based and Airborne Instrumentation for
  Astronomy III. p. 773508, \mn@doi{10.1117/12.856027}

\bibitem[\protect\citeauthoryear{{Bacon} et~al.,}{{Bacon}
  et~al.}{2015}]{Bacon2015}
{Bacon} R.,  et~al., 2015, \mn@doi [\aap] {10.1051/0004-6361/201425419}, \href
  {http://cdsads.u-strasbg.fr/abs/2015A%26A...575A..75B} {575, A75}

\bibitem[\protect\citeauthoryear{{Baldry} et~al.,}{{Baldry}
  et~al.}{2014}]{Baldry2014}
{Baldry} I.~K.,  et~al., 2014, \mn@doi [\mnras] {10.1093/mnras/stu727}, \href
  {http://adsabs.harvard.edu/abs/2014MNRAS.441.2440B} {441, 2440}

\bibitem[\protect\citeauthoryear{{Bertin}}{{Bertin}}{2006}]{Scamp}
{Bertin} E.,  2006, in {Gabriel} C.,  {Arviset} C.,  {Ponz} D.,   {Enrique} S.,
   eds,  Astronomical Society of the Pacific Conference Series Vol. 351,
  Astronomical Data Analysis Software and Systems XV. p.~112

\bibitem[\protect\citeauthoryear{{Bertin} \& {Arnouts}}{{Bertin} \&
  {Arnouts}}{1996}]{SEx}
{Bertin} E.,  {Arnouts} S.,  1996, \mn@doi [\aaps] {10.1051/aas:1996164}, \href
  {http://adsabs.harvard.edu/abs/1996A%26AS..117..393B} {117, 393}

\bibitem[\protect\citeauthoryear{{Bina} et~al.,}{{Bina}
  et~al.}{2016}]{Bina2016}
{Bina} D.,  et~al., 2016, \mn@doi [\aap] {10.1051/0004-6361/201527913}, \href
  {http://adsabs.harvard.edu/abs/2016A%26A...590A..14B} {590, A14}

\bibitem[\protect\citeauthoryear{{Boschin}, {Girardi}, {Spolaor}  \&
  {Barrena}}{{Boschin} et~al.}{2006}]{Boschin2006}
{Boschin} W.,  {Girardi} M.,  {Spolaor} M.,   {Barrena} R.,  2006, \mn@doi
  [\aap] {10.1051/0004-6361:20054408}, \href
  {http://adsabs.harvard.edu/abs/2006A%26A...449..461B} {449, 461}

\bibitem[\protect\citeauthoryear{{Bouwens}, {Oesch}, {Illingworth}, {Ellis}  \&
  {Stefanon}}{{Bouwens} et~al.}{2016}]{Bouwens2016}
{Bouwens} R.~J.,  {Oesch} P.~A.,  {Illingworth} G.~D.,  {Ellis} R.~S.,
  {Stefanon} M.,  2016, preprint, \href
  {http://adsabs.harvard.edu/abs/2016arXiv161000283B} {} (\mn@eprint {arXiv}
  {1610.00283})

\bibitem[\protect\citeauthoryear{{Brada{\v c}} et~al.,}{{Brada{\v c}}
  et~al.}{2008}]{Bradac2008}
{Brada{\v c}} M.,  et~al., 2008, \mn@doi [\apj] {10.1086/588377}, \href
  {http://adsabs.harvard.edu/abs/2008ApJ...681..187B} {681, 187}

\bibitem[\protect\citeauthoryear{{Broadhurst} et~al.,}{{Broadhurst}
  et~al.}{2005}]{Broadhurst2005}
{Broadhurst} T.,  et~al., 2005, \mn@doi [\apj] {10.1086/426494}, \href
  {http://adsabs.harvard.edu/abs/2005ApJ...621...53B} {621, 53}

\bibitem[\protect\citeauthoryear{{Caminha} et~al.,}{{Caminha}
  et~al.}{2016}]{Caminha2016}
{Caminha} G.~B.,  et~al., 2016, preprint, \href
  {http://adsabs.harvard.edu/abs/2016arXiv160703462C} {} (\mn@eprint {arXiv}
  {1607.03462})

\bibitem[\protect\citeauthoryear{{Castellano} et~al.,}{{Castellano}
  et~al.}{2016}]{ASTRODEEP2}
{Castellano} M.,  et~al., 2016, \mn@doi [\aap] {10.1051/0004-6361/201527514},
  \href {http://adsabs.harvard.edu/abs/2016A%26A...590A..31C} {590, A31}

\bibitem[\protect\citeauthoryear{{Coe} et~al.,}{{Coe} et~al.}{2013}]{Coe2013}
{Coe} D.,  et~al., 2013, \mn@doi [\apj] {10.1088/0004-637X/762/1/32}, \href
  {http://adsabs.harvard.edu/abs/2013ApJ...762...32C} {762, 32}

\bibitem[\protect\citeauthoryear{{Couch} \& {Newell}}{{Couch} \&
  {Newell}}{1984}]{CouchNewell1984}
{Couch} W.~J.,  {Newell} E.~B.,  1984, \mn@doi [\apjs] {10.1086/190979}, \href
  {http://adsabs.harvard.edu/abs/1984ApJS...56..143C} {56, 143}

\bibitem[\protect\citeauthoryear{{Diego}, {Broadhurst}, {Wong}, {Silk}, {Lim},
  {Zheng}, {Lam}  \& {Ford}}{{Diego} et~al.}{2016}]{Diego2016}
{Diego} J.~M.,  {Broadhurst} T.,  {Wong} J.,  {Silk} J.,  {Lim} J.,  {Zheng}
  W.,  {Lam} D.,   {Ford} H.,  2016, \mn@doi [\mnras] {10.1093/mnras/stw865},
  \href {http://adsabs.harvard.edu/abs/2016MNRAS.459.3447D} {459, 3447}

\bibitem[\protect\citeauthoryear{{Drake} et~al.,}{{Drake}
  et~al.}{2016}]{Drake2016}
{Drake} A.~B.,  et~al., 2016, preprint, \href
  {http://adsabs.harvard.edu/abs/2016arXiv160902920D} {} (\mn@eprint {arXiv}
  {1609.02920})

\bibitem[\protect\citeauthoryear{{Eckert} et~al.,}{{Eckert}
  et~al.}{2015}]{Eckert2015}
{Eckert} D.,  et~al., 2015, \mn@doi [\nat] {10.1038/nature16058}, \href
  {http://adsabs.harvard.edu/abs/2015Natur.528..105E} {528, 105}

\bibitem[\protect\citeauthoryear{{El{\'{\i}}asd{\'o}ttir}
  et~al.,}{{El{\'{\i}}asd{\'o}ttir} et~al.}{2007}]{Eliasdottir2007}
{El{\'{\i}}asd{\'o}ttir} {\'A}.,  et~al., 2007, preprint, \href
  {http://cdsads.u-strasbg.fr/abs/2007arXiv0710.5636E} {} (\mn@eprint {arXiv}
  {0710.5636})

\bibitem[\protect\citeauthoryear{{Grillo} et~al.,}{{Grillo}
  et~al.}{2015}]{Grillo2015}
{Grillo} C.,  et~al., 2015, \mn@doi [\apj] {10.1088/0004-637X/800/1/38}, \href
  {http://adsabs.harvard.edu/abs/2015ApJ...800...38G} {800, 38}

\bibitem[\protect\citeauthoryear{{Harvey}, {Massey}, {Kitching}, {Taylor}  \&
  {Tittley}}{{Harvey} et~al.}{2015}]{Harvey2015}
{Harvey} D.,  {Massey} R.,  {Kitching} T.,  {Taylor} A.,   {Tittley} E.,  2015,
  \mn@doi [Science] {10.1126/science.1261381}, \href
  {http://cdsads.u-strasbg.fr/abs/2015Sci...347.1462H} {347, 1462}

\bibitem[\protect\citeauthoryear{{Hoag} et~al.,}{{Hoag}
  et~al.}{2016}]{Hoag2016}
{Hoag} A.,  et~al., 2016, \mn@doi [\apj] {10.3847/0004-637X/831/2/182}, \href
  {http://adsabs.harvard.edu/abs/2016ApJ...831..182H} {831, 182}

\bibitem[\protect\citeauthoryear{{Horne}}{{Horne}}{1986}]{Horne86}
{Horne} K.,  1986, \mn@doi [\pasp] {10.1086/131801}, \href
  {http://adsabs.harvard.edu/abs/1986PASP...98..609H} {98, 609}

\bibitem[\protect\citeauthoryear{{Ishigaki}, {Kawamata}, {Ouchi}, {Oguri},
  {Shimasaku}  \& {Ono}}{{Ishigaki} et~al.}{2015}]{Ishigaki2015}
{Ishigaki} M.,  {Kawamata} R.,  {Ouchi} M.,  {Oguri} M.,  {Shimasaku} K.,
  {Ono} Y.,  2015, \mn@doi [\apj] {10.1088/0004-637X/799/1/12}, \href
  {http://adsabs.harvard.edu/abs/2015ApJ...799...12I} {799, 12}

\bibitem[\protect\citeauthoryear{{Jauzac} et~al.,}{{Jauzac}
  et~al.}{2014}]{Jauzac2014}
{Jauzac} M.,  et~al., 2014, \mn@doi [\mnras] {10.1093/mnras/stu1355}, \href
  {http://adsabs.harvard.edu/abs/2014MNRAS.443.1549J} {443, 1549}

\bibitem[\protect\citeauthoryear{{Jauzac} et~al.,}{{Jauzac}
  et~al.}{2015}]{Jauzac2015}
{Jauzac} M.,  et~al., 2015, \mn@doi [\mnras] {10.1093/mnras/stv1402}, \href
  {http://adsabs.harvard.edu/abs/2015MNRAS.452.1437J} {452, 1437}

\bibitem[\protect\citeauthoryear{{Jauzac} et~al.,}{{Jauzac}
  et~al.}{2016a}]{Jauzac2016}
{Jauzac} M.,  et~al., 2016a, \mn@doi [\mnras] {10.1093/mnras/stw2251}, \href
  {http://adsabs.harvard.edu/abs/2016MNRAS.tmp.1357J} {}

\bibitem[\protect\citeauthoryear{{Jauzac} et~al.,}{{Jauzac}
  et~al.}{2016b}]{Jauzac2016b}
{Jauzac} M.,  et~al., 2016b, \mn@doi [\mnras] {10.1093/mnras/stw069}, \href
  {http://adsabs.harvard.edu/abs/2016MNRAS.457.2029J} {457, 2029}

\bibitem[\protect\citeauthoryear{{Johnson} \& {Sharon}}{{Johnson} \&
  {Sharon}}{2016}]{Johnson2016}
{Johnson} T.~L.,  {Sharon} K.,  2016, preprint, \href
  {http://adsabs.harvard.edu/abs/2016arXiv160808713J} {} (\mn@eprint {arXiv}
  {1608.08713})

\bibitem[\protect\citeauthoryear{{Johnson}, {Sharon}, {Bayliss}, {Gladders},
  {Coe}  \& {Ebeling}}{{Johnson} et~al.}{2014}]{Johnson2014}
{Johnson} T.~L.,  {Sharon} K.,  {Bayliss} M.~B.,  {Gladders} M.~D.,  {Coe} D.,
   {Ebeling} H.,  2014, \mn@doi [\apj] {10.1088/0004-637X/797/1/48}, \href
  {http://adsabs.harvard.edu/abs/2014ApJ...797...48J} {797, 48}

\bibitem[\protect\citeauthoryear{{Jouvel} et~al.,}{{Jouvel}
  et~al.}{2014}]{Jouvel2014}
{Jouvel} S.,  et~al., 2014, \mn@doi [\aap] {10.1051/0004-6361/201322419}, \href
  {http://adsabs.harvard.edu/abs/2014A%26A...562A..86J} {562, A86}

\bibitem[\protect\citeauthoryear{{Jullo}, {Kneib}, {Limousin},
  {El{\'{\i}}asd{\'o}ttir}, {Marshall}  \& {Verdugo}}{{Jullo}
  et~al.}{2007}]{Jullo2007}
{Jullo} E.,  {Kneib} J.-P.,  {Limousin} M.,  {El{\'{\i}}asd{\'o}ttir} {\'A}.,
  {Marshall} P.~J.,   {Verdugo} T.,  2007, \mn@doi [New Journal of Physics]
  {10.1088/1367-2630/9/12/447}, \href
  {http://cdsads.u-strasbg.fr/abs/2007NJPh....9..447J} {9, 447}

\bibitem[\protect\citeauthoryear{{Jullo}, {Natarajan}, {Kneib}, {D'Aloisio},
  {Limousin}, {Richard}  \& {Schimd}}{{Jullo} et~al.}{2010}]{Jullo2010}
{Jullo} E.,  {Natarajan} P.,  {Kneib} J.-P.,  {D'Aloisio} A.,  {Limousin} M.,
  {Richard} J.,   {Schimd} C.,  2010, \mn@doi [Science]
  {10.1126/science.1185759}, \href
  {http://adsabs.harvard.edu/abs/2010Sci...329..924J} {329, 924}

\bibitem[\protect\citeauthoryear{{Karman} et~al.,}{{Karman}
  et~al.}{2016}]{Karman2016}
{Karman} W.,  et~al., 2016, \mn@doi [\aap] {10.1051/0004-6361/201527443}, \href
  {http://adsabs.harvard.edu/abs/2016A%26A...585A..27K} {585, A27}

\bibitem[\protect\citeauthoryear{{Kawamata}, {Ishigaki}, {Shimasaku}, {Oguri}
  \& {Ouchi}}{{Kawamata} et~al.}{2015}]{Kawamata2015}
{Kawamata} R.,  {Ishigaki} M.,  {Shimasaku} K.,  {Oguri} M.,   {Ouchi} M.,
  2015, \mn@doi [\apj] {10.1088/0004-637X/804/2/103}, \href
  {http://adsabs.harvard.edu/abs/2015ApJ...804..103K} {804, 103}

\bibitem[\protect\citeauthoryear{{Kawamata}, {Oguri}, {Ishigaki}, {Shimasaku}
  \& {Ouchi}}{{Kawamata} et~al.}{2016}]{Kawamata2016}
{Kawamata} R.,  {Oguri} M.,  {Ishigaki} M.,  {Shimasaku} K.,   {Ouchi} M.,
  2016, \mn@doi [\apj] {10.3847/0004-637X/819/2/114}, \href
  {http://adsabs.harvard.edu/abs/2016ApJ...819..114K} {819, 114}

\bibitem[\protect\citeauthoryear{{Kneib}, {Ellis}, {Smail}, {Couch}  \&
  {Sharples}}{{Kneib} et~al.}{1996}]{Kneib1996}
{Kneib} J.-P.,  {Ellis} R.~S.,  {Smail} I.,  {Couch} W.~J.,   {Sharples} R.~M.,
   1996, \mn@doi [\apj] {10.1086/177995}, \href
  {http://adsabs.harvard.edu/abs/1996ApJ...471..643K} {471, 643}

\bibitem[\protect\citeauthoryear{{Lagattuta} et~al.,}{{Lagattuta}
  et~al.}{2016}]{Lagattuta2016}
{Lagattuta} D.~J.,  et~al., 2016, preprint, \href
  {http://adsabs.harvard.edu/abs/2016arXiv161101513L} {} (\mn@eprint {arXiv}
  {1611.01513})

\bibitem[\protect\citeauthoryear{{Lam}, {Broadhurst}, {Diego}, {Lim}, {Coe},
  {Ford}  \& {Zheng}}{{Lam} et~al.}{2014}]{Lam2014}
{Lam} D.,  {Broadhurst} T.,  {Diego} J.~M.,  {Lim} J.,  {Coe} D.,  {Ford}
  H.~C.,   {Zheng} W.,  2014, \mn@doi [\apj] {10.1088/0004-637X/797/2/98},
  \href {http://adsabs.harvard.edu/abs/2014ApJ...797...98L} {797, 98}

\bibitem[\protect\citeauthoryear{{Limousin} et~al.,}{{Limousin}
  et~al.}{2007}]{Limousin2007}
{Limousin} M.,  et~al., 2007, \mn@doi [\apj] {10.1086/521293}, \href
  {http://adsabs.harvard.edu/abs/2007ApJ...668..643L} {668, 643}

\bibitem[\protect\citeauthoryear{{Limousin} et~al.,}{{Limousin}
  et~al.}{2016}]{Limousin2016}
{Limousin} M.,  et~al., 2016, \mn@doi [\aap] {10.1051/0004-6361/201527638},
  \href {http://cdsads.u-strasbg.fr/abs/2016A%26A...588A..99L} {588, A99}

\bibitem[\protect\citeauthoryear{{Livermore}, {Finkelstein}  \&
  {Lotz}}{{Livermore} et~al.}{2016}]{Livermore2016}
{Livermore} R.~C.,  {Finkelstein} S.~L.,   {Lotz} J.~M.,  2016, preprint, \href
  {http://adsabs.harvard.edu/abs/2016arXiv160406799L} {} (\mn@eprint {arXiv}
  {1604.06799})

\bibitem[\protect\citeauthoryear{{Lotz} et~al.,}{{Lotz}
  et~al.}{2016}]{Lotz2016}
{Lotz} J.~M.,  et~al., 2016, preprint, \href
  {http://adsabs.harvard.edu/abs/2016arXiv160506567L} {} (\mn@eprint {arXiv}
  {1605.06567})

\bibitem[\protect\citeauthoryear{{Markevitch}, {Gonzalez}, {Clowe},
  {Vikhlinin}, {Forman}, {Jones}, {Murray}  \& {Tucker}}{{Markevitch}
  et~al.}{2004}]{Markevitch2004}
{Markevitch} M.,  {Gonzalez} A.~H.,  {Clowe} D.,  {Vikhlinin} A.,  {Forman} W.,
   {Jones} C.,  {Murray} S.,   {Tucker} W.,  2004, \mn@doi [\apj]
  {10.1086/383178}, \href {http://adsabs.harvard.edu/abs/2004ApJ...606..819M}
  {606, 819}

\bibitem[\protect\citeauthoryear{{Maurogordato}, {Sauvageot}, {Bourdin},
  {Cappi}, {Benoist}, {Ferrari}, {Mars}  \& {Houairi}}{{Maurogordato}
  et~al.}{2011}]{Maurogordato2011}
{Maurogordato} S.,  {Sauvageot} J.~L.,  {Bourdin} H.,  {Cappi} A.,  {Benoist}
  C.,  {Ferrari} C.,  {Mars} G.,   {Houairi} K.,  2011, \mn@doi [\aap]
  {10.1051/0004-6361/201014415}, \href
  {http://adsabs.harvard.edu/abs/2011A%26A...525A..79M} {525, A79}

\bibitem[\protect\citeauthoryear{{Meneghetti} et~al.,}{{Meneghetti}
  et~al.}{2016}]{Meneghetti2016}
{Meneghetti} M.,  et~al., 2016, preprint, \href
  {http://adsabs.harvard.edu/abs/2016arXiv160604548M} {} (\mn@eprint {arXiv}
  {1606.04548})

\bibitem[\protect\citeauthoryear{{Merlin} et~al.,}{{Merlin}
  et~al.}{2016}]{ASTRODEEP1}
{Merlin} E.,  et~al., 2016, \mn@doi [\aap] {10.1051/0004-6361/201527513}, \href
  {http://adsabs.harvard.edu/abs/2016A%26A...590A..30M} {590, A30}

\bibitem[\protect\citeauthoryear{Merten et~al.,}{Merten
  et~al.}{2011}]{Merten2011}
Merten J.,  et~al., 2011, \mn@doi [Monthly Notices of the Royal Astronomical
  Society] {10.1111/j.1365-2966.2011.19266.x}, 417, 333

\bibitem[\protect\citeauthoryear{{Montes} \& {Trujillo}}{{Montes} \&
  {Trujillo}}{2014}]{Montes2014}
{Montes} M.,  {Trujillo} I.,  2014, \mn@doi [\apj]
  {10.1088/0004-637X/794/2/137}, \href
  {http://adsabs.harvard.edu/abs/2014ApJ...794..137M} {794, 137}

\bibitem[\protect\citeauthoryear{{Natarajan} et~al.,}{{Natarajan}
  et~al.}{2017}]{Natarajan2017}
{Natarajan} P.,  et~al., 2017, preprint, \href
  {http://adsabs.harvard.edu/abs/2017arXiv170204348N} {} (\mn@eprint {arXiv}
  {1702.04348})

\bibitem[\protect\citeauthoryear{{Newman}, {Treu}, {Ellis}, {Sand}, {Nipoti},
  {Richard}  \& {Jullo}}{{Newman} et~al.}{2013a}]{Newman1}
{Newman} A.~B.,  {Treu} T.,  {Ellis} R.~S.,  {Sand} D.~J.,  {Nipoti} C.,
  {Richard} J.,   {Jullo} E.,  2013a, \mn@doi [\apj]
  {10.1088/0004-637X/765/1/24}, \href
  {http://adsabs.harvard.edu/abs/2013ApJ...765...24N} {765, 24}

\bibitem[\protect\citeauthoryear{{Newman}, {Treu}, {Ellis}  \& {Sand}}{{Newman}
  et~al.}{2013b}]{Newman2}
{Newman} A.~B.,  {Treu} T.,  {Ellis} R.~S.,   {Sand} D.~J.,  2013b, \mn@doi
  [\apj] {10.1088/0004-637X/765/1/25}, \href
  {http://adsabs.harvard.edu/abs/2013ApJ...765...25N} {765, 25}

\bibitem[\protect\citeauthoryear{{Oke}}{{Oke}}{1974}]{Oke1974}
{Oke} J.~B.,  1974, \mn@doi [\apjs] {10.1086/190287}, \href
  {http://adsabs.harvard.edu/abs/1974ApJS...27...21O} {27, 21}

\bibitem[\protect\citeauthoryear{{Owers}, {Randall}, {Nulsen}, {Couch}, {David}
   \& {Kempner}}{{Owers} et~al.}{2011}]{Owers2011}
{Owers} M.~S.,  {Randall} S.~W.,  {Nulsen} P.~E.~J.,  {Couch} W.~J.,  {David}
  L.~P.,   {Kempner} J.~C.,  2011, \mn@doi [\apj] {10.1088/0004-637X/728/1/27},
  \href {http://adsabs.harvard.edu/abs/2011ApJ...728...27O} {728, 27}

\bibitem[\protect\citeauthoryear{{Patr{\'{\i}}cio} et~al.,}{{Patr{\'{\i}}cio}
  et~al.}{2016}]{Patricio2016}
{Patr{\'{\i}}cio} V.,  et~al., 2016, \mn@doi [\mnras] {10.1093/mnras/stv2859},
  \href {http://adsabs.harvard.edu/abs/2016MNRAS.456.4191P} {456, 4191}

\bibitem[\protect\citeauthoryear{{Postman} et~al.,}{{Postman}
  et~al.}{2012}]{Postman2012}
{Postman} M.,  et~al., 2012, \mn@doi [\apjs] {10.1088/0067-0049/199/2/25},
  \href {http://adsabs.harvard.edu/abs/2012ApJS..199...25P} {199, 25}

\bibitem[\protect\citeauthoryear{{Priewe}, {Williams}, {Liesenborgs}, {Coe}  \&
  {Rodney}}{{Priewe} et~al.}{2017}]{Priewe2017}
{Priewe} J.,  {Williams} L.~L.~R.,  {Liesenborgs} J.,  {Coe} D.,   {Rodney}
  S.~A.,  2017, \mn@doi [\mnras] {10.1093/mnras/stw2785}, \href
  {http://adsabs.harvard.edu/abs/2017MNRAS.465.1030P} {465, 1030}

\bibitem[\protect\citeauthoryear{{Richard}, {Kneib}, {Ebeling}, {Stark},
  {Egami}  \& {Fiedler}}{{Richard} et~al.}{2011}]{Richard2011}
{Richard} J.,  {Kneib} J.-P.,  {Ebeling} H.,  {Stark} D.~P.,  {Egami} E.,
  {Fiedler} A.~K.,  2011, \mn@doi [\mnras] {10.1111/j.1745-3933.2011.01050.x},
  \href {http://adsabs.harvard.edu/abs/2011MNRAS.414L..31R} {414, L31}

\bibitem[\protect\citeauthoryear{{Richard} et~al.,}{{Richard}
  et~al.}{2014}]{Richard2014}
{Richard} J.,  et~al., 2014, \mn@doi [\mnras] {10.1093/mnras/stu1395}, \href
  {http://adsabs.harvard.edu/abs/2014MNRAS.444..268R} {444, 268}

\bibitem[\protect\citeauthoryear{{Richard} et~al.,}{{Richard}
  et~al.}{2015}]{Richard2015}
{Richard} J.,  et~al., 2015, \mn@doi [\mnras] {10.1093/mnrasl/slu150}, \href
  {http://adsabs.harvard.edu/abs/2015MNRAS.446L..16R} {446, L16}

\bibitem[\protect\citeauthoryear{{Rodney} et~al.,}{{Rodney}
  et~al.}{2015}]{Rodney2015}
{Rodney} S.~A.,  et~al., 2015, \mn@doi [\apj] {10.1088/0004-637X/811/1/70},
  \href {http://adsabs.harvard.edu/abs/2015ApJ...811...70R} {811, 70}

\bibitem[\protect\citeauthoryear{{Schmidt} et~al.,}{{Schmidt}
  et~al.}{2014}]{GLASS2}
{Schmidt} K.~B.,  et~al., 2014, \mn@doi [\apjl] {10.1088/2041-8205/782/2/L36},
  \href {http://adsabs.harvard.edu/abs/2014ApJ...782L..36S} {782, L36}

\bibitem[\protect\citeauthoryear{{Schmidt} et~al.,}{{Schmidt}
  et~al.}{2016}]{Schmidt2016}
{Schmidt} K.~B.,  et~al., 2016, \mn@doi [\apj] {10.3847/0004-637X/818/1/38},
  \href {http://adsabs.harvard.edu/abs/2016ApJ...818...38S} {818, 38}

\bibitem[\protect\citeauthoryear{{Schwinn}, {Jauzac}, {Baugh}, {Bartelmann},
  {Eckert}, {Harvey}, {Natarajan}  \& {Massey}}{{Schwinn}
  et~al.}{2016}]{Schwinn2016}
{Schwinn} J.,  {Jauzac} M.,  {Baugh} C.~M.,  {Bartelmann} M.,  {Eckert} D.,
  {Harvey} D.,  {Natarajan} P.,   {Massey} R.,  2016, preprint, \href
  {http://adsabs.harvard.edu/abs/2016arXiv161102790S} {} (\mn@eprint {arXiv}
  {1611.02790})

\bibitem[\protect\citeauthoryear{{Smith}, {Kneib}, {Smail}, {Mazzotta},
  {Ebeling}  \& {Czoske}}{{Smith} et~al.}{2005}]{Smith2005}
{Smith} G.~P.,  {Kneib} J.-P.,  {Smail} I.,  {Mazzotta} P.,  {Ebeling} H.,
  {Czoske} O.,  2005, \mn@doi [\mnras] {10.1111/j.1365-2966.2005.08911.x},
  \href {http://adsabs.harvard.edu/abs/2005MNRAS.359..417S} {359, 417}

\bibitem[\protect\citeauthoryear{{Soto}, {Lilly}, {Bacon}, {Richard}  \&
  {Conseil}}{{Soto} et~al.}{2016}]{Soto2016}
{Soto} K.~T.,  {Lilly} S.~J.,  {Bacon} R.,  {Richard} J.,   {Conseil} S.,
  2016, \mn@doi [\mnras] {10.1093/mnras/stw474}, \href
  {http://adsabs.harvard.edu/abs/2016MNRAS.458.3210S} {458, 3210}

\bibitem[\protect\citeauthoryear{{Soucail}, {Mellier}, {Fort}, {Mathez}  \&
  {Cailloux}}{{Soucail} et~al.}{1988}]{Soucail1988}
{Soucail} G.,  {Mellier} Y.,  {Fort} B.,  {Mathez} G.,   {Cailloux} M.,  1988,
  \aap, \href {http://adsabs.harvard.edu/abs/1988A%26A...191L..19S} {191, L19}

\bibitem[\protect\citeauthoryear{{Treu} et~al.,}{{Treu} et~al.}{2015}]{GLASS1}
{Treu} T.,  et~al., 2015, \mn@doi [\apj] {10.1088/0004-637X/812/2/114}, \href
  {http://adsabs.harvard.edu/abs/2015ApJ...812..114T} {812, 114}

\bibitem[\protect\citeauthoryear{{Umetsu} et~al.,}{{Umetsu}
  et~al.}{2009}]{Umetsu2009}
{Umetsu} K.,  et~al., 2009, \mn@doi [\apj] {10.1088/0004-637X/694/2/1643},
  \href {http://adsabs.harvard.edu/abs/2009ApJ...694.1643U} {694, 1643}

\bibitem[\protect\citeauthoryear{{Verdugo}, {Motta}, {Mu{\~n}oz}, {Limousin},
  {Cabanac}  \& {Richard}}{{Verdugo} et~al.}{2011}]{Verdugo2011}
{Verdugo} T.,  {Motta} V.,  {Mu{\~n}oz} R.~P.,  {Limousin} M.,  {Cabanac} R.,
  {Richard} J.,  2011, \mn@doi [\aap] {10.1051/0004-6361/201014965}, \href
  {http://adsabs.harvard.edu/abs/2011A%26A...527A.124V} {527, A124}

\bibitem[\protect\citeauthoryear{{Wang} et~al.,}{{Wang}
  et~al.}{2015}]{Wang2015}
{Wang} X.,  et~al., 2015, \mn@doi [\apj] {10.1088/0004-637X/811/1/29}, \href
  {http://adsabs.harvard.edu/abs/2015ApJ...811...29W} {811, 29}

\bibitem[\protect\citeauthoryear{{Weilbacher}, {Streicher}, {Urrutia}, {Jarno},
  {P{\'e}contal-Rousset}, {Bacon}  \& {B{\"o}hm}}{{Weilbacher}
  et~al.}{2012}]{Weilbacher2012}
{Weilbacher} P.~M.,  {Streicher} O.,  {Urrutia} T.,  {Jarno} A.,
  {P{\'e}contal-Rousset} A.,  {Bacon} R.,   {B{\"o}hm} P.,  2012, in Software
  and Cyberinfrastructure for Astronomy II. p. 84510B,
  \mn@doi{10.1117/12.925114}

\bibitem[\protect\citeauthoryear{{Weilbacher}, {Streicher}, {Urrutia},
  {P{\'e}contal-Rousset}, {Jarno}  \& {Bacon}}{{Weilbacher}
  et~al.}{2014}]{Weilbacher2014}
{Weilbacher} P.~M.,  {Streicher} O.,  {Urrutia} T.,  {P{\'e}contal-Rousset} A.,
   {Jarno} A.,   {Bacon} R.,  2014, in {Manset} N.,  {Forshay} P.,  eds,
  Astronomical Society of the Pacific Conference Series Vol. 485, Astronomical
  Data Analysis Software and Systems XXIII. p.~451 (\mn@eprint {arXiv}
  {1507.00034})

\bibitem[\protect\citeauthoryear{{Zitrin} et~al.,}{{Zitrin}
  et~al.}{2014}]{Zitrin2014}
{Zitrin} A.,  et~al., 2014, \mn@doi [\apjl] {10.1088/2041-8205/793/1/L12},
  \href {http://adsabs.harvard.edu/abs/2014ApJ...793L..12Z} {793, L12}

\bibitem[\protect\citeauthoryear{{Zitrin} et~al.,}{{Zitrin}
  et~al.}{2015}]{Zitrin2015}
{Zitrin} A.,  et~al., 2015, \mn@doi [\apj] {10.1088/0004-637X/801/1/44}, \href
  {http://adsabs.harvard.edu/abs/2015ApJ...801...44Z} {801, 44}

\makeatother
\end{thebibliography}

% Alternatively you could enter them by hand, like this:
% This method is tedious and prone to error if you have lots of references
%\begin{thebibliography}{99}
%\bibitem[\protect\citeauthoryear{Author}{2012}]{Author2012}
%Author A.~N., 2013, Journal of Improbable Astronomy, 1, 1
%\bibitem[\protect\citeauthoryear{Others}{2013}]{Others2013}
%Others S., 2012, Journal of Interesting Stuff, 17, 198
%\end{thebibliography}

%%%%%%%%%%%%%%%%%%%%%%%%%%%%%%%%%%%%%%%%%%%%%%%%%%

%%%%%%%%%%%%%%%%% APPENDICES %%%%%%%%%%%%%%%%%%%%%

\appendix
\onecolumn
\section{list of multiple images}
\label{apendix:mul}

%\begin{table*}

%\onecolumn
\begin {table}
\begin{tabular}{cc}
\, & \,\\
\end{tabular}
\caption{ Multiply imaged systems considered in this work. In the column $z_{\rm ref}$, the letter refers to previous studies reporting spectroscopic redshifts in agreement with our detection: J for  \citep{Johnson2014}, R for \citep{Richard2014}, and W for  \citep{Wang2015}. M refers to this study. Column {\it conf}  corresponds to the confidence level attached to the spectroscopic identification of the redshift. {\it emline} refers to emission lines detected in the spectrum and {\it absline} refers to  absorption features.  Columns rms$_{x}$ refer to the rms (in arcsec.) of the predicted image positions according to models runs with the related set of constraints (g for \gold, s \silver, b \bronze, and c {\it copper}). Column {\it category} refers to the category of confidence level in which each image belongs, see Sect. \ref{sect:reliability} for a detailed description of each category.
}
\label{multipletable}
\end{table}
\tiny
\begin{center}

%\begin{tabular}{ccccccccc}
%\begin{longtable*}{ccccccccc}
\par
\tablefirsthead{\hline           \hline      \multicolumn{1}{c}{\textbf{ID}} &
                                \multicolumn{1}{c}{\textbf{R.A.}} &
                                \multicolumn{1}{c}{\textbf{Decl.}} &
                                \multicolumn{1}{c}{\textbf{$z_{\rm ref}$}} &
                                \multicolumn{1}{c}{\textbf{$z_{\rm spec}$}} &
                                \multicolumn{1}{c}{\textbf{confi}} &
                                 \multicolumn{1}{c}{\textbf{$z_{\rm model}$}} &
                                  \multicolumn{1}{c}{\textbf{absline}} &
                                   \multicolumn{1}{c}{\textbf{emline}} &
			                    \multicolumn{1}{l}{\textbf{rms$_{g}$}}&
			                    \multicolumn{1}{l}{\textbf{rms$_{s}$}}&
			                    \multicolumn{1}{l}{\textbf{rms$_{b}$}}&
			                    \multicolumn{1}{l}{\textbf{rms$_{c}$}}&
			                    \multicolumn{1}{l}{\textbf{category}}
                                \\ \hline }
                                
\tablehead{\hline \multicolumn{9}{l}{\small\sl continued from previous page}\\
                         \hline      \multicolumn{1}{c}{\textbf{ID}} &
                                \multicolumn{1}{c}{\textbf{R.A.}} &
                                \multicolumn{1}{c}{\textbf{Decl.}} &
                                \multicolumn{1}{c}{\textbf{$z_{\rm ref}$}} &
                                \multicolumn{1}{c}{\textbf{$z_{\rm spec}$}} &
                                \multicolumn{1}{c}{\textbf{confi}} &
                                 \multicolumn{1}{c}{\textbf{$z_{\rm model}$}} &
                                  \multicolumn{1}{c}{\textbf{absline}} &
                                   \multicolumn{1}{c}{\textbf{emline}} &
			                    \multicolumn{1}{l}{\textbf{rms$_{g}$}}&
			                    \multicolumn{1}{l}{\textbf{rms$_{s}$}}&
			                    \multicolumn{1}{l}{\textbf{rms$_{b}$}}&
			                    \multicolumn{1}{l}{\textbf{rms$_{c}$}}&
			                    \multicolumn{1}{l}{\textbf{category}}
                                \\ \hline }
\tabletail{\hline\multicolumn{9}{r}{\small\sl continued on next page}\\\hline}
\tablelasttail{\hline}
\par
\begin{supertabular}{lccccccccccccl}
1.1 & 3.5975477 & -30.403918 & M & 1.688 & 2 &  & FeII & CIII],FeII & 0.43 & 0.36 & 0.66 & 0.67 & gsbc\\ 
1.2 & 3.5959510 & -30.406813 & M & 1.688 & 2 &  & -- & CIII] & 0.36 & 0.57 & 0.70 & 0.52 & gsbc\\ 
1.3 & 3.5862330 & -30.409989 & M & 1.688 & 2 &  & -- & CIII] & 0.28 & 0.17 & 0.22 & 0.26 & gsbc\\ 
2.1 & 3.5832588 & -30.403351 & M & 1.8876 & 2 &  & -- & CIII] & 0.95 & 1.10 & 1.02 & 0.90 & gsbc\\ 
2.2 & 3.5972752 & -30.396724 & M & 1.8876 & 1 &  & -- & CIII] & 0.37 & 0.54 & 0.53 & 0.77 & gsbc\\ 
2.3 & 3.5854036 & -30.399898 & M & 1.8876 & 1 &  & -- & CIII] & 1.59 & 1.36 & 1.74 & 2.37 & gsbc\\ 
2.4 & 3.5864275 & -30.402128 & M & 1.8876 & 1 &  & -- & CIII] & 0.87 & 0.69 & 0.85 & 1.48 & gsbc\\ 
3.1 & 3.5893714 & -30.393864 & MJ & 3.9803 & 3 &  & LyB,SiII,O & HeII,OIII] & 0.38 & 0.47 & 0.66 & 0.53 & gsbc\\ 
3.2 & 3.5887908 & -30.393806 & MJ & 3.9803 & 3 &  & LyB,SiII,O & HeII,OIII] & 0.20 & 0.21 & 0.32 & 0.32 & gsbc\\ 
3.3 & 3.5766250 & -30.401813 & -- & -- & -- &  & -- & -- & -- & -- & -- & 0.69 & c\\ 
4.1 & 3.5921145 & -30.402634 & M & 3.5769 & 3 &  & SiII & Ly-a & 0.62 & 0.17 & 0.46 & 0.97 & gsbc\\ 
4.2 & 3.5956434 & -30.401623 & M & 3.5769 & 3 &  & SiII & Ly-a & 0.45 & 0.62 & 1.11 & 0.84 & gsbc\\ 
4.3 & 3.5804331 & -30.408926 & MR & 3.5769 & 3 &  & SiII & Ly-a & 0.62 & 0.55 & 1.29 & 1.16 & gsbc\\ 
4.4 & 3.5931933 & -30.404915 & M & 3.5769 & 3 &  & SiII & Ly-a & 0.54 & 0.97 & 1.45 & 0.51 & gsbc\\ 
4.5 & 3.5935934 & -30.405106 & MJ & 3.5769 & 3 &  & SiII & Ly-a & 0.04 & 0.04 & 0.26 & 0.57 & gsbc\\ 
5.1 & 3.5869257 & -30.390704 & M & 4.0225 & 3 &  & -- & Ly-a & -- & -- & -- & 0.54 & c\\ 
5.2 & 3.5849816 & -30.391374 & M & 4.0225 & 3 &  & -- & Ly-a & 0.10 & 0.76 & 0.43 & 0.70 & gsbc\\ 
5.3 & 3.5799583 & -30.394772 & M & 4.0225 & 3 &  & -- & Ly-a & 0.11 & 0.24 & 0.19 & 2.02 & gsbc\\ 
105.1 & 3.5834304 & -30.392070 & M & 4.0225 & 3 &  & -- & Ly-a & 0.26 & 0.11 & 0.55 & 1.56 & gsbc\\ 
105.2 & 3.5822917 & -30.392789 & M & 4.0225 & 3 &  & -- & Ly-a & 0.35 & 0.20 & 0.26 & 0.13 & gsbc\\ 
105.3 & 3.5804118 & -30.394316 & M & 4.0225 & 3 &  & -- & Ly-a & -- & -- & -- & 0.83 & c\\ 
105.4 & 3.5810603 & -30.393624 & M & 4.0225 & 3 &  & -- & Ly-a & -- & -- & -- & 0.27 & c\\ 
6.1 & 3.5985340 & -30.401800 & MRW & 2.016 & 3 &  & MgII & CIII] & 0.45 & 0.40 & 0.12 & 0.18 & gsbc\\ 
6.2 & 3.5940518 & -30.408011 & MW & 2.016 & 3 &  & MgII & CIII] & 0.21 & 0.38 & 0.39 & 0.75 & gsbc\\ 
6.3 & 3.5864225 & -30.409371 & MW & 2.016 & 3 &  & MgII & CIII] & 0.21 & 0.23 & 0.22 & 0.26 & gsbc\\ 
7.1 & 3.5982604 & -30.402326 & -- & -- & -- & 2.5791$^{+0.1065}_{-0.1103}$ & -- & -- & -- & 0.30 & 0.35 & 0.40 & sbc\\ 
7.2 & 3.5952195 & -30.407412 & -- & -- & -- &  & -- & -- & -- & 0.34 & 0.47 & 0.49 & sbc\\ 
7.3 & 3.5845989 & -30.409822 & -- & -- & -- &  & -- & -- & -- & 0.07 & 0.27 & 0.20 & sbc\\ 
8.1 & 3.5897088 & -30.394339 & M & 3.975 & 2 &  & LyB,OI,CII & -- & 0.38 & 0.49 & 0.48 & 0.75 & gsbc\\ 
8.2 & 3.5888225 & -30.394210 & M & 3.975 & 2 &  & LyB,OI,CII & -- & 0.29 & 0.37 & 0.40 & 0.13 & gsbc\\ 
8.3 & 3.5763966 & -30.402554 & -- & -- & -- &  & -- & -- & -- & -- & -- & 1.03 & c\\ 
9.1 & 3.5883900 & -30.405272 & -- & -- & -- & 2.3556$^{+0.4538}_{-0.0603}$ & -- & -- & -- & 0.48 & 0.70 & 0.96 & sbc\\ 
9.2 & 3.5871362 & -30.406229 & -- & -- & -- &  & -- & -- & -- & 0.36 & 1.00 & 0.89 & sbc\\ 
9.3 & 3.6001511 & -30.397153 & -- & -- & -- &  & -- & -- & -- & -- & 1.47 & 1.95 & bc\\ 
10.1 & 3.5884011 & -30.405880 & M & 2.6565 & 3 &  & -- & CIII] & 0.77 & 0.70 & 0.63 & 0.24 & gsbc\\ 
10.2 & 3.5873776 & -30.406485 & M & 2.6565 & 3 &  & -- & CIII] & 1.27 & 1.89 & 2.27 & 3.42 & gsbc\\ 
10.3 & 3.6007208 & -30.397095 & -- & -- & -- &  & -- & -- & -- & -- & -- & 4.91 & c\\ 
11.1 & 3.5913930 & -30.403847 & -- & -- & -- & 2.4508$^{+0.0942}_{-0.0488}$ & -- & -- & -- & -- & 0.19 & 0.35 & bc\\ 
11.2 & 3.5972708 & -30.401435 & -- & -- & -- &  & -- & -- & -- & -- & 0.16 & 0.49 & bc\\ 
11.3 & 3.5828051 & -30.408910 & -- & -- & -- &  & -- & -- & -- & -- & 0.16 & 0.33 & bc\\ 
11.4 & 3.5945298 & -30.406546 & -- & -- & -- &  & -- & -- & -- & -- & 0.14 & 0.25 & bc\\ 
12.1 & 3.5936156 & -30.404464 & -- & -- & -- & 3.6388$^{+0.6205}_{-0.2953}$ & -- & -- & -- & -- & 0.82 & 0.28 & bc\\ 
12.2 & 3.5932349 & -30.403259 & -- & -- & -- &  & -- & -- & -- & -- & 0.21 & 0.41 & bc\\ 
12.3 & 3.5945646 & -30.402986 & -- & -- & -- &  & -- & -- & -- & -- & 0.76 & 0.32 & bc\\ 
12.4 & 3.5795731 & -30.410258 & -- & -- & -- &  & -- & -- & -- & -- & -- & 0.98 & c\\ 
13.1 & 3.5923985 & -30.402536 & -- & -- & -- & 1.3905$^{+0.0385}_{-0.0378}$ & -- & -- & -- & 0.24 & 0.28 & 0.31 & sbc\\ 
13.2 & 3.5937700 & -30.402170 & -- & -- & -- &  & -- & -- & -- & 0.20 & 0.14 & 0.19 & sbc\\ 
13.3 & 3.5827578 & -30.408035 & -- & -- & -- &  & -- & -- & -- & 0.56 & 0.29 & 0.55 & sbc\\ 
14.1 & 3.5897344 & -30.394638 & -- & -- & -- & 2.571$^{+0.0348}_{-0.5613}$ & -- & -- & -- & 0.22 & 0.19 & 0.22 & sbc\\ 
14.2 & 3.5884245 & -30.394434 & -- & -- & -- &  & -- & -- & -- & 0.22 & 0.23 & 0.20 & sbc\\ 
14.3 & 3.5775729 & -30.401688 & -- & -- & -- &  & -- & -- & -- & -- & -- & 0.08 & c\\ 
18.1 & 3.5761227 & -30.404485 & M & 5.6625 & 3 &  & -- & Ly-a & 0.84 & 0.93 & 1.00 & 1.00 & gsbc\\ 
18.2 & 3.5883786 & -30.395646 & M & 5.6625 & 3 &  & -- & Ly-a & 0.66 & 0.53 & 0.34 & 0.65 & gsbc\\ 
18.3 & 3.5907250 & -30.395557 & M & 5.6625 & 3 &  & -- & Ly-a & 0.50 & 0.63 & 0.95 & 0.18 & gsbc\\ 
19.1 & 3.5889167 & -30.397439 & -- & -- & -- & 1.9902$^{+0.0966}_{-0.0791}$ & -- & -- & -- & -- & -- & 1.29 & c\\ 
19.2 & 3.5914427 & -30.396684 & -- & -- & -- &  & -- & -- & -- & -- & -- & 1.44 & c\\ 
19.3 & 3.5787177 & -30.404017 & -- & -- & -- &  & -- & -- & -- & -- & -- & 0.35 & c\\ 
20.1 & 3.5962413 & -30.402970 & -- & -- & -- & 2.5127$^{+0.4213}_{-0.1922}$ & -- & -- & -- & 0.25 & 0.05 & 0.38 & sbc\\ 
20.2 & 3.5951992 & -30.405437 & -- & -- & -- &  & -- & -- & -- & 0.35 & 0.27 & 0.30 & sbc\\ 
20.3 & 3.5820007 & -30.409552 & -- & -- & -- &  & -- & -- & -- & -- & 0.41 & 0.76 & bc\\ 
21.1 & 3.5961754 & -30.403112 & -- & -- & -- & 2.564$^{+0.2476}_{-0.3263}$ & -- & -- & -- & 0.29 & 0.16 & 0.62 & sbc\\ 
21.2 & 3.5952536 & -30.405340 & -- & -- & -- &  & -- & -- & -- & 0.28 & 0.24 & 0.32 & sbc\\ 
21.3 & 3.5819601 & -30.409610 & -- & -- & -- &  & -- & -- & -- & -- & 0.44 & 1.14 & bc\\ 
22.1 & 3.5879067 & -30.411612 & M & 5.2845 & 3 &  & -- & Ly-a & 1.37 & 1.18 & 0.94 & 0.43 & gsbc\\ 
22.2 & 3.6000458 & -30.404417 & M & 5.2845 & 3 &  & -- & Ly-a & 1.64 & 1.53 & 1.96 & 1.72 & gsbc\\ 
22.3 & 3.5965885 & -30.408983 & M & 5.2845 & 3 &  & -- & Ly-a & 1.27 & 1.01 & 0.88 & 0.15 & gsbc\\ 
23.1 & 3.5881623 & -30.410545 & -- & -- & -- & 4.4156$^{+0.3567}_{-0.1839}$ & -- & -- & -- & 0.29 & 0.18 & 0.66 & sbc\\ 
23.2 & 3.5935338 & -30.409717 & -- & -- & -- &  & -- & -- & -- & 0.42 & 0.38 & 0.64 & sbc\\ 
23.3 & 3.6005416 & -30.401831 & -- & -- & -- &  & -- & -- & -- & 0.36 & 0.05 & 0.07 & sbc\\ 
24.1 & 3.5959003 & -30.404480 & M & 1.043 & 3 &  & -- & [OII] & 0.58 & 0.69 & 0.46 & 0.08 & gsbc\\ 
24.2 & 3.5951250 & -30.405933 & M & 1.043 & 3 &  & -- & [OII] & 0.73 & 0.82 & 0.72 & 0.58 & gsbc\\ 
24.3 & 3.5873333 & -30.409102 & M & 1.043 & 1 &  & -- & [OII] & 0.75 & 1.04 & 0.81 & 0.25 & gsbc\\ 
25.1 & 3.5944626 & -30.402732 & -- & -- & -- & 1.2168$^{+0.0316}_{-0.0317}$ & -- & -- & -- & -- & 0.32 & 0.30 & bc\\ 
25.2 & 3.592150 & -30.403318 & -- & -- & -- &  & -- & -- & -- & -- & 0.14 & 0.46 & bc\\ 
25.3 & 3.5842145 & -30.408281 & -- & -- & -- &  & -- & -- & -- & -- & 0.42 & 0.71 & bc\\ 
26.1 & 3.5938976 & -30.409731 & M & 3.0537 & 1 &  & -- & Ly-a & 0.48 & 0.54 & 0.66 & 1.52 & gsbc\\ 
26.2 & 3.5903464 & -30.410581 & M & 3.0537 & 2 &  & -- & Ly-a & 0.61 & 0.49 & 0.64 & 1.68 & gsbc\\ 
26.3 & 3.6001103 & -30.402942 & M & 3.0537 & 2 &  & -- & Ly-a & 0.35 & 0.40 & 0.97 & 0.76 & gsbc\\ 
27.1 & 3.5807266 & -30.403137 & -- & -- & -- & 2.485$^{+0.1006}_{-0.086}$ & -- & -- & -- & -- & 0.50 & 0.96 & bc\\ 
27.2 & 3.5956979 & -30.396153 & -- & -- & -- &  & -- & -- & -- & -- & 0.24 & 1.20 & bc\\ 
27.3 & 3.5854978 & -30.397653 & -- & -- & -- &  & -- & -- & -- & -- & 0.64 & 0.55 & bc\\ 
28.1 & 3.5804479 & -30.405051 & -- & -- & -- & 6.5166$^{+0.0332}_{-0.6791}$ & -- & -- & -- & -- & 1.19 & 1.75 & bc\\ 
28.2 & 3.5978333 & -30.395964 & -- & -- & -- &  & -- & -- & -- & -- & 0.20 & 0.99 & bc\\ 
28.3 & 3.5853176 & -30.397958 & -- & -- & -- &  & -- & -- & -- & -- & 0.60 & 0.63 & bc\\ 
28.4 & 3.5874511 & -30.401372 & -- & -- & -- &  & -- & -- & -- & -- & 1.03 & 1.76 & bc\\ 
29.1 & 3.5824475 & -30.397567 & -- & -- & -- & 1.9859$^{+0.0557}_{-0.0424}$ & -- & -- & -- & -- & -- & 1.42 & c\\ 
29.2 & 3.5805261 & -30.400464 & -- & -- & -- &  & -- & -- & -- & -- & -- & 5.01 & c\\ 
29.3 & 3.5836000 & -30.396581 & -- & -- & -- &  & -- & -- & -- & -- & -- & 2.67 & c\\ 
30.1 & 3.5910104 & -30.397440 & M & 1.0252 & 3 &  & -- & [OII] & 1.08 & 0.75 & 0.91 & 1.72 & gsbc\\ 
30.2 & 3.5866771 & -30.398188 & M & 1.0252 & 3 &  & -- & [OII] & 1.30 & 1.31 & 1.53 & 1.35 & gsbc\\ 
30.3 & 3.5819245 & -30.401700 & M & 1.0252 & 3 &  & -- & [OII] & 0.51 & 0.18 & 0.37 & 1.53 & gsbc\\ 
31.1 & 3.5859340 & -30.403159 & M & 4.7594 & 3 &  & -- & Ly-a & 0.45 & 0.32 & 0.32 & 0.60 & gsbc\\ 
31.2 & 3.5837083 & -30.404105 & M & 4.7594 & 3 &  & -- & Ly-a & 0.59 & 0.44 & 0.49 & 1.18 & gsbc\\ 
31.3 & 3.5998296 & -30.395522 & -- & -- & 3 &  & -- & -- & -- & -- & -- & 0.78 & c\\ 
32.1 & 3.5835963 & -30.404705 & -- & -- & -- & 5.6678$^{+0.4284}_{-0.4721}$ & -- & -- & -- & -- & 0.56 & 0.79 & bc\\ 
32.2 & 3.5866709 & -30.403345 & -- & -- & -- &  & -- & -- & -- & -- & 0.15 & 0.70 & bc\\ 
32.3 & 3.5997765 & -30.395981 & -- & -- & -- &  & -- & -- & -- & -- & 0.98 & 0.56 & bc\\ 
33.1 & 3.5847083 & -30.403152 & M & 5.7255 & 3 &  & -- & Ly-a & 0.53 & 0.48 & 0.65 & 0.24 & gsbc\\ 
33.2 & 3.5843959 & -30.403400 & M & 5.7255 & 3 &  & -- & Ly-a & 0.46 & 0.30 & 0.35 & 0.70 & gsbc\\ 
33.3 & 3.6004183 & -30.395110 & M & 5.7255 & 3 &  & -- & Ly-a & 1.39 & 1.34 & 1.30 & 1.21 & gsbc\\ 
34.1 & 3.5934255 & -30.410842 & M & 3.785 & 3 &  & -- & Ly-a & -- & -- & -- & -- & gsbc\\ 
34.2 & 3.5938141 & -30.410718 & M & 3.785 & 3 &  & -- & Ly-a & -- & -- & -- & -- & gsbc\\ 
34.3 & 3.6007050 & -30.404605 & M & 3.785 & 3 &  & -- & Ly-a & -- & -- & -- & -- & gsbc\\ 
35.1 & 3.58111058 & -30.400215 & M & 2.656 & 3 &  & -- & 0,Ly-a & -- & -- & 0.23 & 0.43 & bc\\ 
35.2 & 3.5815417 & -30.399392 & -- & -- & -- &  & -- & -- & -- & -- & 0.20 & 0.40 & bc\\ 
35.3 & 3.5978333 & -30.395542 & -- & -- & -- &  & -- & -- & -- & -- & 0.13 & 0.73 & bc\\ 
36.1 & 3.5894583 & -30.394408 & -- & -- & -- & 3.9291$^{+1.3075}_{-1.1899}$ & -- & -- & -- & 0.06 & 0.12 & 0.19 & sbc\\ 
36.2 & 3.5886666 & -30.394300 & -- & -- & -- &  & -- & -- & -- & 0.06 & 0.17 & 0.17 & sbc\\ 
36.3 & 3.5774792 & -30.401508 & -- & -- & -- &  & -- & -- & -- & -- & -- & 0.20 & c\\ 
37.1 & 3.5890417 & -30.394913 & M & 2.6501 & 3 &  & -- & CIII],Ly-a & 0.16 & 0.10 & 0.11 & 2.35 & gsbc\\ 
37.2 & 3.5887083 & -30.394852 & M & 2.6501 & 3 &  & -- & CIII],Ly-a & 0.21 & 0.12 & 0.16 & 1.06 & gsbc\\ 
37.3 & 3.5794427 & -30.400275 & -- & -- & -- &  & -- & -- & -- & -- & -- & 9.62 & c\\ 
38.1 & 3.5894166 & -30.394100 & -- & -- & -- & 5.4665$^{+1.2035}_{-0.8191}$ & -- & -- & -- & -- & 0.01 & 0.15 & bc\\ 
38.2 & 3.5889396 & -30.394044 & -- & -- & -- &  & -- & -- & -- & -- & 0.01 & 0.17 & bc\\ 
38.3 & 3.5763968 & -30.402128 & -- & -- & -- &  & -- & -- & -- & -- & -- & 1.11 & c\\ 
39.1 & 3.5887917 & -30.392530 & M & 4.015 & 3 &  & LyB & -- & 0.24 & 0.07 & 0.51 & 0.38 & gsbc\\ 
39.2 & 3.5885417 & -30.392508 & M & 4.015 & 3 &  & LyB & -- & 0.38 & 0.45 & 0.53 & 0.34 & gsbc\\ 
39.3 & 3.5774787 & -30.399568 & M & 4.015 & -- &  & LyB & -- & 0.75 & 0.89 & 1.20 & 0.84 & gsbc\\ 
40.1 & 3.5890859 & -30.392668 & M & 4.0 & 3 &  & LyB & -- & 0.23 & 0.12 & 0.09 & 0.49 & gsbc\\ 
40.2 & 3.5881935 & -30.392551 & M & 4.0 & 3 &  & LyB & -- & 0.21 & 0.13 & 0.12 & 0.26 & gsbc\\ 
40.3 & 3.5775443 & -30.399376 & -- & -- & -- &  & -- & -- & -- & -- & -- & 1.83 & c\\ 
41.1 & 3.5991758 & -30.399582 & M & 4.9113 & 1 &  & -- & Ly-a & 0.36 & 0.38 & 0.27 & 0.26 & gsbc\\ 
41.2 & 3.5935571 & -30.407769 & M & 4.9113 & 1 &  & -- & Ly-a & 0.61 & 0.53 & 0.37 & 0.12 & gsbc\\ 
41.3 & 3.5834467 & -30.408500 & M & 4.9113 & 1 &  & -- & Ly-a & 0.51 & 0.23 & 0.20 & 0.40 & gsbc\\ 
41.4 & 3.5906170 & -30.404459 & M & 4.9113 & 1 &  & -- & Ly-a & 0.85 & 0.61 & 0.44 & 0.46 & gsbc\\ 
42.1 & 3.5973056 & -30.400612 & M & 3.6915 & 3 &  & -- & Ly-a & 0.24 & 0.22 & 0.24 & 0.82 & gsbc\\ 
42.2 & 3.5909609 & -30.403255 & M & 3.6915 & 3 &  & -- & Ly-a & 1.21 & 0.50 & 0.36 & 1.35 & gsbc\\ 
42.3 & 3.5815842 & -30.408635 & M & 3.6915 & 3 &  & -- & Ly-a & 0.58 & 0.22 & 0.63 & 0.50 & gsbc\\ 
42.4 & 3.5942281 & -30.406390 & M & 3.6915 & 3 &  & -- & Ly-a & 0.21 & 0.20 & 0.22 & 0.55 & gsbc\\ 
42.5 & 3.5924125 & -30.405194 & M & 3.6915 & 3 &  & -- & Ly-a & 1.05 & 0.47 & 0.67 & 1.98 & gsbc\\ 
43.1 & 3.5978359 & -30.402507 &  & -- & -- & 5.8998$^{+0.2275}_{-0.139}$ & -- & -- & -- & -- & -- & 3.69 & c\\ 
43.2 & 3.5839609 & -30.409811 & & -- & -- &  & -- & -- & -- & -- & -- & 3.75 & c\\ 
44.1 & 3.5834655 & -30.406964 & & -- & -- & 2.224$^{+0.514}_{-0.1725}$ & -- & -- & -- & -- & -- & 1.09 & c\\ 
44.2 & 3.5966979 & -30.399755 & & -- & -- &  & -- & -- & -- & -- & -- & 1.21 & c\\ 
45.1 & 3.5848425 & -30.398474 &  & -- & -- & 5.9555$^{+0.1575}_{-0.119}$ & -- & -- & -- & -- & -- & 1.42 & c\\ 
45.2 & 3.5814059 & -30.403962 &  & -- & -- &  & -- & -- & -- & -- & -- & 1.03 & c\\ 
45.3 & 3.5869000 & -30.401299 &  & -- & -- &  & -- & -- & -- & -- & -- & 2.84 & c\\ 
45.4 & 3.5974146 & -30.396146 & & -- & -- &  & -- & -- & -- & -- & -- & 2.17 & c\\ 
46.1 & 3.5950222 & -30.400755 &  & -- & -- & 4.9368$^{+0.2411}_{-0.1549}$ & -- & -- & -- & -- & -- & 3.88 & c\\ 
46.2 & 3.5925108 & -30.401486 & & -- & -- &  & -- & -- & -- & -- & -- & 0.31 & c\\ 
46.3 & 3.5775195 & -30.408704 &  & -- & -- &  & -- & -- & -- & -- & -- & 2.82 & c\\ 
47.1 & 3.5901625 & -30.392181 & M & 4.0225 & 3 &  & -- & Ly-a & 0.35 & 0.20 & 0.20 & 0.98 & gsbc\\ 
47.2 & 3.5858417 & -30.392244 & M & 4.0225 & 3 &  & -- & Ly-a & 0.49 & 0.35 & 1.32 & 0.59 & gsbc\\ 
47.3 & 3.5783292 & -30.398133 & M & 4.0225 & 3 &  & -- & Ly-a & 0.53 & 0.49 & 0.90 & 0.57 & gsbc\\ 
147.1 & 3.5896792 & -30.392136 & M & 4.0225 & 3 &  & -- & Ly-a & 0.22 & 0.22 & 0.48 & 1.05 & gsbc\\ 
147.2 & 3.5864542 & -30.392128 & M & 4.0225 & 3 &  & -- & Ly-a & 0.27 & 0.37 & 0.57 & 1.17 & gsbc\\ 
147.3 & 3.5780083 & -30.398392 & M & 4.0225 & 3 &  & -- & Ly-a & 0.29 & 0.18 & 0.19 & 0.10 & gsbc\\ 
48.1 & 3.5942500 & -30.402845 &  & -- & -- & 1.7775$^{+0.305}_{-0.2038}$ & -- & -- & -- & -- & 0.02 & 0.87 & bc\\ 
48.2 & 3.5927667 & -30.403138 &  & -- & -- &  & -- & -- & -- & -- & 0.03 & 0.15 & bc\\ 
48.3 & 3.5820469 & -30.408594 &  & -- & -- &  & -- & -- & -- & -- & -- & 1.46 & c\\ 
49.1 & 3.5926320 & -30.408274 &  & -- & -- & 1.1172$^{+0.0269}_{-0.0331}$ & -- & -- & -- & -- & -- & 2.11 & c\\ 
49.2 & 3.5902300 & -30.408802 &  & -- & -- &  & -- & -- & -- & -- & -- & 1.22 & c\\ 
49.3 & 3.5975108 & -30.403160 &  & -- & -- &  & -- & -- & -- & -- & -- & 0.74 & c\\ 
50.1 & 3.5779770 & -30.401607 &  & -- & -- & 4.9179$^{+0.3544}_{-0.1974}$ & -- & -- & -- & -- & 0.22 & 0.18 & bc\\ 
50.2 & 3.5939583 & -30.394281 &  & -- & -- &  & -- & -- & -- & -- & 0.19 & 0.71 & bc\\ 
50.3 & 3.5851100 & -30.393739 &  & -- & -- &  & -- & -- & -- & -- & -- & 1.26 & c\\ 
51.1 & 3.5868774 & -30.405662 &  & -- & -- & 4.7621$^{+0.361}_{-0.317}$ & -- & -- & -- & -- & 0.26 & 0.53 & bc\\ 
51.2 & 3.5864583 & -30.405662 &  & -- & -- &  & -- & -- & -- & -- & 0.59 & 0.65 & bc\\ 
51.3 & 3.5990000 & -30.398303 &  & -- & -- &  & -- & -- & -- & -- & -- & 4.42 & c\\ 
52.1 & 3.5865069 & -30.397039 &  & -- & -- & 1.0097$^{+0.0135}_{-0.0014}$ & -- & -- & -- & -- & 0.07 & 2.05 & bc\\ 
52.2 & 3.5861430 & -30.397133 & & -- & -- &  & -- & -- & -- & -- & 0.08 & 3.22 & bc\\ 
52.3 & 3.5884301 & -30.396822 &  & -- & -- &  & -- & -- & -- & -- & -- & 3.97 & c\\ 
53.1 & 3.5798420 & -30.401592 &  & -- & -- & 6.8098$^{+0.0234}_{-0.2741}$ & -- & -- & -- & -- & 1.44 & 0.95 & bc\\ 
53.2 & 3.5835495 & -30.396703 &  & -- & -- &  & -- & -- & -- & -- & 1.30 & 1.81 & bc\\ 
53.3 & 3.5970416 & -30.394547 &  & -- & -- &  & -- & -- & -- & -- & -- & 1.52 & c\\ 
54.1 & 3.592345 & -30.409895 &  & -- & -- & 5.4223$^{+0.3005}_{-0.1579}$ & -- & -- & -- & -- & -- & 1.35 & c\\ 
54.2 & 3.5882578 & -30.410328 &-- & -- & -- &  & -- & -- & -- & -- & -- & 0.98 & c\\ 
54.3 & 3.5884037 & -30.410295 & -- & -- & -- &  & -- & -- & -- & -- & -- & 0.52 & c\\ 
54.4 & 3.5901058 & -30.410259 & -- & -- & -- &  & -- & -- & -- & -- & -- & 2.01 & c\\ 
54.5 & 3.60092196 & -30.400831 & -- & -- & -- &  & -- & -- & -- & -- & -- & 1.78 & c\\ 
59.1 & 3.5842840 & -30.408924 & -- & -- & -- & 4.0575$^{+0.646}_{-0.1358}$ & -- & -- & -- & -- & -- & 2.34 & c\\ 
59.2 & 3.5981200 & -30.400983 & -- & -- & -- &  & -- & -- & -- & -- & -- & 2.28 & c\\ 
60.1 & 3.5980780 & -30.403990 & -- & -- & -- & 1.6981$^{+0.0564}_{-0.046}$ & -- & -- & -- & 0.17 & 0.70 & 0.85 & sbc\\ 
60.2 & 3.5957235 & -30.407549 & -- & -- & -- &  & -- & -- & -- & 0.54 & 0.56 & 0.71 & sbc\\ 
60.3 & 3.5873816 & -30.410162 & -- & -- & -- &  & -- & -- & -- & 0.21 & 0.14 & 0.27 & sbc\\ 
61.1 & 3.5955330 & -30.403499 & M & 2.951 & 3 &  & -- & Ly-a & 0.16 & 0.13 & 0.19 & 0.40 & gsbc\\ 
61.2 & 3.5951427 & -30.404495 & M & 2.951 & 3 &  & -- & Ly-a & 0.14 & 0.12 & 0.18 & 0.34 & gsbc\\ 
62.1 & 3.5913260 & -30.398643 & M & 4.1935 & 3 &  & -- & Ly-a & 0.30 & 0.50 & 0.65 & 0.73 & gsbc\\ 
62.2 & 3.5905821 & -30.398918 & M & 4.1935 & 3 &  & -- & Ly-a & 0.24 & 0.34 & 0.41 & 0.43 & gsbc\\ 
63.1 & 3.5822614 & -30.407166 & M & 5.6616 & 3 &  & -- & Ly-a & 0.59 & 0.38 & 0.58 & 0.80 & gsbc\\ 
63.2 & 3.5927578 & -30.407022 & M & 5.6616 & 3 &  & -- & Ly-a & 0.57 & 0.56 & 0.20 & 0.25 & gsbc\\ 
63.3 & 3.5891334 & -30.403419 & M & 5.6616 & 3 &  & -- & Ly-a & 0.85 & 0.47 & 0.33 & 1.00 & gsbc\\ 
63.4 & 3.5988055 & -30.398279 & M & 5.6616 & 3 &  & -- & Ly-a & 0.40 & 0.53 & 0.47 & 0.31 & gsbc\\ 
64.1 & 3.5811967 & -30.398708 & M & 3.4087 & 3 &  & -- & Ly-a & 0.49 & 0.41 & 0.13 & 0.73 & gsbc\\ 
64.3 & 3.5963329 & -30.394232 & M & 3.4087 & 3 &  & -- & Ly-a & 1.81 & 1.75 & 1.04 & 1.99 & gsbc\\

\end{supertabular}
\end{center}

\onecolumn

\section{Redshift comparison with previous spectroscopic catalogs}
\label{sect:redshifts_comparison}
In this Appendix we compare discrepant MUSE redshift with corresponding values from the literature. The details are summarised in Table~\ref{tab:comparison_redshifts}. 

 \begin{table*}
 	\centering
 	\caption{Summary of the redshifts comparison with previous studies. The first subdivision use the synthesised catalog of \citet{Boschin2006} and keep the original identification of object made by \citet{CouchNewell1984}. The second and the third subdivisions summarise the cross-match between the MUSE redshift measurements presented in the current study and the publicly available GLASS redshift catalog (\citealt{GLASS2,GLASS1}) and the strong lensing analysis \citep{Wang2015}. Column C refers to the confidence level associated with the MUSE redshifts, while Q refers to the redshift quality provided in the GLASS catalogs.
Arrows ($\rightarrow$) indicate updates in redshift catalogs based on comparison.
   %\footnote{\url{https://archive.stsci.edu/prepds/glass/}} ).
   %If the difference in redshifts was lower than 0.01 I considered that the measurement agree together.
   }
 	\label{tab:comparison_redshifts}
 	\begin{tabular}{lcccccp{90mm}} 
 		\hline
 		\hline
 \multicolumn{7}{l}{Comparison of MUSE redshifts with Couch \& Newell (1984) redshifts}\\
 ID$_{\rm MUSE}$ &  ${z}_{\rm MUSE}$ &C& ID$_{\rm CN}$ & ${z}_{\rm CN}$ && Description \\
 [1.5ex] 
\hline
     5693  & 0.2986 &3& 47      &0.2896 && Multiple absorption features (incl. K, H and G) are detected in the MUSE spectra.
   \\
   9778  & 0.6011 &3& 33  &0.4982 &&  strong [\ion{O}{II}] doublet emission is detected in the MUSE spectra. \\
   10059&0.3204&3&3&0.31&& Multiple absorption features (incl. K, H and G) are detected in the MUSE spectra.
    \\ 
   10508 & 0.1900 &3& 5 &0.0631 && We securely identified a very strong \ion{H}{$\alpha$} emission. 
%also recorded in \citep{LEDA} as LEDA 138141
\\
 [2.5ex] 
\hline
\hline
\multicolumn{7}{l}{Comparison of MUSE redshifts with GLASS v001 redshift catalog.} \\
\multicolumn{7}{l}{Redshift updates are included in the GLASS v002 redshift catalog (\url{https://archive.stsci.edu/prepds/glass/}) released with this paper.}  \\
 [1.5ex] 
 ID$_{\rm MUSE}$ &  ${z}_{\rm MUSE}$ &C& ID$_{\rm GLASS}$ & ${z}_{\rm GLASS}$ &Q& Description \\
 	\hline

   9910&1.3397&3&793&2.090$\rightarrow$1.340&3$\rightarrow$4& GLASS emission mis-identified as [\ion{O}{III}] instead of \ion{H}{$\alpha$}. The \ion{H}{$\alpha$} agrees with strong [\ion{O}{II}] emission and multiple absorption features in the MUSE spectrum.
   %They thought measuring [OIII] at 15372 angstrom but it was Halpha at 15355 . 
    \\
  5838 &2.5809&2&1346&1.03 0$\rightarrow$2.581&3$\rightarrow$4& GLASS emission was mis-identified as \ion{H}{$\alpha$} instead of [\ion{O}{II}]. The [\ion{O}{II}] agrees with strong \ion{Si}{II} emission and \ion{C}{IV} absorption features in the MUSE spectrum.
  %MUSE SiII and CIV absorption. GLASS thought it was Halpha but acording to our redshifts the line they found is OII 
  \\
     3361&2.7416&1$\rightarrow$2&1740&1.130$\rightarrow$2.742&2$\rightarrow$3& GLASS detection of the [\ion{O}{II}] emission line confirm the faint \ion{C}{III}] emission detected by MUSE.\\

  7692&2.0700&1$\rightarrow$2&1144&2.081$\rightarrow$2.070&2.5$\rightarrow$3& GLASS detection of [\ion{O}{III}] emission confirms multiple faint absorption lines (MgII, FeII, AlIII)
found in the MUSE spectra. \\
   10999&1.1425&1$\rightarrow$2&322&1.1425&4& Strong \ion{H}{$\alpha$} and \ion{Si}{II} emission detected in the GLASS spectra confirms the multiple faint absorption lines found in the MUSE spectra.\\
  14412 &1.6750&1$\rightarrow$2&169&1.6750&4& Strong [\ion{O}{III}], \ion{H}{$\beta$} and [\ion{O}{II}] emission detected in the GLASS spectra confirms faint \ion{Al}{III} absorptions and faint \ion{C}{III}] emission in the MUSE spectra.\\
  14675&1.8925&1$\rightarrow$2&263&1.8925&4& Strong [\ion{O}{III}] and \ion{H}{$\beta$} emission detected in the GLASS spectra confirms multiple faint UV absorption features found in the MUSE spectra.  \\
 1.3 &1.688&2&1760&1.8630&3& The MUSE redshifts for the multiply-imaged system 1 was measured based on the \ion{C}{III]} doublet emission and multiple UV absorption features in the stacked spectra on all multiple images. No spectral feature was detected around the GLASS redshift. Redshift disagreement unresolved.\\
  8400&...&...&997&1.1750&3& The \ion{H}{$\alpha$} based redshift from the GLASS spectra is not matched by any prominent emission in the MUSE spectra (e.g.[\ion{O}{II}]). \\
56.1 &1.8876&2&1467&1.20$\rightarrow$1.8876&3$\rightarrow$4& The multiply imaged system 56 is physically related to the multiply imaged system 2. Detected emission in the GLASS spectra was identified as \ion{H}{$\alpha$} by \citet{Wang2015}. However, correcting this to [\ion{O}{III}] agrees with the MUSE \ion{C}{III}] detection. In the MUSE data cube we performed a manual extraction of image 2.1/56.1 due to the high level of contamination of the three counter images. \\
3402& 1.6480 &3&    1773 &  1.660$\rightarrow$1.6480 &  4 &
The MUSE redshift is based on multiple absorption feature (\ion{Al}{III}, \ion{Fe}{II} and \ion{Mg}{II}) and faint \ion{C}{III}] emission. Discrepancy with the GLASS redshift is attributed to the lower resolution of the HST grisms and the source morphology convolution when extracting 1D GLASS spectra. The clear [\ion{O}{II}] and [\ion{O}{III}] detections by GLASS match the MUSE redshift.\\
11419 &  0.3213   &3&   435 &  1.0500& 3 &The MUSE redshifts based on multiple faint absorption features and the continuum level of flux clearly identified a cluster member. The [\ion{O}{III}] based redshift from the GLASS spectra is not matched by any prominent emission at a different redshift. Redshift disagreement unresolved.
\end{tabular}
 \end{table*}
 
 \setcounter{table}{0}
 \begin{table*}
 \caption{(continued)}
 	\begin{tabular}{lcccccp{90mm}} 
\hline
\hline
 \multicolumn{7}{l}{Comparison of MUSE redshifts with \citet{Wang2015} redshifts (if source is not listed in GLASS v001/v002 redshift catalog)} \\
 [1.5ex] 
 ID$_{\rm MUSE}$ &  ${z}_{\rm MUSE}$ &C &ID$_{\rm Wang\ et\ al.}$ & ${z}_{\rm Wang\ et\ al.}$ &Q& Description \\
  		\hline
  22.2&5.283&3&807&4.84&2& MUSE securely identified the Ly$\alpha$ emission line in all multiple images of this system. \\
6261 
 &0.546&3&996&1.14&2& The GLASS redshift was mis-identified as [\ion{O}{II}] instead of [\ion{O}{III}], which was realised because of a six different emission line detection in the MUSE spectrum.\\
7007&...&...&1064&1.17&3& The \ion{H}{$\alpha$} based redshift from the GLASS spectra is not matched by any prominent [\ion{O}{II}] or MgII emission in the MUSE spectra.  \\
 \hline
 \end{tabular}
 \end{table*}

%%%%%%%%%%%%%%%%%%%%%%%%%%%%%%%%%%%%%%%%%%%%%%%%%%%%%
%%%%%%%%%%%%%%%%%%%%%%%%%%%%%%%%%%%%%%%%%%%%%SEction SExtr param
% \section{{\sc sextractor} parameters}
% \label{sexparam}
% \begin{table}
% 	\centering
% 	\caption{{\sc SExtractor} parameters used in dual mode to detect faint sources }
% 	\label{tab:sexparam}
% 	\begin{tabular}{lc} 
% 		\hline
% 		\hline
% Parameter name & Value \\
% \hline
% DETECT\_MINAREA &  $10$  \\
% DETECT\_THRESH   & $1$ \\
% ANALYSIS\_THRESH  & $0.8$ \\
% FILTER     &      Y            \\  
% FILTER\_NAME    &  tophat\_3.0\_3x3.conv \\
% DEBLEND\_NTHRESH  &$8$  \\
% DEBLEND\_MINCONT  &$0.00001$ \\
% SEEING\_FWHM    &  $0.07$ \\
% %Put the background parameters   &  $XXX$ \\
% 	\hline
% \end{tabular}
% \end{table}

\clearpage

\section{image multiple}
\label{sect:img_mul}
%newpage 
\begin{figure*} 
\centerline{ 
\includegraphics[width=0.5\textwidth]{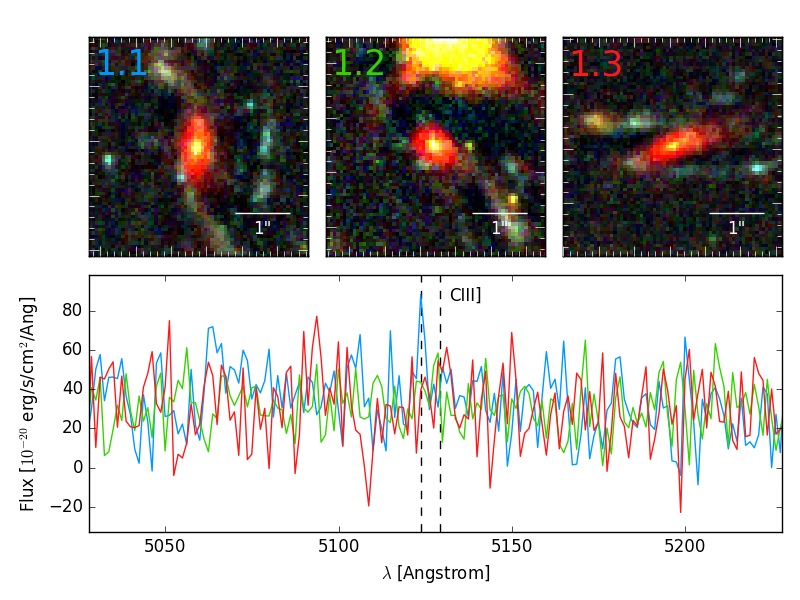} 
\includegraphics[width=0.5\textwidth]{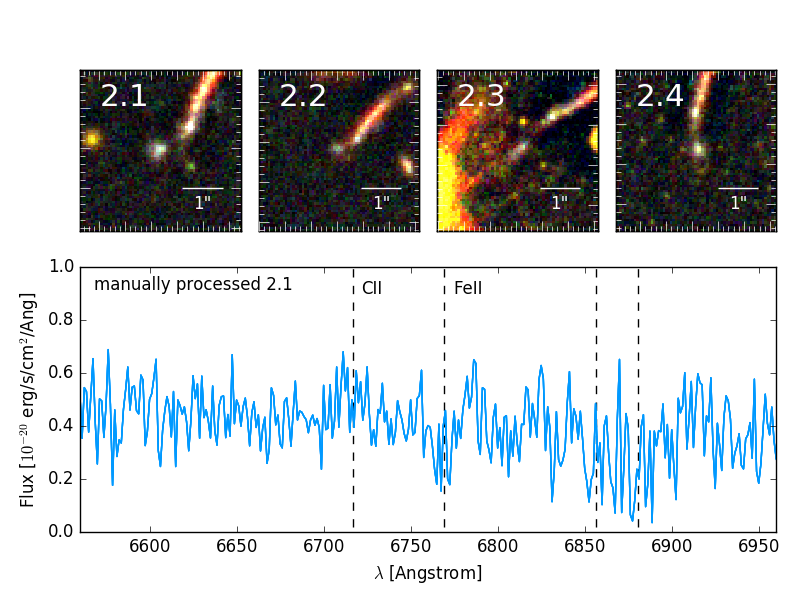} } 
\centerline{ 
\includegraphics[width=0.5\textwidth]{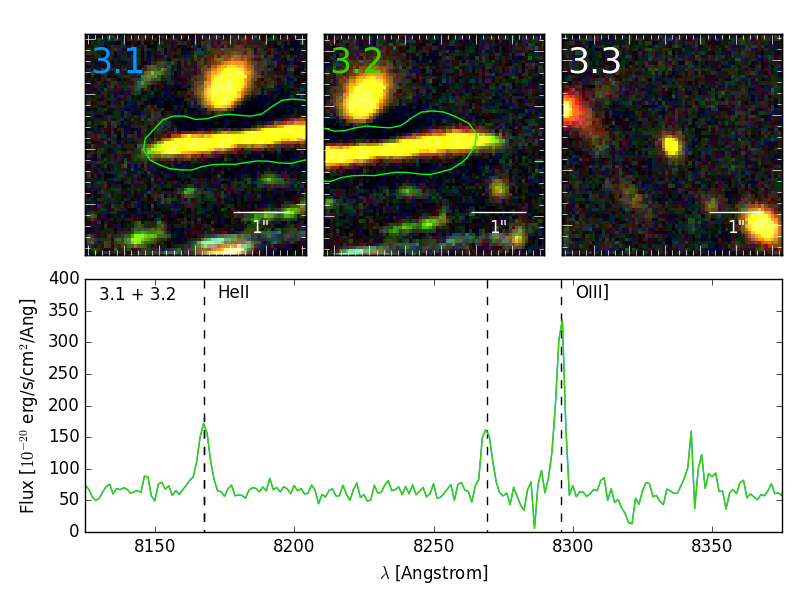} 
\includegraphics[width=0.5\textwidth]{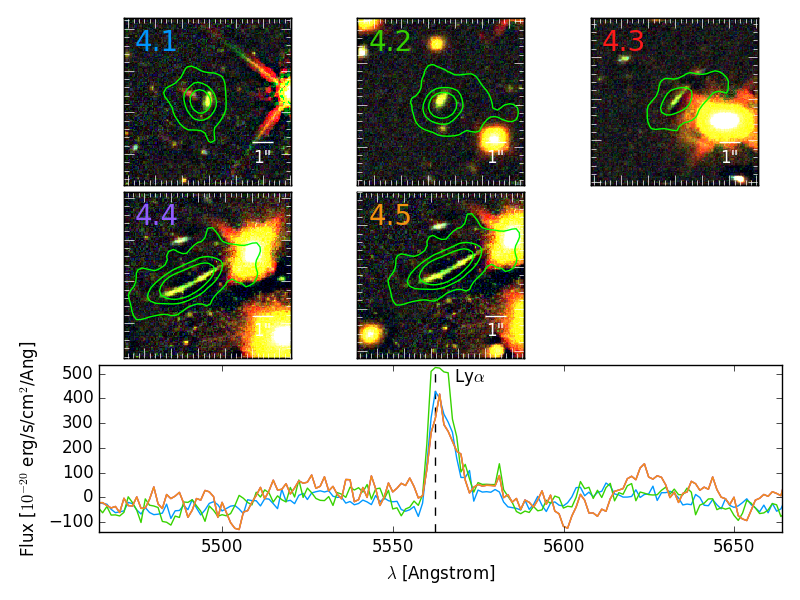} } 
\centerline{ 
\includegraphics[width=0.5\textwidth]{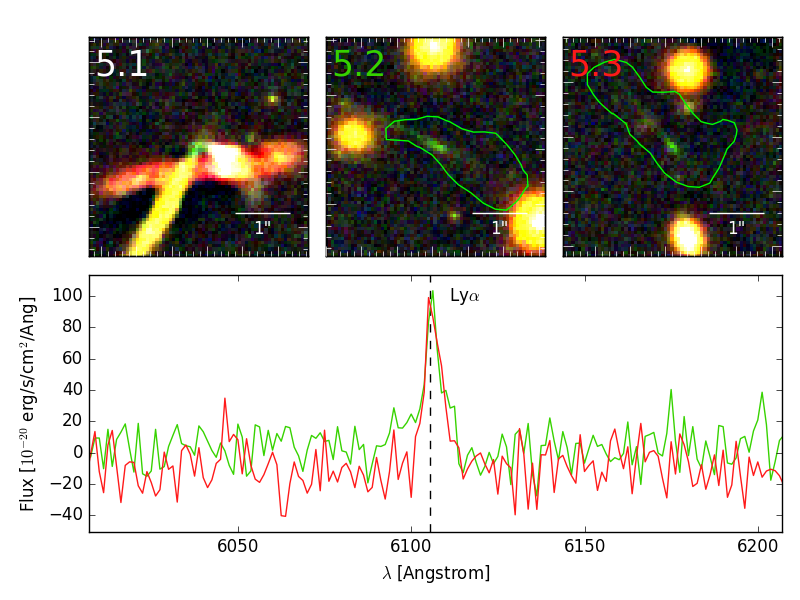} 
\includegraphics[width=0.5\textwidth]{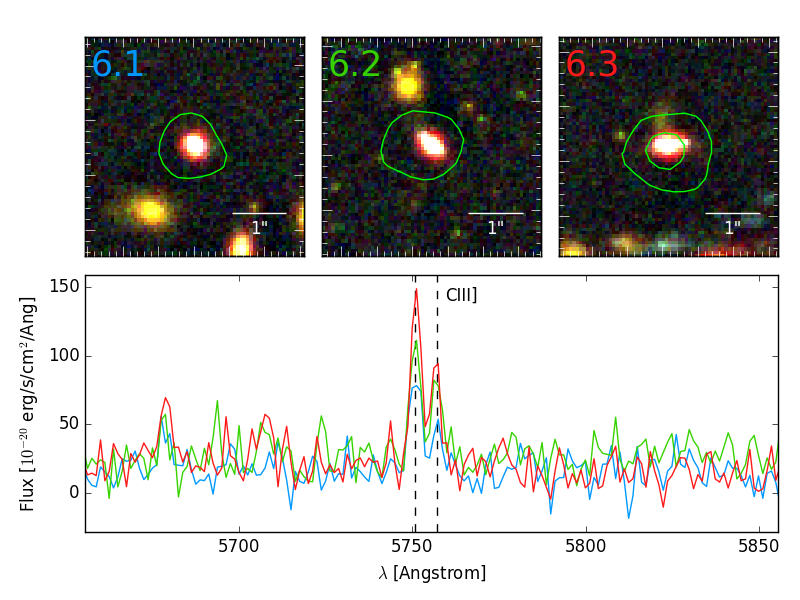} } 
\caption{ Each panel presents multiply-imaged systems which contains at least two images with spectroscopy. On the bottom of each panel a spectrum presents the most obvious spectral features used to measure the redshift of the system. On top, HST RGB images made from the median subtracted images used for the photometry-based spectral extraction. Special cases are made for system 1 and 2. System 1 was detected only on the stacked spectra of the three images. Due to his large contamination by cluster members, the redshifts for system 2 was only measured based on image 2.1.} 
\end{figure*} 
\begin{figure*}\ContinuedFloat 
\centerline{ 
\includegraphics[width=0.5\textwidth]{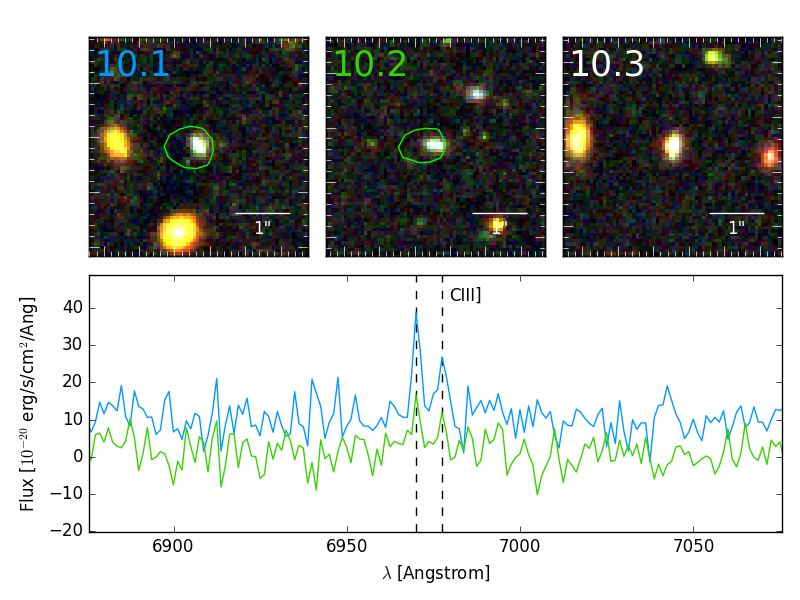} 
\includegraphics[width=0.5\textwidth]{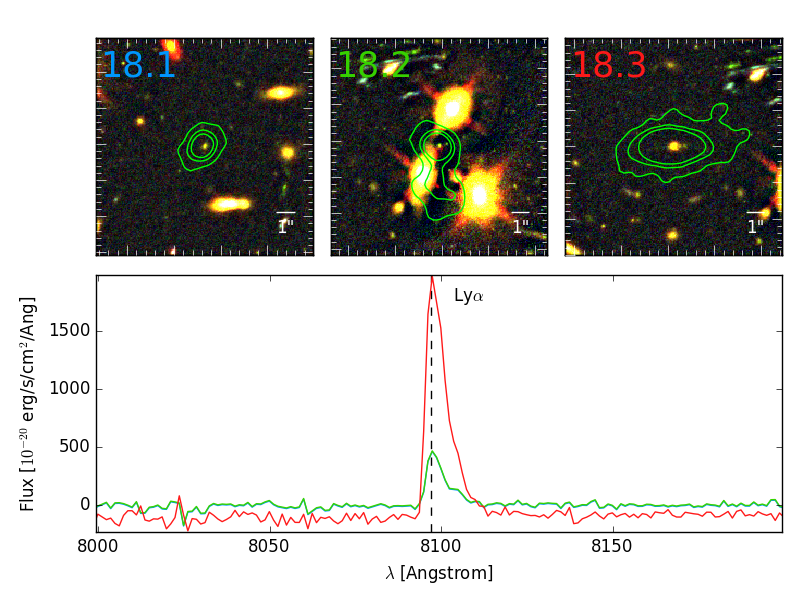} } 
\centerline{ 
\includegraphics[width=0.5\textwidth]{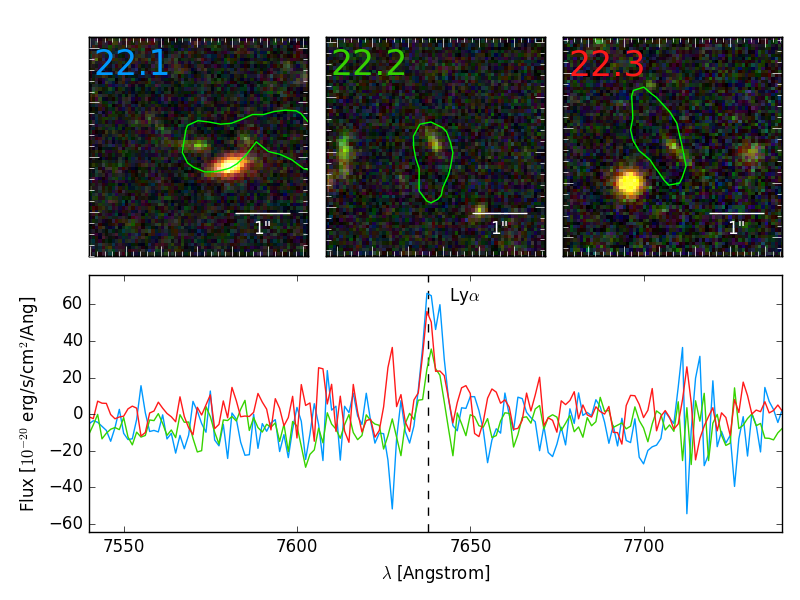} 
\includegraphics[width=0.5\textwidth]{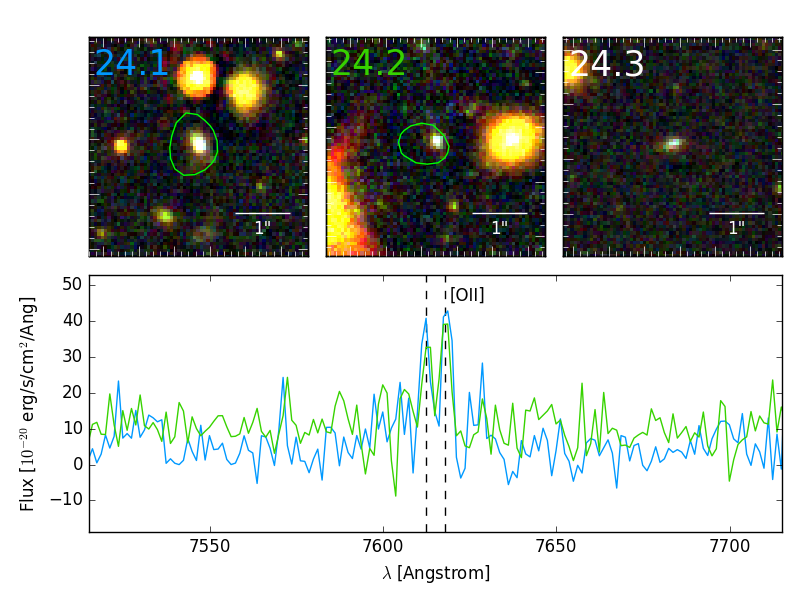} } 
\centerline{ 
\includegraphics[width=0.5\textwidth]{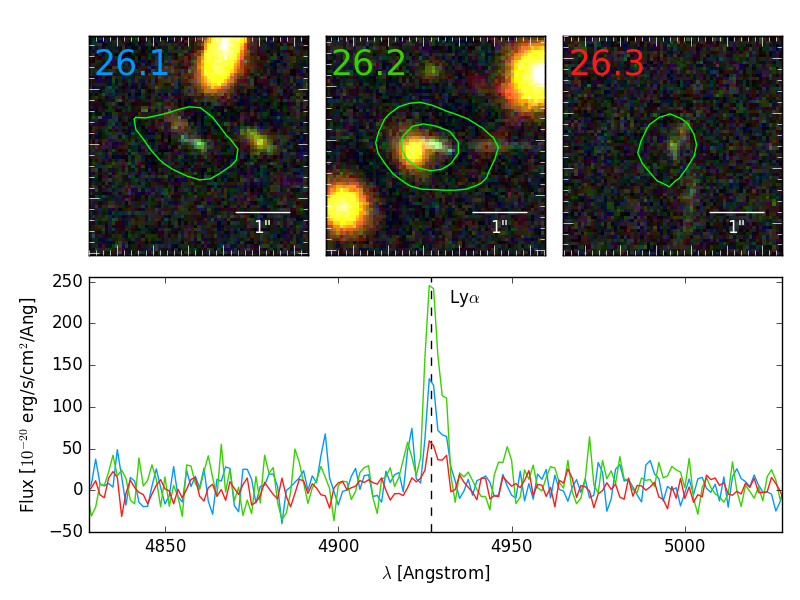} 
\includegraphics[width=0.5\textwidth]{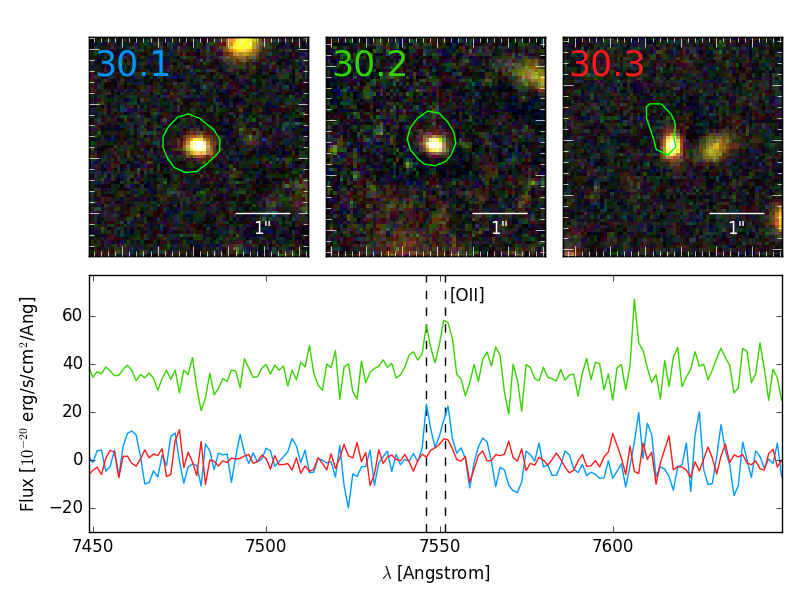} } 
\caption{(continued) Multiply-imaged systems. Image 10.2 shows an extracted region at 0.5$\sigma$ } 
\end{figure*} 
\begin{figure*}\ContinuedFloat 
\centerline{ 
\includegraphics[width=0.5\textwidth]{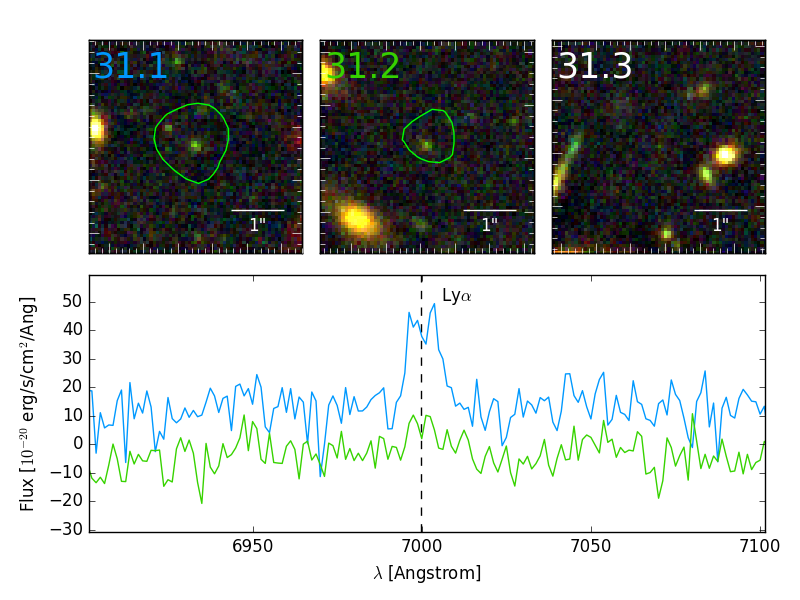} 
\includegraphics[width=0.5\textwidth]{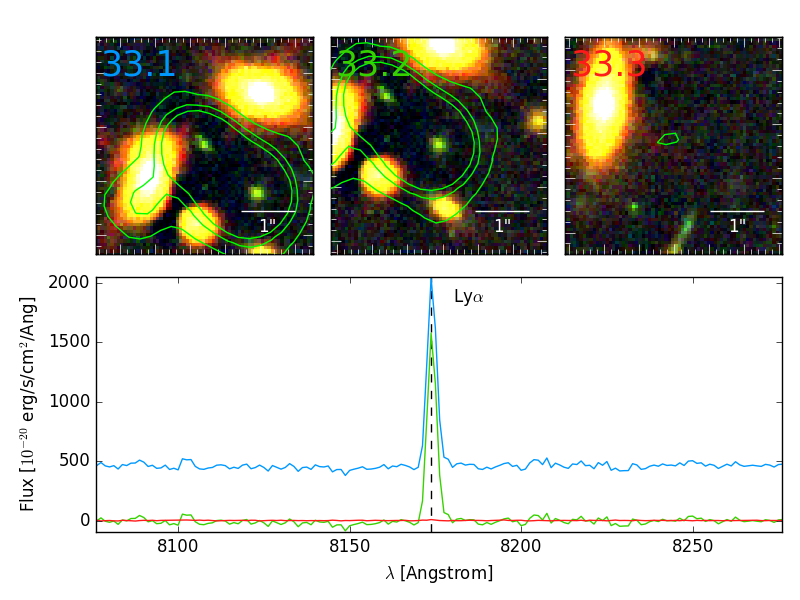} } 
\centerline{ 
\includegraphics[width=0.5\textwidth]{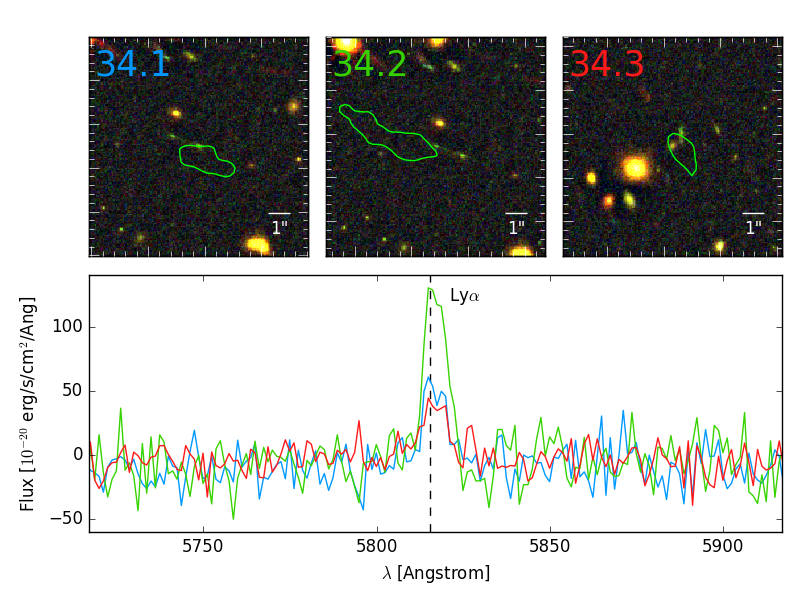} 
\includegraphics[width=0.5\textwidth]{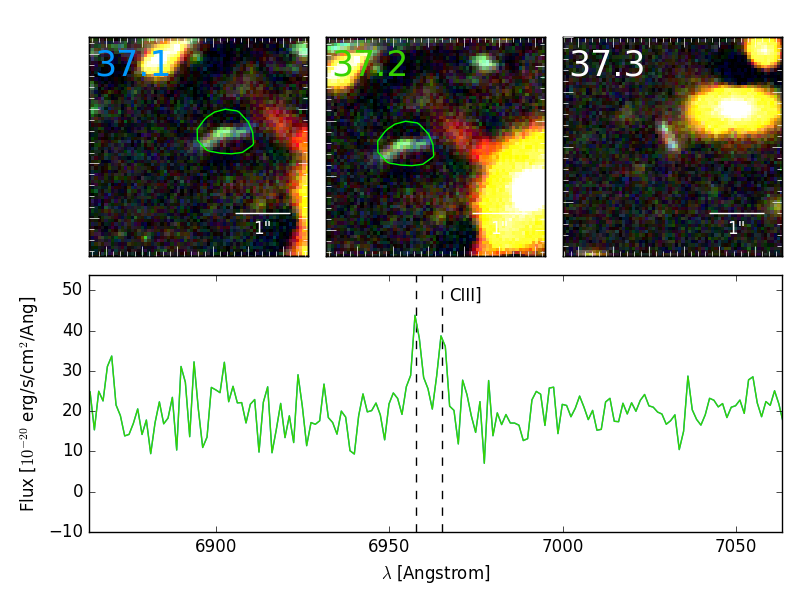} } 
\centerline{ 
\includegraphics[width=0.5\textwidth]{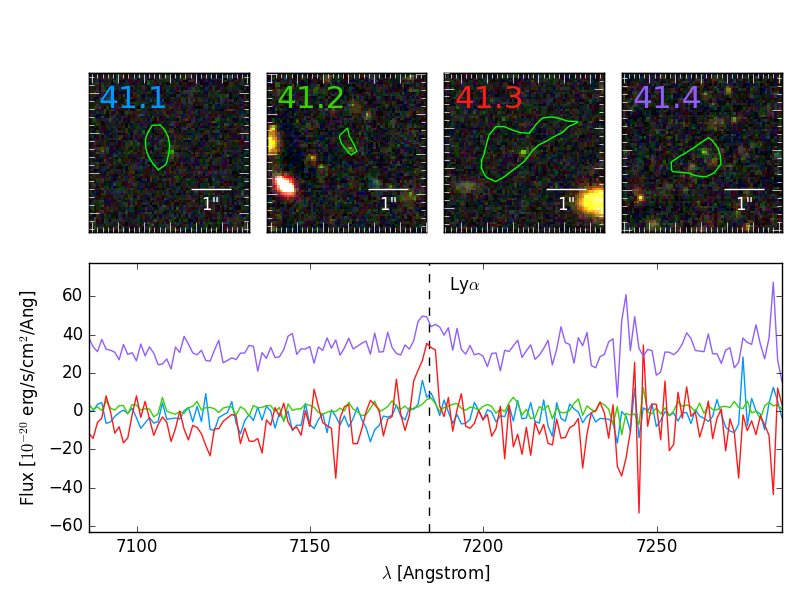} 
\includegraphics[width=0.5\textwidth]{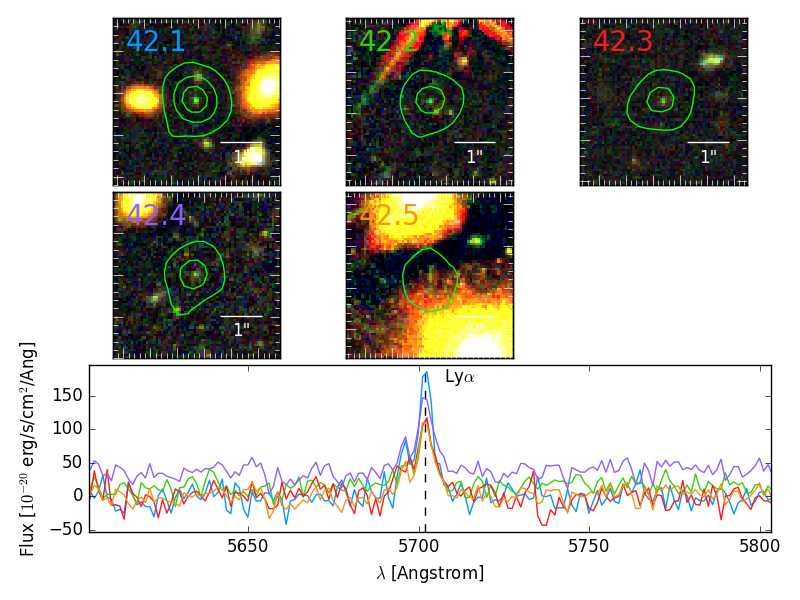} } 
\caption{(continued) Multiply-imaged systems. } 
\end{figure*} 
\begin{figure*}\ContinuedFloat 
\centerline{ 
\includegraphics[width=0.5\textwidth]{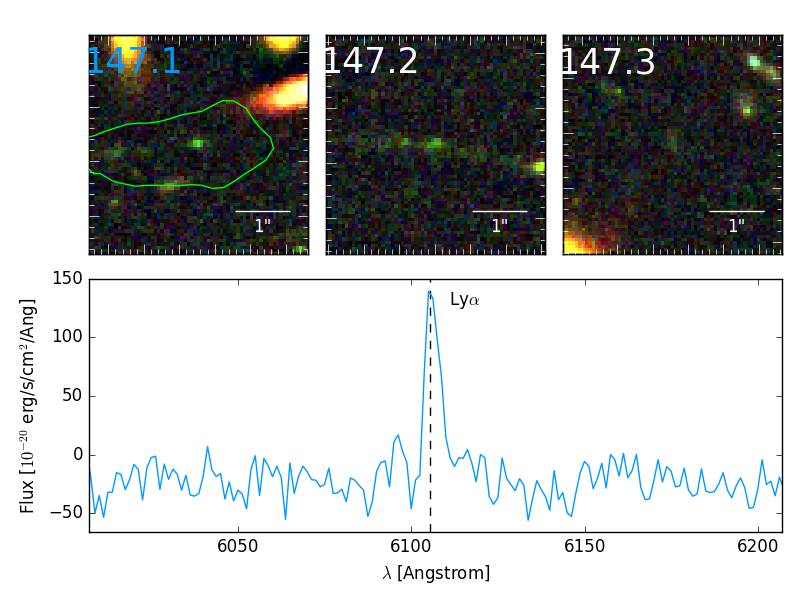}  
\includegraphics[width=0.5\textwidth]{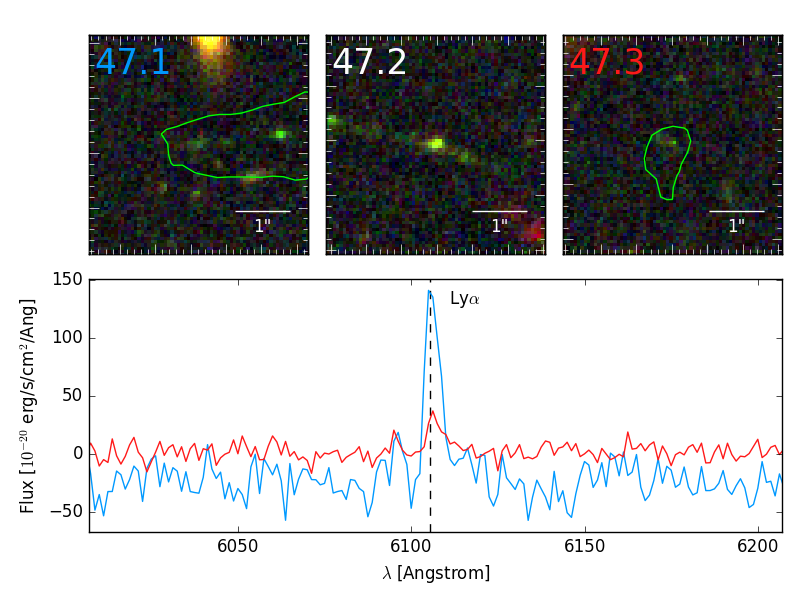} } 
\centerline{ 
\includegraphics[width=0.5\textwidth]{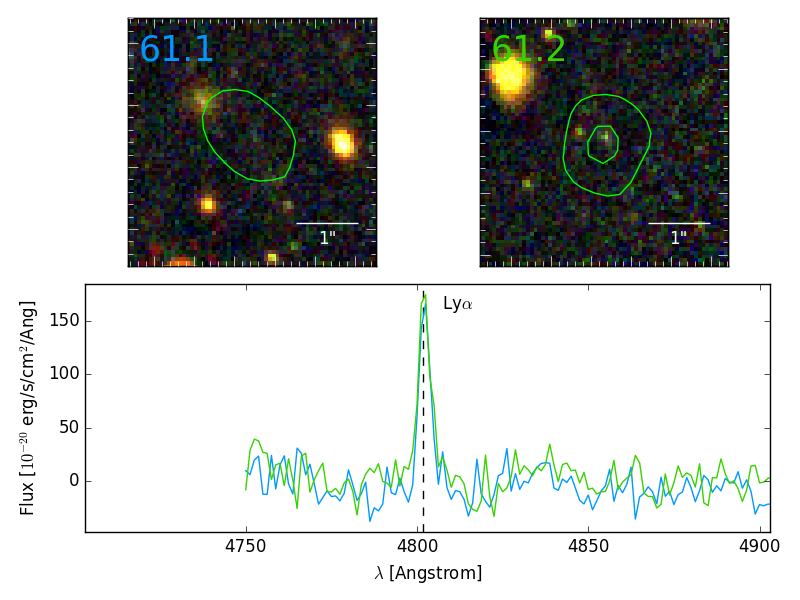} 
\includegraphics[width=0.5\textwidth]{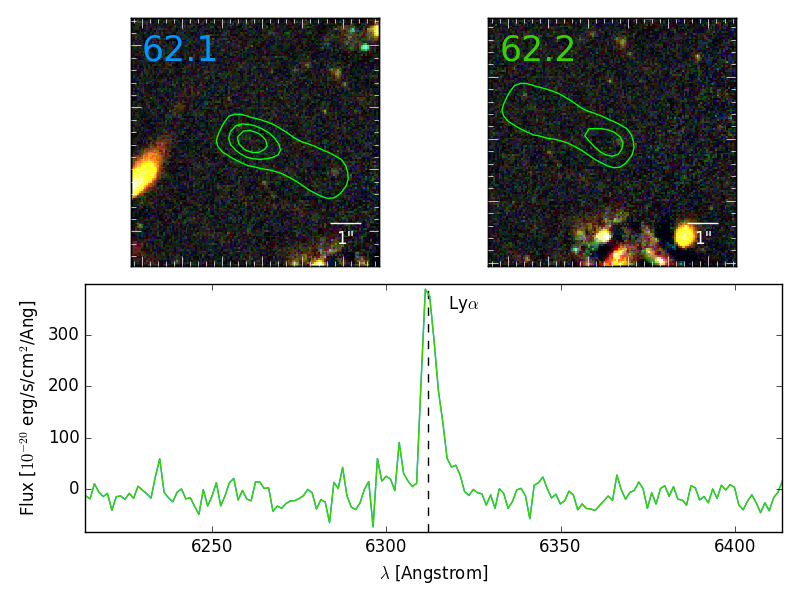} } 
\centerline{ 
\includegraphics[width=0.5\textwidth]{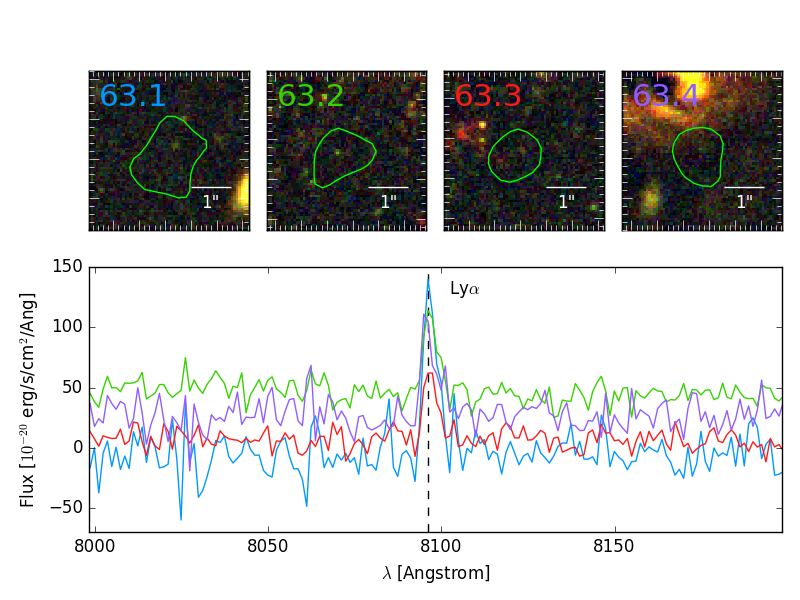} 
\includegraphics[width=0.5\textwidth]{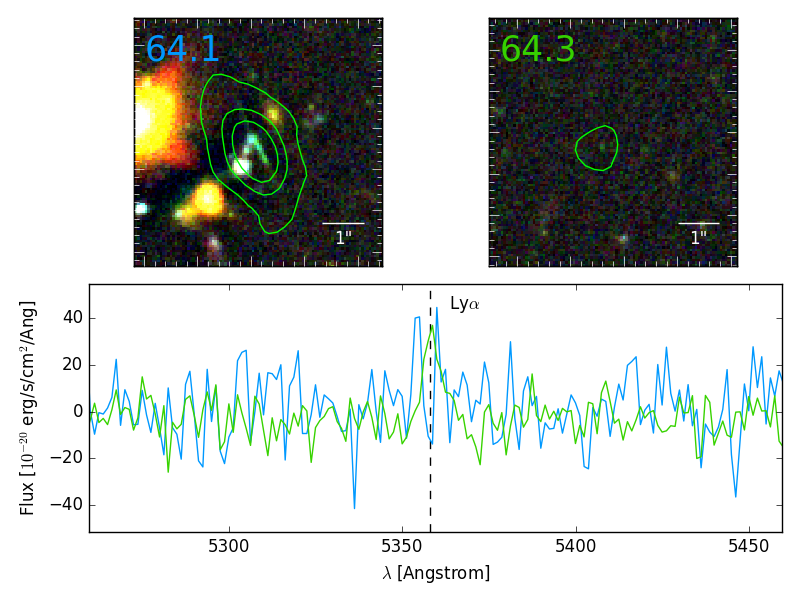} } 
\caption{(continued) Multiply-imaged systems. } 
\end{figure*} 
\begin{figure*}\ContinuedFloat 
\centerline{ 
\includegraphics[width=0.5\textwidth]{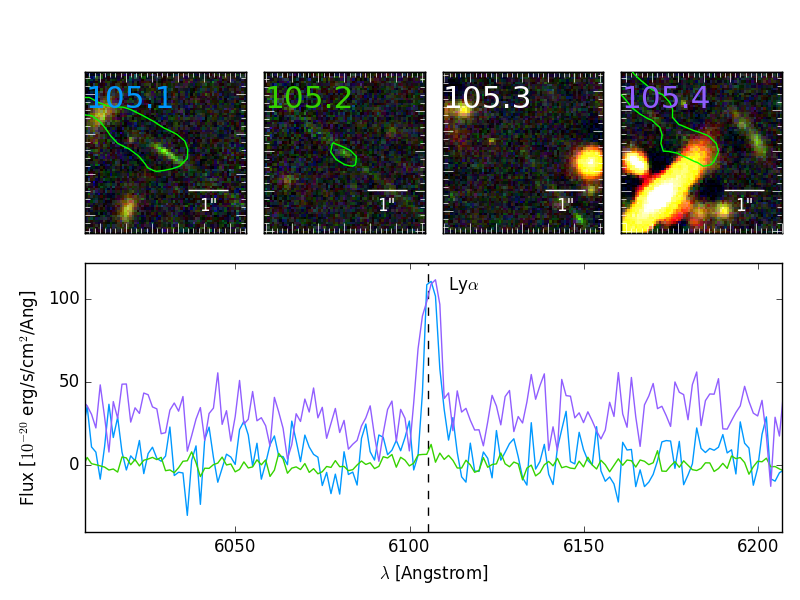} 
} 
\caption{(continued) Multiply-imaged systems. } 
\end{figure*}

%%%%%%%%%%%%%%%%%%%%%%%%%%%%%%%%%%%%%%%%%%%%%%%%%%

% Don't change these lines
\bsp	% typesetting comment
\label{lastpage}
\end{document}